\documentclass[12pt,aps,preprint,nofootinbib]{revtex4}
\pdfoutput=1
\usepackage{graphicx,amsmath}
\usepackage{amssymb}
\usepackage{amsfonts}
\usepackage{xcolor}
\usepackage{float}

\def\h{h^0}
\def\H{H^0}
\def\A{A^0}

\begin{document}

\title{Constraining Type II 2HDM in Light of LHC Higgs Searches}

\author{Baradhwaj Coleppa, Felix Kling, Shufang Su}
\email {baradhwa@email.arizona.edu, kling@email.arizona.edu, shufang@email.arizona.edu}

\affiliation {Department of Physics,
University of Arizona,
Tucson, AZ 85721, USA}

\begin{abstract}
We study the implication of the LHC Higgs search results on the Type II Two Higgs-Doublet Model.  In particular, we explore the scenarios in which the observed 126 GeV Higgs signal is interpreted as either the light CP-even Higgs $h^0$ or the heavy CP-even Higgs $H^0$.  Imposing both   theoretical and experimental constraints, we analyze the surviving parameter regions in $m_H$ ($m_h$), $m_A$, $m_{H^\pm}$, $\tan\beta$ and $\sin(\beta - \alpha)$.  
We further identify the regions that could accommodate a 126 GeV Higgs with cross sections consistent with the observed Higgs signal.
We find that in the $\h$-126 case,  we are restricted to narrow regions of $\sin(\beta-\alpha) \approx \pm 1$ with $\tan\beta$ up to 4, or   an extended region with $0.55 < \sin(\beta-\alpha) < 0.9$   and $1.5 < \tan\beta < 4$.   The values of  $m_H$, $m_A$ and $m_{H^\pm}$, however, are   relatively unconstrained.  In the $\H$-126 case, we are restricted to a narrow region of $\sin(\beta-\alpha) \sim 0$ with $\tan\beta$ up to about 8, or an extended region of $\sin(\beta-\alpha) $ between $-0.8$ to $-0.05$, 
 with   $\tan\beta$  extended to 30 or higher.  $m_A$ and $m_{H^\pm}$ are  nearly degenerate due to $\Delta\rho$ constraints.
   Imposing flavor constraints shrinks the surviving parameter space  significantly for the $\H$-126 case, limiting $\tan\beta \lesssim 10$, but has little effect in the $\h$-126 case.  We also investigate the correlation between $\gamma\gamma$, $VV$  and $bb/\tau\tau$ channels.  $\gamma\gamma$ and $VV$ channels are most likely to be highly correlated with $\gamma\gamma:VV \sim 1$ for the normalized cross sections.

\end{abstract}

\maketitle

\section{Introduction}

The discovery of a resonance at 126 GeV with properties consistent with the Standard Model (SM) Higgs boson in both the ATLAS \cite{aad:2012gk, ATLAS-CONF-2013-014} and CMS experiments \cite{Chatrchyan:2012ufa, CMS-PAS-HIG-13-005} is undoubtedly the most significant experimental triumph of the Large Hadron Collider (LHC) to date. The nature of this particle, as regards its CP properties and couplings, are currently being established \cite{CMS-PAS-HIG-13-005, ATLAS_spin_coupling,Aad:2013wqa}. Though further data would undoubtedly point us in the right direction, at this point it is useful to explore the implication of the current Higgs search results on models beyond the SM.  There are quite a few models that admit a scalar particle in their spectrum and many of them can have couplings and decays consistent with the SM Higgs boson. Thus it behooves us to constrain these models as much as possible with the Higgs search results at hand. 
 
One of the simplest extensions of the SM involves enlarged Higgs sectors. This can be done by simply adding more scalar doublets, or considering Higgs sectors with more complicated representations. In the work, we will study the Two Higgs-Doublet Models (2HDM) that  involve two scalar doublets both charged under the SM ${\rm SU}(2)_L\times {\rm U}(1)_Y$ gauge symmetries \cite{review,type1,hallwise,type2}.  The neutral components of both the Higgs fields develop vacuum expectation values (vev), breaking ${\rm SU}(2)_L\times {\rm U}(1)_Y$ down to ${\rm U}(1)_{\textrm{em}}$. Assuming no CP-violation in the Higgs sector, the resulting physical spectrum for the scalars is enlarged relative to the SM and includes light and heavy neutral CP-even Higgses ($h^0$ and  $H^0$), charged Higgses ($H^{\pm}$), and a pseudoscalar $A^0$.  In addition to the masses,   two additional parameters are introduced in the theory: the ratio of the vevs of the two Higgs fields 
($\tan\beta$), and the mixing of the two neutral CP-even Higgses ($\sin\alpha$). 

There are many types of 2HDMs,  each differing in the way the two Higgs doublets couple to the fermions (for a comprehensive review, see \cite{review}). In this work, we will be concentrating on the Type II case, in which one Higgs doublet couples to the up-type quarks, while the other Higgs doublet couples to the down-type quarks and leptons.  This model is of  particular interest  as it shares many of the features of the Higgs sector of the Minimal Supersymmetric Standard Model (MSSM). This enables us to translate existing LHC MSSM results to this case. Before proceeding, we point out that over the last few months, there have been various studies on the 2HDM based on the recent discovery \cite{Cheon:2012rh,Ferreira:2011aa,Ferreira:2012my,Drozd:2012vf,  Chang:2012ve, Chen:2013kt, Grinstein:2013npa, Chiang:2013ixa, Craig:2012vn, Basso:2012st, Ferreira:2012nv, Burdman:2011ki,Cervero:2012cx,Shu:2013uua}.   While  most studies concentrated on finding regions of parameter space that admit $\sigma \times$ Br 
values reported by the LHC experiments in  various channels, some also looked at correlations between the various decay channels.   The authors of  Ref.~\cite{Ferreira:2011aa} and Ref.~\cite{Ferreira:2012my} did the initial study of looking at  the $\tan\beta-\sin\alpha$ plane where the observed Higgs signal is feasible, interpreting the discovered scalar as either the light or the heavy  CP-even Higgs boson.  Ref.~\cite{Cheon:2012rh, Drozd:2012vf, Chang:2012ve, Chen:2013kt, Grinstein:2013npa, Chiang:2013ixa} 
fit the observed Higgs signals in various 2HDM scenarios, taken into account theoretical and experimental constraints.
Ref.~\cite{Craig:2012vn}  also paid careful attention to various Higgs production modes.   Ref.~\cite{Basso:2012st}  focused on the CP-violating Type II 2HDM.  Ref.~\cite{Ferreira:2012nv} studied the case of nearly degenerate Higgs bosons.  In addition,  Ref. \cite{Burdman:2011ki,Cervero:2012cx} investigated the possibility that the signal could correspond to the pseudoscalar $A^0$ - in this context, it is worth remarking that Ref.~\cite{Coleppa:2012eh} considered the pseudoscalar interpretation of the observed 126 GeV resonance and found that while it is strongly disfavored, the possibility is not yet ruled out at the $5 \sigma$ level.\footnote{The latest experimental results indicate that the pseudoscalar interpretation of the 126 GeV excess is disfavored \cite{CMS-PAS-HIG-13-005, ATLAS_spin_coupling}. }

In the present paper, we extended the above analyses by combining all the known experimental constraints (the LEP,  Tevatron and the LHC Higgs search bounds, and precision observables) with the theoretical ones (perturbativity, unitarity, and vacuum stability), as well as flavor constraints. A unique aspect of the present work is that our analysis looks at combinations of all parameters of the theory to identify regions that survive all the theoretical and experimental constraints. We further focus on regions that could accommodate the observed Higgs signal as either the light or the heavy CP-even Higgs, and are thus interesting from a collider study perspective. This enables us to draw conclusions about correlations between   different masses and   mixing angles to help identify aspects of the model that warrant future study. 

We start by briefly introducing the  structure and parameters of the Type II 2HDM   in Section~\ref{sec:model}.  In Sec.~\ref{sec:analyses}, we discuss the theoretical constraints and experimental bounds,  and outline our analysis methodology.  In Sec.~\ref{sec:h126}, we present our results for the light CP-even Higgs being the observed 126 GeV SM-like Higgs boson,  looking at surviving regions in various combinations of free parameters. In Sec.~\ref{sec:H126}, we do the same for the heavy CP-even Higgs as the observed 126 GeV SM-like Higgs boson.  In Sec.~\ref{sec:future}, we explore the implications for the  Vector Boson Fusion (VBF) or $VH$ associated production, and decays of Higgs into $bb$ and $\tau\tau$ channels. We conclude in Section~\ref{sec:conclusions}.

%%%%%%%%%%%%%%
\section{Type II 2HDM}
\label{sec:model}
%%%%%%%%%%%%%%
In this section, we   briefly describe the Type II 2HDM, focusing on the particle content, Higgs couplings, and model parameters.   For more details about the model, see   Ref. \cite{review} for a recent review of the theory and phenomenology of 2HDM.

\subsection{Potential, Masses and Mixing Angles}
\label{sub:massangle}

Labeling the two ${\rm SU(2)}_L$ doublet scalar fields $\Phi_1$ and $\Phi_2$, the most general potential for the Higgs sector can be written down in the following form:   
\begin{eqnarray} 
V(\Phi_1,\Phi_2)= \,&& m^2 _{11} \Phi^\dagger _1 \Phi_1 + m^2 _{22} \Phi^\dagger _2 \Phi_2 - m^2 _{12}( \Phi^\dagger _1 \Phi_2 + \textrm{h.c.}) \nonumber \\
&& + \frac{1}{2}\lambda_1 (\Phi^\dagger _1 \Phi_1)^2 + \frac{1}{2}\lambda_2 (\Phi^\dagger _2 \Phi_2 )^2 + \lambda_3 (\Phi^\dagger _1 \Phi_1) (\Phi^\dagger _2 \Phi_2)+ \lambda_4 (\Phi^\dagger_1 \Phi_2 ) (\Phi^\dagger _2 \Phi_1)  \nonumber \\
&&  + \frac{1}{2}\Big \{   \lambda_5(\Phi^\dagger _1 \Phi_2 )^2+ \textrm{h.c.} \Big \} +\Big \{\left[ \lambda_6(\Phi^\dagger_1 \Phi_1)+\lambda_7 (\Phi^\dagger_2 \Phi_2)  \right](\Phi^\dagger_1 \Phi_2 )+\textrm{h.c.} \Big \}.
 \label{eq:potential}
\end{eqnarray}
We   impose a discrete $Z_2$ symmetry on the Lagrangian, the effect of which is to render $m_{12},\lambda_6,\lambda_7=0$\footnote{Ref.~\cite{Drozd:2012vf}, which also addresses similar issues as in this paper, allowed for a soft breaking of the $Z_2$ symmetry with $m_{12}^2\neq 0$. In this paper, we don't consider such soft-breaking terms.}.    Note that one consequence of requiring $m_{12}=0$ is that there is no so called decoupling limit in which only one SM-like Higgs appears at low energy while all other Higgses are heavy and decoupled from the low energy spectrum.   After electroweak symmetry breaking (EWSB): $\langle \phi_1^0 \rangle = v_1/\sqrt{2}$,  $\langle \phi_2^0 \rangle = v_2/\sqrt{2}$ with $\sqrt{v_1^2+v_2^2}= $ 246 GeV, we are left with six free parameters, which can be chosen as  the four Higgs masses ($m_h$, $m_H$, $m_A$, $m_{H^{\pm}}$), a mixing angle $\sin\alpha$ between the two CP-even Higgses, and the ratio of the two vacuum expectation values, $\tan\beta=v_2/v_1$. 
 
 Writing the two Higgs fields as: 
\begin{equation}
\Phi_{i}=\begin{pmatrix} 
  \phi_i^{+}    \\ 
  (v_i+\phi^{0}_i+iG_i)/\sqrt{2}  
\end{pmatrix},
\end{equation}
the mass eigenstates of the physical scalars can be written as:   
\begin{equation}
\left(\begin{array}{c}
\H\\ \h
\end{array}
\right)
=\left(
\begin{array}{cc}
\cos\alpha &\sin\alpha\\
-\sin\alpha&\cos\alpha
\end{array}
\right)  \left(
\begin{array}{c}
\phi_1^0\\\phi_2^0
\end{array}
\right),\ \ \ 
\begin{array}{c}
 \A \\H^\pm
 \end{array}
 \begin{array}{l}
 =  -G_1\sin\beta+G_2\cos\beta\\
 =-\phi_1^{\pm}\sin\beta+\phi_2^{\pm} \cos\beta
 \end{array}.
 \end{equation}
 For our purposes, it is useful to express the  quartic couplings $\lambda_{1 \ldots 5}$ in terms of the physical Higgs masses, $\tan\beta$ and the mixing angle $\alpha$:
\begin{eqnarray}
\lambda_1= \frac{m_{H}^2\cos^2\alpha+m_{h}^2\sin^2\alpha}{v^2\cos^2\beta},\ \ \ 
&&\lambda_2= \frac{m_{H}^2\sin^2\alpha+m_{h}^2\cos^2\alpha}{v^2\cos^2\beta}  \label{eq:lambda12}\\
\lambda_3=\frac{\sin2\alpha(m_{H}^2-m_{h}^2)+2\,\sin2\beta\,m_{H^{\pm}}^2}{v^2\,\sin2\beta} 
  ,\ \ \  
&&\lambda_4=\frac{m_{A}^2-2 m_{H^{\pm}}^2}{v^2}  ,\ \ \ 
\lambda_5=-\frac{m_{A}^2}{v^2}.
\label{eq:lambda}
\end{eqnarray}
Imposing the   perturbativity and unitarity bounds, as explained below in Sec.~\ref{sub: constraints}, typically leads to an upper bound on the masses of $H^0$, $A^0$ and $H^\pm$.
 The couplings of the  CP-even Higgses and CP-odd Higgs to the SM gauge bosons and fermions are scaled by a factor $\xi$ relative to the SM   value -- these are presented in Table~\ref{tab:couplings}.
%%%%%%%%%%%%%%
\begin{table}
\begin{center}
  \begin{tabular}{| l | p{2.5cm} || l | p{2.5cm} || l | p{1.5cm}| }
    \hline
    $\xi^{VV}_{h}$ & $\sin(\beta-\alpha)$ & $\xi^{VV}_{H}$ &$\cos(\beta-\alpha)$& $\xi^{VV}_{A}$ &$0$ \\ \hline
    $\xi^{u}_{h}$ & $\cos\alpha/\sin\beta$ & $\xi^{u}_{H}$ & $\sin\alpha/\sin\beta$& $\xi^{u}_{A}$ & $\cot\beta$\\ \hline
    $\xi^{d,l}_{h}$ & $-\sin\alpha/\cos\beta$ & $\xi^{d,l}_{H}$ & $\cos\alpha/\cos\beta$& $\xi^{d,l}_{A}$ & $\tan\beta$\\ \hline
  \end{tabular}
\end{center}
\caption{The multiplicative factor $\xi$ by which the couplings of the CP-even Higgses and the CP-odd Higgs  to the gauge bosons and fermions scale with respect to the SM value.  The superscripts $u,d,l$ and $VV$ refer to the up-type quarks, down-type quarks, leptons, and $WW/ZZ$ respectively. }
\label{tab:couplings}
\end{table}
%%%%%%%%%%%
In order to translate the ATLAS and CMS limits, we need to pay particular attention to the couplings of the light (heavy) CP-even Higgs to the SM gauge bosons (controlling the partial decay width to $WW$, $ZZ$ as well as $\gamma\gamma$ channels) and to the top quark (controlling the gluon fusion production cross section), as well as to the bottom quark (controlling the $bb$ partial decay width, which enters   the total decay width as well).  From Table~\ref{tab:couplings}, we see that the relevant couplings are proportional to $\sin(\beta-\alpha)$ ($\cos(\beta-\alpha)$),  $1/\sin\beta$ and $1/\cos\beta$. Thus, even though it is customary to look at the combination of parameters $(\sin\alpha, \tan\beta)$, we present our results in Sec.~\ref{sec:h126}  and \ref{sec:H126} using $\sin(\beta-\alpha)$ and $\tan\beta$ as the independent parameters (in addition to the masses of the physical Higgses) to manifest the effects on the Higgs couplings to gauge bosons.   Using $\sin(\beta-\alpha)$ instead of $\sin\alpha$ 
has the additional advantage of being basis-independent, as explained in Ref.~\cite{Davidson:2005cw,Ginzburg:2004vp}.

%%%%%%%%%%%%%%%%%%%%%%%%%%%%
\section{Constraints and analyses}
\label{sec:analyses}

\subsection{Theoretical and Experimental Constraints}
\label{sub: constraints}
%%%%%%%%%%%%%%%%%%%%%%%%%%%%
To implement the various experimental and theoretical constraints, we have employed two   programs: the 2HDM Calculator (2HDMC) \cite{Eriksson:2009ws}  to calculate the Higgs couplings,   compute all the decay branching fractions of the Higgses,   and implement all the theoretical constraints;  and HiggsBounds 3.8 \cite{HiggsBounds} to consistently put in all the experimental constraints on the model. Here, we briefly describe the list of theoretical and experimental bounds that are of interest.

\textbf{Theoretical Constraints}:  
\begin{itemize}
 \item Vacuum Stability: This   implies that the potential should be bounded from below, which is translated to various conditions for  the quartic couplings in the Higgs potential \cite{Deshpande:1977rw}: $\lambda_1>0$, $\lambda_2>0$, $\lambda_3>-\sqrt{\lambda_1 \lambda_2}$, and $\lambda_3+\lambda_4-|\lambda_5|>-\sqrt{\lambda_1 \lambda_2}$.  With Eqs.~(\ref{eq:lambda12}) and (\ref{eq:lambda}), the above requirements   serve to constrain the Higgs masses and  angles.
 
 \item Perturbativity: 2HDMC imposes constraints on the physical Higgs quartic couplings, specifically demanding that $\lambda_{h_ih_jh_kh_l}<4\pi$ to  stay inside the perturbative regime. Note that even though these are different from the $\lambda$s in the Higgs potential in Eq.~(\ref{eq:potential}), we can still use Eqs.~(\ref{eq:lambda12}) and (\ref{eq:lambda}) as  rough guides to understand the perturbative bounds, as we will do in later sections to   explain the features of our results. The top yukawa coupling $y_t$ could also become nonperturbative for very small $\tan\beta$.  We require the perturbativity of $y_t$ at scales below 1 TeV, which results in  $\tan\beta \gtrsim 0.35$ \cite{Bijnens:2011gd}.
 
 \item Unitarity:  It is well known that in the SM, the scattering cross section for the longitudinal $W$ modes is unitary only if the Higgs exchange diagrams are included. Since the couplings of the Higgs are modified in the 2HDM, we need to ensure unitarity by demanding that the $S$ matrix of \emph{all} scattering cross sections of Higgs$-$Higgs and Higgs$-V_L$ (where $V_L$ is either $W_L$ or $Z_L$) have eigenvalues bounded by $16\pi$ \cite{Ginzburg:2005dt}. 
 
\end{itemize} 

\textbf{Experimental Constraints}: 
The LHC experiments have searched for the SM Higgs in   $\gamma\gamma$, $ZZ$, $WW$, $\tau\tau$ and $bb$ channels.   Both the ATLAS and CMS collaboration have reported the observation of a new resonance at a mass of around 126 GeV with more than 5$\sigma$ significance \cite{aad:2012gk, Chatrchyan:2012ufa, ATLAS-CONF-2013-014, CMS-PAS-HIG-13-005, ATLAS_spin_coupling, Aad:2013wqa, ATLAS_moriond2013_channel, CMS_moriond2013_channel}.
The production cross sections and partial decay widths of the 2HDM Higgses to the various SM final states differ from that of the SM Higgs,  which can be obtained using  the coupling scaling factors  $\xi$   from Table~\ref{tab:couplings}.  Thus, we can   identify the regions in parameter space where the signal cross sections are compatible with the Higgs signal observed at the ATLAS and CMS collaborations.  We can also translate the exclusion bounds on the Higgs search to the ones in the 2HDM.  We used HiggsBounds 3.8  to impose the exclusion limits from Higgs searches at the LEP and the Tevatron \cite{lep98, lep98b, lepcharged0, lepcharged1,Group:2012zca}.  We also incorporated the latest   Higgs search results at the LHC \cite{ATLAS-CONF-2013-014, CMS-PAS-HIG-13-005, ATLAS_moriond2013_channel, CMS_moriond2013_channel, ATLAS_MSSM, CMS_MSSM}. 
 
$Z$-pole precision observables, in particular, the oblique parameters $S$, $T$ (or equivalently, $\Delta \rho$, which is the deviation of $\rho\equiv \frac{m_W^2}{m_Z^2\cos^2\theta_W}$ from the SM value), and $U$ \cite{Peskin:1990zt} constrain any new physics model that couples to  the $W$ and $Z$.  In particular, $T$  imposes a   strong constraint on the amount of custodial symmetry breaking in the new physics sector.  In the case of 2HDM, the mass difference between   the various Higgses are therefore highly constrained \cite{Hunter}, which   leads to interesting correlations between some of the masses, as will be demonstrated in Sec.~\ref{sec:h126} and Sec.~\ref{sec:H126}.   In our analysis, we require the contribution from extra Higgses to $S$ and $T$ to fall within the 90\% C.L. $S-T$ contour, for a SM Higgs reference mass of 126 GeV \cite{Nakamura:2010zzi}. In addition, the charged Higgs contributes to $Zbb$ coupling \cite{Logan:1999if}, which has been measured precisely at the LEP via the observable $R_b=\Gamma(Z\rightarrow b\bar{b})/\Gamma(Z\rightarrow{\rm hadrons})$ \cite{ALEPH:2010aa}.  Imposing bounds from $R_b$ rules out small $\tan\beta$ regions for a light charged Higgs.

\begin{figure}[h!]
  \centering
	\includegraphics[scale=0.4]{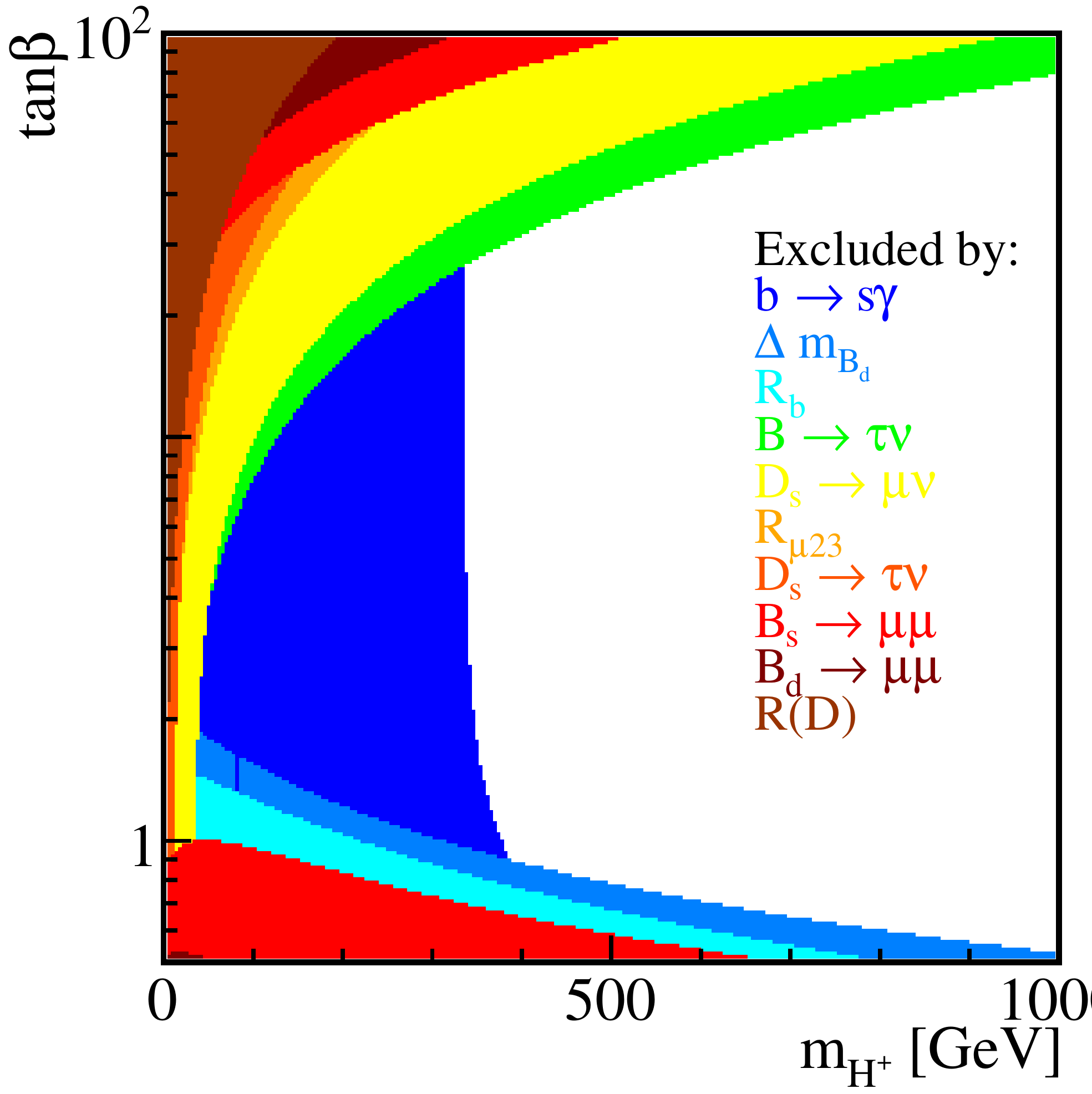}
	\includegraphics[scale=0.4]{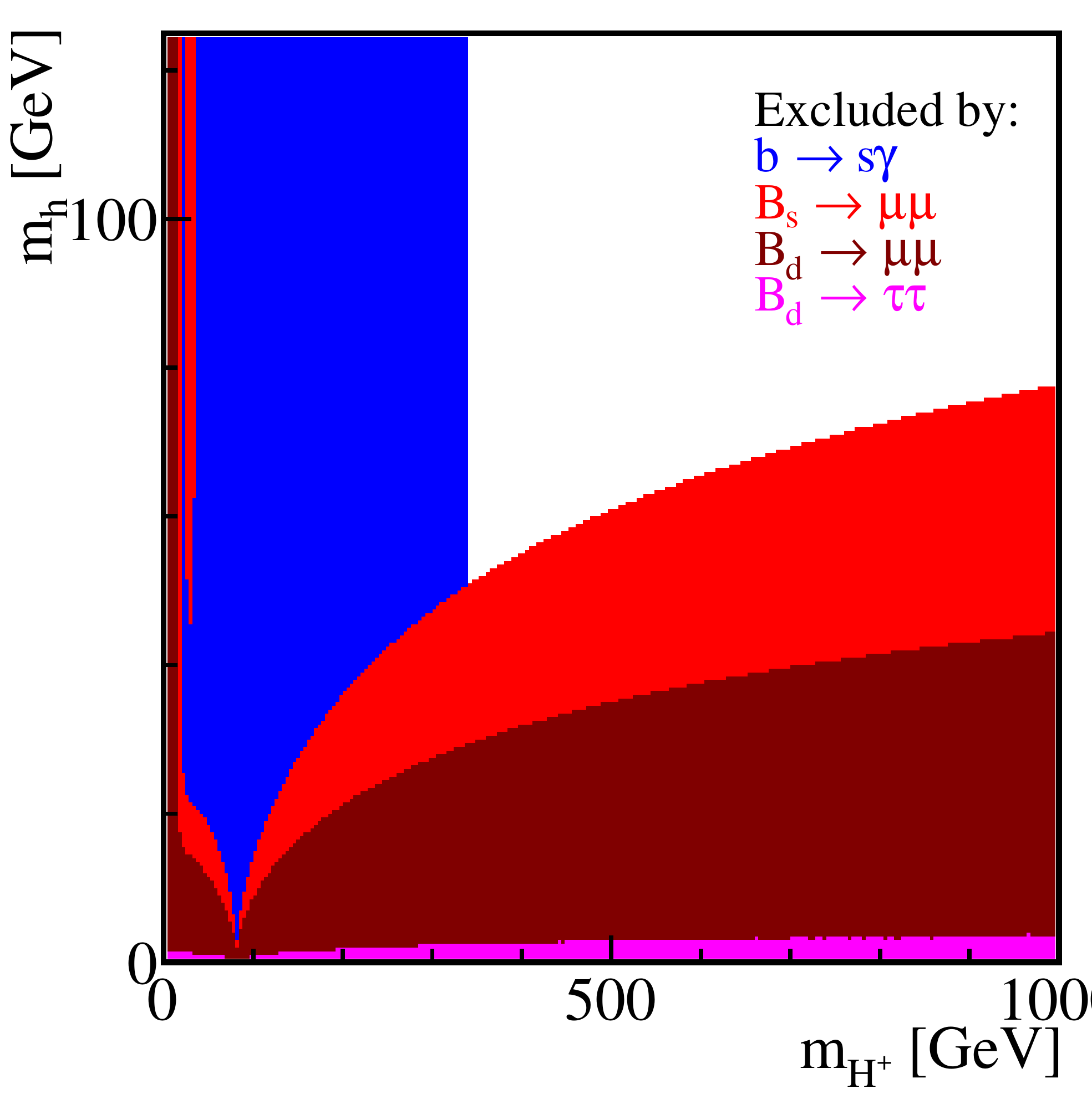}
  \caption{Regions of parameter space excluded by various flavor constraints. The left plot shows the  $m_{H^{\pm}}$ versus $\tan\beta$ plane for fixed $m_{h}=125$ GeV, $m_H=400$ GeV, $m_A$ = 200 GeV and $\sin (\beta - \alpha) = -0.1$. The right plot shows the $m_{H^{\pm}}$ versus $m_h$ plane for $m_{A}=m_{H^{\pm}}$, $m_H=125$ GeV, $\tan \beta = 5$ and $\sin (\beta - \alpha) = -0.01$. }
  \label{fig:Flav_new}
  \end{figure}

We  also show the effect on the available parameter spaces once bounds from flavor sector are imposed in addition to the ones described. To do this, we employed the program SuperIso 3.3 \cite{Mahmoudi:2008tp}, which incorporates, among other things, bounds   from $B\to X_s\gamma,\, \Delta M_{B_d},$   $B^- \rightarrow \tau^- \bar\nu_\tau$, $D_s^\pm\rightarrow \tau^\pm(\mu^\pm)\nu$, $B\rightarrow \tau^+\tau^-$ and  $B_{d,s} \to \mu^+ \mu^-$  \cite{PDG2013,Amhis:2012bh,Adachi:2012mm,delAmoSanchez:2010jg,Aaij:2013aka,Aubert:2005qw}. 
 A summary of flavor bounds can be found in Ref.~\cite{Mahmoudi:2009zx}.  We have used the latest bounds either from PDG  \cite{PDG2013} or from individual experiment.
To show the impact of the flavor constraints on the  2HDM parameter space, in Fig.~\ref{fig:Flav_new}, we present the regions excluded by various flavor constraints in the $m_{H^{\pm}}$ versus $\tan \beta$ plane (left panel) and the  $m_{H^{\pm}}$ versus $m_h$ plane (right panel).  While $B\rightarrow X_s \gamma$ excludes $m_{H^\pm}$ up to about 300  GeV for all $\tan\beta$, $B^- \rightarrow \tau^- \bar\nu_\tau$ and $\Delta M_{B_d}$ provide the strongest constraints at large and small $\tan\beta$, respectively.  The strongest bound on the neutral Higgs mass comes from $B_{s} \to \mu^+ \mu^-$, which excludes $m_h$ at about 50 GeV or lower.

In addition, we included the latest results from BaBar on  $\bar{B}\rightarrow D \tau \bar\nu_\tau$ and $\bar{B}\rightarrow D^* \tau \bar\nu_\tau$ \cite{Lees:2012xj}, which observed   excesses over the SM prediction at about 2 $\sigma$ level.  We treat the observed excesses as  upper bounds and take the 95\% C.L. range as $R(D)<0.58$ and $R(D^*)<0.39$.  Note that as pointed out in Ref.~\cite{Lees:2012xj}, the excesses in both $R(D)$ and $R(D^*)$ can not be simultaneously explained by the Type II 2HDM \cite{Tanaka:2010se,Barger:1989fj}.  Other new physics contributions have to enter if the excesses in both $R(D)$ and $R(D^*)$ stay in the future.
 Flavor constraints on the Higgs sector are, however, typically more model-dependent.   Therefore, our focus in this work is mainly on the implication of the Higgs search results on the Type II 2HDM, and we only impose the flavor bounds at the last step to indicate how the surviving regions further shrink.

\subsection{Analysis Method}
\label{sub:methods}

In our analysis, we considered two scenarios:
\begin{itemize}
\item $\h$-126 case where $m_h=126$ GeV with $m_H>126$ GeV,
\item $\H$-126 case  where $m_H=126$ GeV with $m_h<126$ GeV
\end{itemize} 
and scanned  over the entire remaining parameter space varying $m_H$ (or $m_h$), $m_A, m_{H^\pm}$, $\tan \beta$ and $\sin (\beta-\alpha)$:  
\begin{eqnarray}
  20 {\rm\  GeV} \leq &\ \ m_A,m_{H^\pm}\ \ & \leq 900  {\rm\  GeV}   \hspace{0.5 in} {\rm\ in\  steps\  of\  20\ GeV},
 \label{eq:range1}\\
-1 \leq& \sin(\beta-\alpha)&\leq 1   \hspace{1.05 in} {\rm\ in \ steps \ of  \ 0.05},
\label{eq:range2}\\
{\bf \h-126\ case}: 0.25 \leq &\tan \beta& \leq 5  \hspace{1.05 in} {\rm \ in\  steps\  of\  0.25, }
\label{eq:range3}\\
126 {\rm \ GeV} \leq &m_H& \leq 900 {\rm\  GeV}   \hspace{0.5 in} {\rm \ in\  steps\  of\  20\ GeV,}
\label{eq:range4}\\
{\bf \H-126\ case}: 1 \leq &\tan \beta& \leq 30   \hspace{0.95 in} {\rm \  in\  steps\  of\  1,}  
\label{eq:range5}\\
  6 {\rm \ GeV} \leq &m_h&< 126  {\rm\  GeV}   \hspace{0.5 in}  {\rm \ in\  steps\  of\  5\  GeV}.  \label{eq:range6}
\end{eqnarray}
In certain regions in which very few points are left after all the constraints are imposed, we generated more points with smaller steps. We used  the 2HDMC 1.2beta \cite{Eriksson:2009ws} which   tested if each parameter point fulfills the theoretical and experimental constraints implemented in HiggsBounds 3.8 \cite{HiggsBounds}. New LHC results that are not included in HiggsBounds 3.8 were implemented in addition.   In particular, the CMS results on MSSM Higgs search in $\tau\tau$ channel \cite{CMS_MSSM} were imposed using the cross section limits reverse-engineered from bounds in $m_{A}-\tan\beta$ plane for $m_h^{\rm max}$ scenario, as provided in HiggsBounds 4.0 \cite{HiggsBounds}.  We also required each parameter point to satisfy the precision constraints, in particular, $S$ and $T$, as well as $R_b$.

We further required either $\h$ or $\H$ to satisfy the dominant gluon fusion cross section requirement for $\gamma\gamma$, $WW$ and $ZZ$   channels to accommodate the observed Higgs signal at 95\% C.L. \cite{CMS-PAS-HIG-13-005,Aad:2013wqa}: 
\begin{equation}
0.7<\frac{\sigma   (gg \rightarrow \h/\H \rightarrow \gamma\gamma)}{\sigma_{\textrm{SM}}}<1.5, \ \ \ 
0.6<\frac{\sigma   (gg \rightarrow  \h/\H \rightarrow WW/ZZ  )}{\sigma_{\textrm{SM}}}<1.3,
\label{eq:sigmabrrange}
\end{equation} 
in which we have taken the tighter limits from the ATLAS and CMS results, as well as the tighter results for the $WW$ and $ZZ$ channel.
In the last step, we imposed the flavor bounds on all   points that satisfy Eq.~(\ref{eq:sigmabrrange}) using the SuperIso 3.3 program to study  the consequence of the flavor constraints.
   
%%%%%%%%%%%%%%%%%%%
\section{Light Higgs at 126 GeV}
\label{sec:h126}

%%%%%%%%%%%%%%%%%%%
\subsection{Cross sections and Correlations}
%%%%%%%%%%%%%%%%%%%%%%%%%%
Before presenting the results of the numerical scanning of parameter regions with all the theoretical and experimental constraints imposed, let us  first study the $\tan\beta$ and $\sin(\beta-\alpha)$ dependence of the cross sections for the   major search channels at the LHC: $gg \rightarrow {\h} \rightarrow \gamma\gamma, WW/ZZ$.   Both production cross sections and decay  branching fractions are modified relative to the SM values:
\begin{equation}
\frac{\sigma\times {\rm Br} (gg\rightarrow \h\rightarrow XX)}{ {\rm SM}}= \frac{\sigma(gg\rightarrow \h)}{\sigma_{\rm SM}} \times
 \frac{\textrm{Br}(\h\rightarrow XX)}{\textrm{Br}(h_{\rm SM}\rightarrow XX)},
 \label{eq:sigmabr}
\end{equation}
for $XX=\gamma\gamma, VV$.   Note that since the $WW$ and $ZZ$ couplings are modified the same way in the Type II 2HDM, we use $VV$ to denote both $WW$ and $ZZ$ channels.

The ratio of the gluon fusion cross section normalized to the SM value can be written
as:
\begin{eqnarray}
&&\frac{\sigma(gg\rightarrow \h)}{\sigma_{\rm SM}}=\frac{\cos^2\alpha}{\sin^2\beta}+\frac{\sin^2\alpha}{\cos^2\beta}\frac{|A_{1/2}(\tau_b)|^2}{|A_{1/2}(\tau_t)|^2} 
\label{eq:sigma_ggh}
\\
&&=\left[  \frac{\cos(\beta-\alpha)}{\tan\beta} +\sin(\beta-\alpha) \right]^2 + \left[  {\cos(\beta-\alpha)}{\tan\beta} -\sin(\beta-\alpha) \right]^2\frac{|A_{1/2}(\tau_b)|^2}{|A_{1/2}(\tau_t)|^2}.
\label{eq:sigma_ggh_detail}
\end{eqnarray}
The expression for the fermion loop functions $A_{1/2}(\tau_{t,b})$ can be found in Ref.~\cite{Hunter}. The first term in  Eq.~(\ref{eq:sigma_ggh}) is the top-loop contribution, and the second term is  the bottom-loop contribution. In the SM,   the top-loop contributes dominantly to the gluon fusion diagram, while the bottom-loop contribution is negligibly small.  The situation alters in type II 2HDM for large $\tan\beta$, when  the bottom-loop contribution can be substantial due to the enhanced bottom Yukawa \cite{Ferreira:2011aa}.   
We also rewrite it in $\sin(\beta-\alpha)$, $\cos(\beta-\alpha)$ and $\tan\beta$ in Eq.~(\ref{eq:sigma_ggh_detail}) to make their dependence explicit.

\begin{figure}[htbp]
	\includegraphics[scale=0.4]{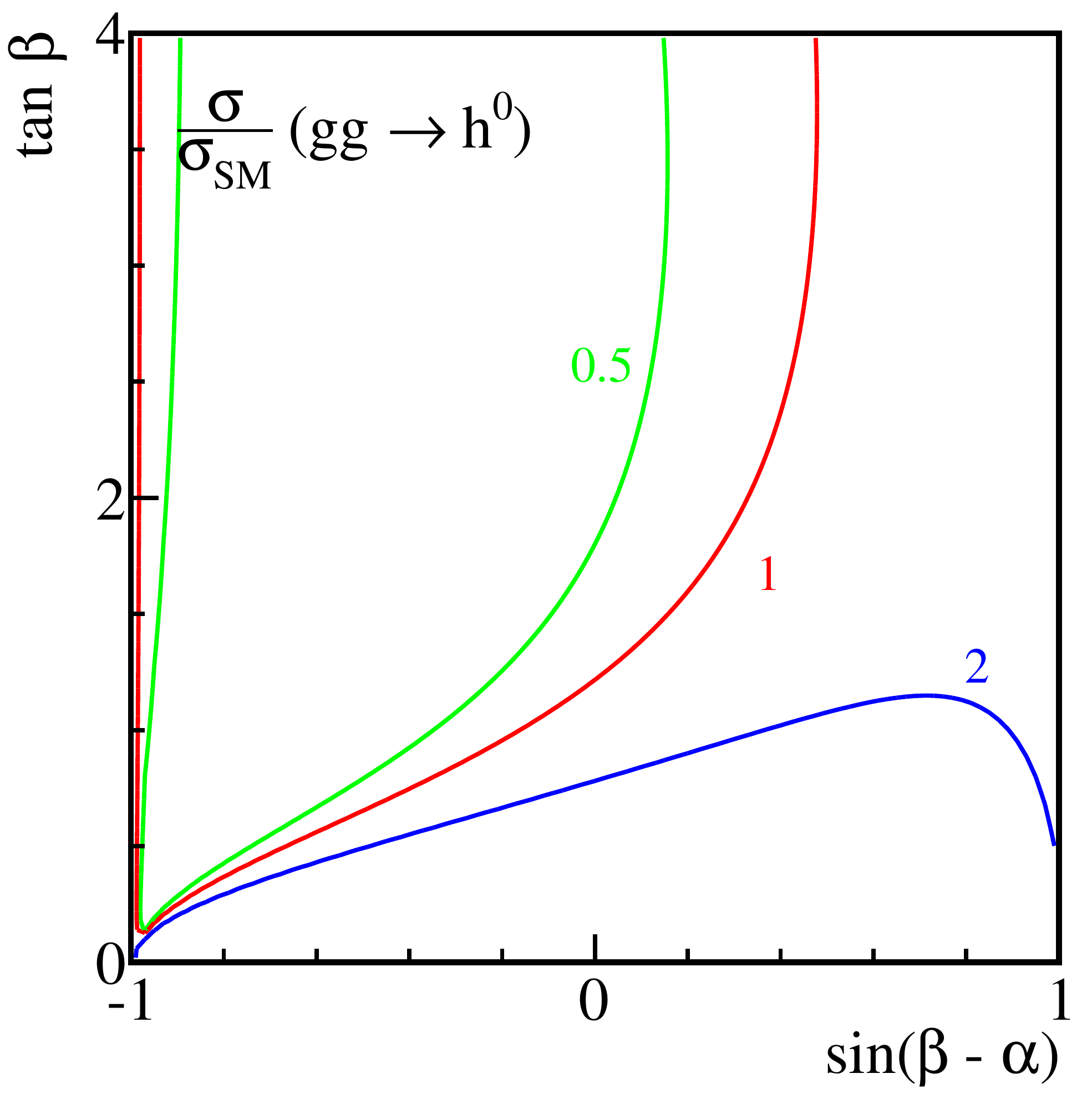}
	\includegraphics[scale=0.4]{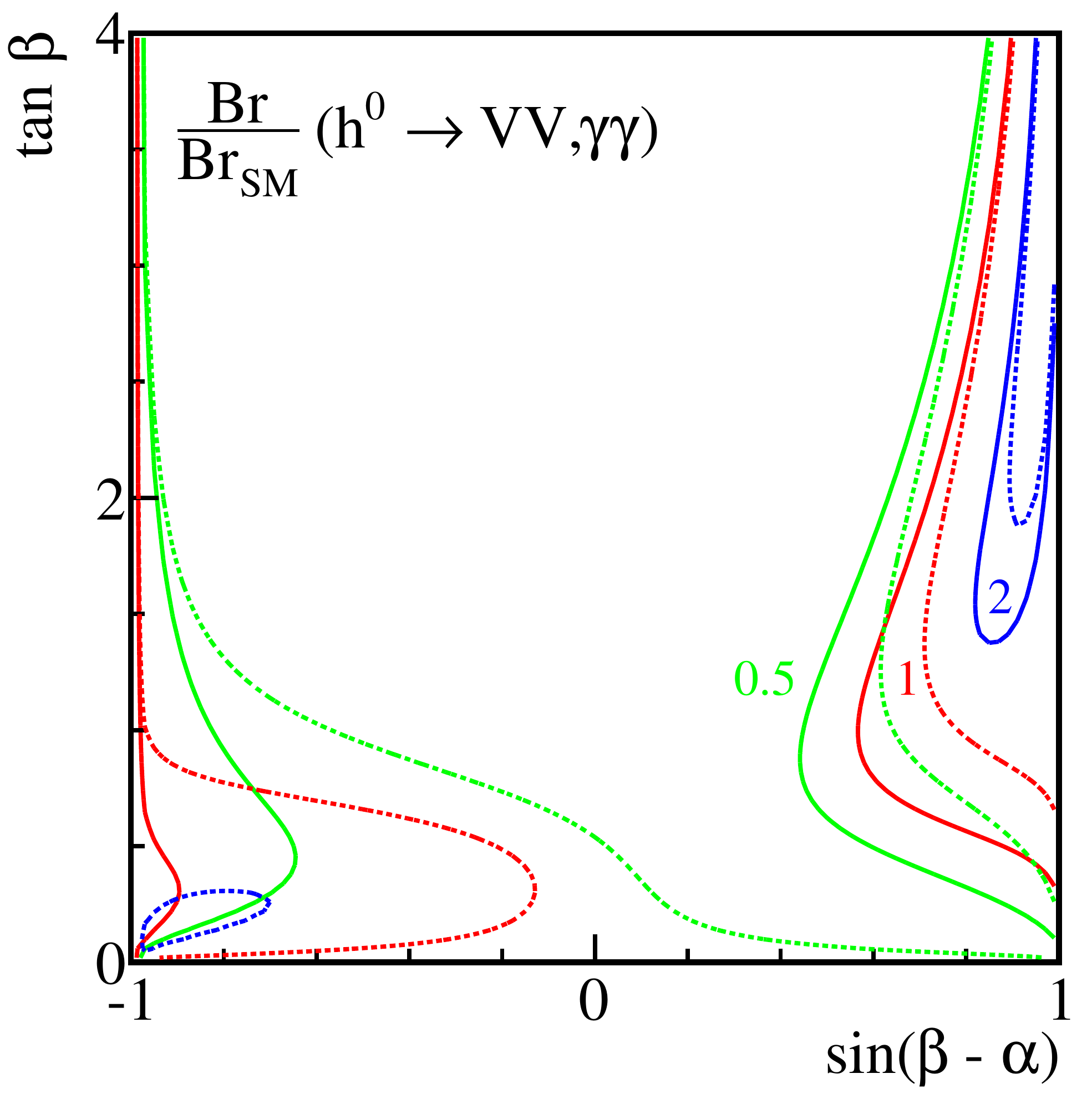}
\caption{The normalized $gg\to \h$ production cross section contours (left panel) and $\h\to VV$ (solid lines of the right panel) and   $\h\to \gamma\gamma$  (dashed lines of the right panel) branching fractions in the  $\h$-126 case.   The contour lines are $\sigma / \sigma_{\rm SM},\ {\rm Br}/{\rm Br}_{\rm SM}  = 0.5$ (green), 1 (red), and 2 (blue).  }
\label{fig:sigma_Br_sep}
\end{figure}

In the left panel of Fig.~\ref{fig:sigma_Br_sep}, we show contours of $\sigma/\sigma_{\rm SM}$ for the gluon fusion: 
$\sigma / \sigma_{\rm SM} = 0.5$ (green), 1 (red), and 2 (blue).   While contours of $\sigma / \sigma_{\rm SM} \geq 1$ accumulate in $\sin(\beta-\alpha)\sim -1$ region,  there is a wide spread of the contours for $\sin(\beta-\alpha) > 0$.  For most regions of 
$\sin(\beta-\alpha) < 0$, $gg\rightarrow h^0$ is suppressed compared to the SM value due to cancellations between the $\cos(\beta-\alpha)$ and $\sin(\beta-\alpha)$ terms in the top Yukawa coupling, as shown in Eq.~(\ref{eq:sigma_ggh_detail}).
 Note that we have shown the plots only for $\tan\beta\leq$ 4 since the  model is perturbatively valid only for $\tan\beta\lesssim$ 4, as  will be demonstrated below in the results of the full analysis.  

The $\h$ decay branching fractions $\h\rightarrow VV, \gamma\gamma$ can be written approximately as 
\begin{equation}
\frac{\textrm{Br}(\h\rightarrow XX)}{\textrm{Br}(h_{\rm SM}\rightarrow XX)}= \frac{\Gamma_{XX}}{\Gamma_{total}}\times \frac{\Gamma_{total}^{\rm SM}}{\Gamma_{XX}^{\rm SM}}  
\approx\left\{
\begin{tabular}{c}
$\frac{\sin^2(\beta-\alpha)}
{\sin^2(\beta-\alpha){\rm Br}(h_{\rm SM}\rightarrow VV)
+\frac{\sin^2\alpha}{\cos^2\beta}{\rm Br}(h_{\rm SM}\rightarrow bb)+\ldots}$\\
$\frac{\Gamma(\h\rightarrow \gamma\gamma)/\Gamma(h_{\rm SM}\rightarrow \gamma\gamma)}{\sin^2(\beta-\alpha){\rm Br}(h_{\rm SM}\rightarrow VV)+\frac{\sin^2\alpha}{\cos^2\beta}{\rm Br}(h_{\rm SM}\rightarrow bb)+\ldots} $
 \end{tabular}
 \right. , 
 \label{eq:BR}
\end{equation}
where we have explicitly listed the dominant $bb$ and $WW/ZZ$ channels and used ``$+\ldots$" to indicate other sub-dominant SM Higgs decay channels.    

In the right panel of Fig.~\ref{fig:sigma_Br_sep}, we show contours of ${\rm Br}/{\rm Br}_{\rm SM}$ for  
$VV$ (solid lines) and   $\gamma\gamma$ (dashed lines) channels.  Both $VV$ and loop induced (dominantly $W$-loop) $\gamma\gamma$ channels exhibit similar parameter dependence on $\tan\beta$ and $\sin(\beta-\alpha)$ since both channels are dominantly controlled by the same $\h VV$ coupling.
While contours of  ${\rm Br}/{\rm Br}_{\rm SM}\gtrsim 1$  appear near $\sin(\beta-\alpha)\sim \pm 1$ for unsuppressed $\h VV$ couplings, $\h\rightarrow \gamma\gamma$ shows some  spread for negative $\sin(\beta-\alpha)$ and small $\tan\beta$ due to the correction to top Yukawa in  the loop-indued $\h\gamma\gamma$ coupling.

 \begin{figure}[htbp]
	\includegraphics[scale=0.4]{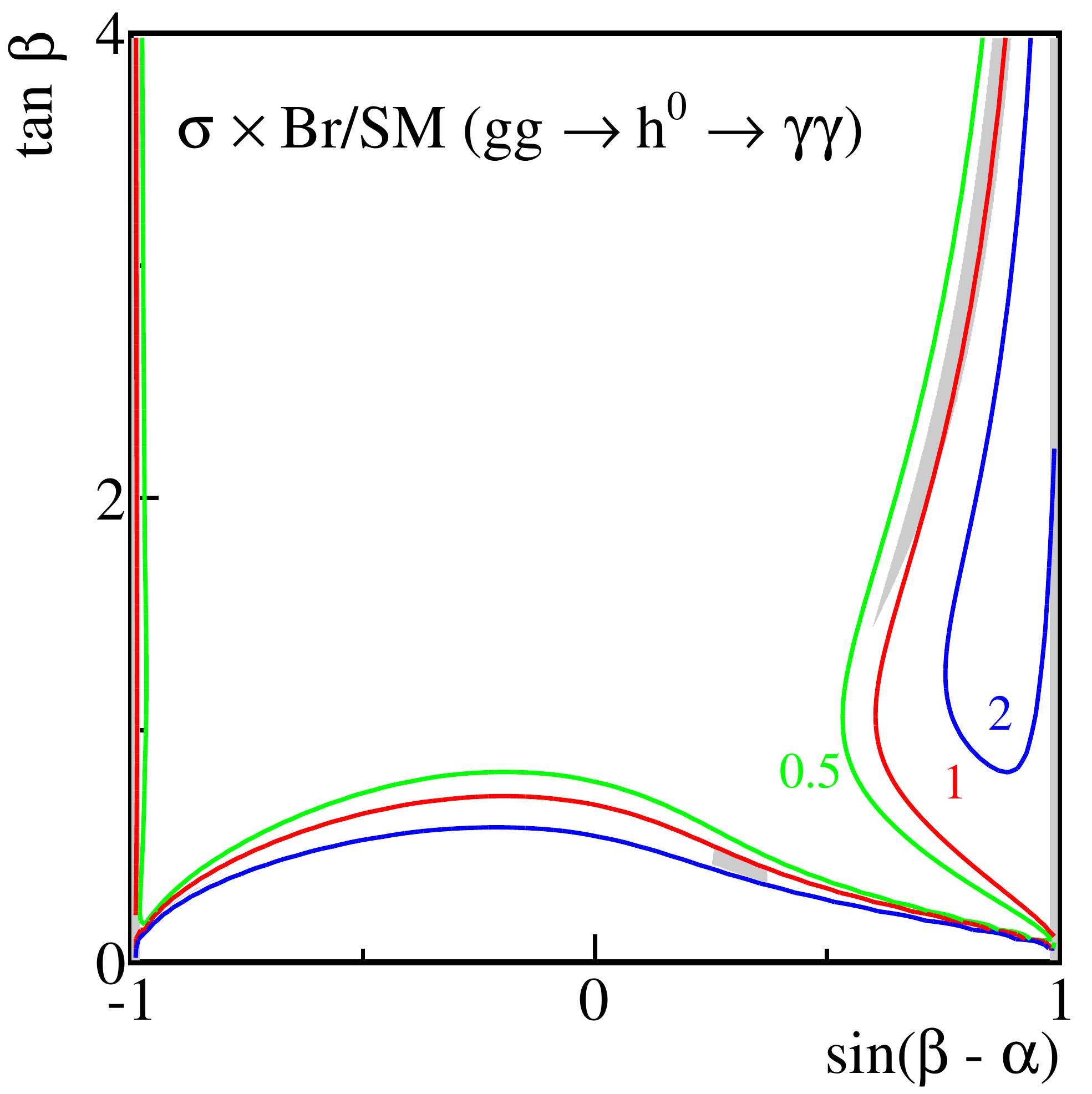}
	\includegraphics[scale=0.4]{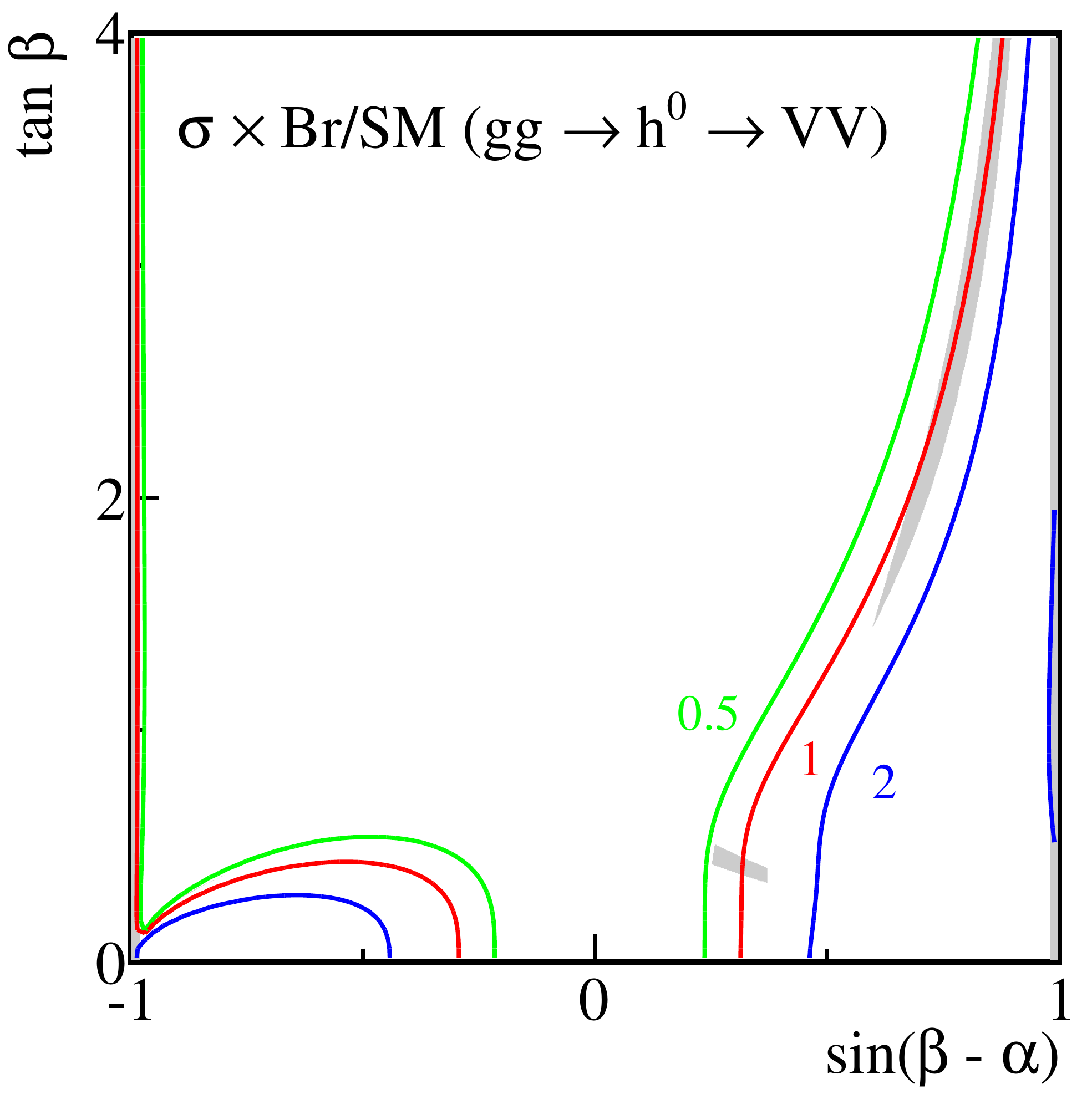}
 \caption{ $\sigma\times{\rm Br} /  {\rm SM}$ for the processes $gg \rightarrow \h \rightarrow \gamma \gamma$ (left), and $gg \rightarrow \h \rightarrow WW/ZZ $ (right) in the $\h$-126 case. The contour lines are  $\sigma\times{\rm Br} /  {\rm SM}= 0.5$ (green), 1 (red), and 2 (blue). The shaded gray are regions where cross sections of $\gamma\gamma$ and $WW/ZZ$ channels satisfy Eq.~(\ref{eq:sigmabrrange}).   }
\label{fig:Contour_Channel_Light}
\end{figure}

Combining both the production and the decay branching fractions, we present the contours of $\sigma\times{\rm Br}/{\rm SM}$ in Fig.~\ref{fig:Contour_Channel_Light} for $\gamma\gamma$ (left panel) and $VV$ (right panel) for  $\sigma\times{\rm Br}/{\rm SM} = 0.5$ (green), 1 (red), and 2 (blue). 
Once we demand that the cross sections for these processes be consistent with the experimental observation of a 126 GeV Higgs, as given in Eq.~(\ref{eq:sigmabrrange}),   the allowed regions of parameter space split into four distinct regions, as indicated by the shaded gray areas.   
There are  two narrow regions one each at  $\sin(\beta-\alpha)=\pm1$ (the gray regions at $\sin(\beta-\alpha) = \pm1$ overlap with the picture frame boundary  and are therefore hard to see),  one extended region of  $0.55 < \sin(\beta-\alpha) < 0.9$,  and one low $\tan\beta$ region around  $\sin(\beta-\alpha)\sim 0.3$ for $\tan\beta\sim 0.5$.  Constraints from $R_b$ disfavor this low $\tan\beta$ region and therefore we will not discuss it further.  In what follows, we will display separate plots for positive and negative $\sin(\beta-\alpha)$ to show the different features that appear in these two cases.

\begin{figure}[htbp]
	\includegraphics[scale=0.4]{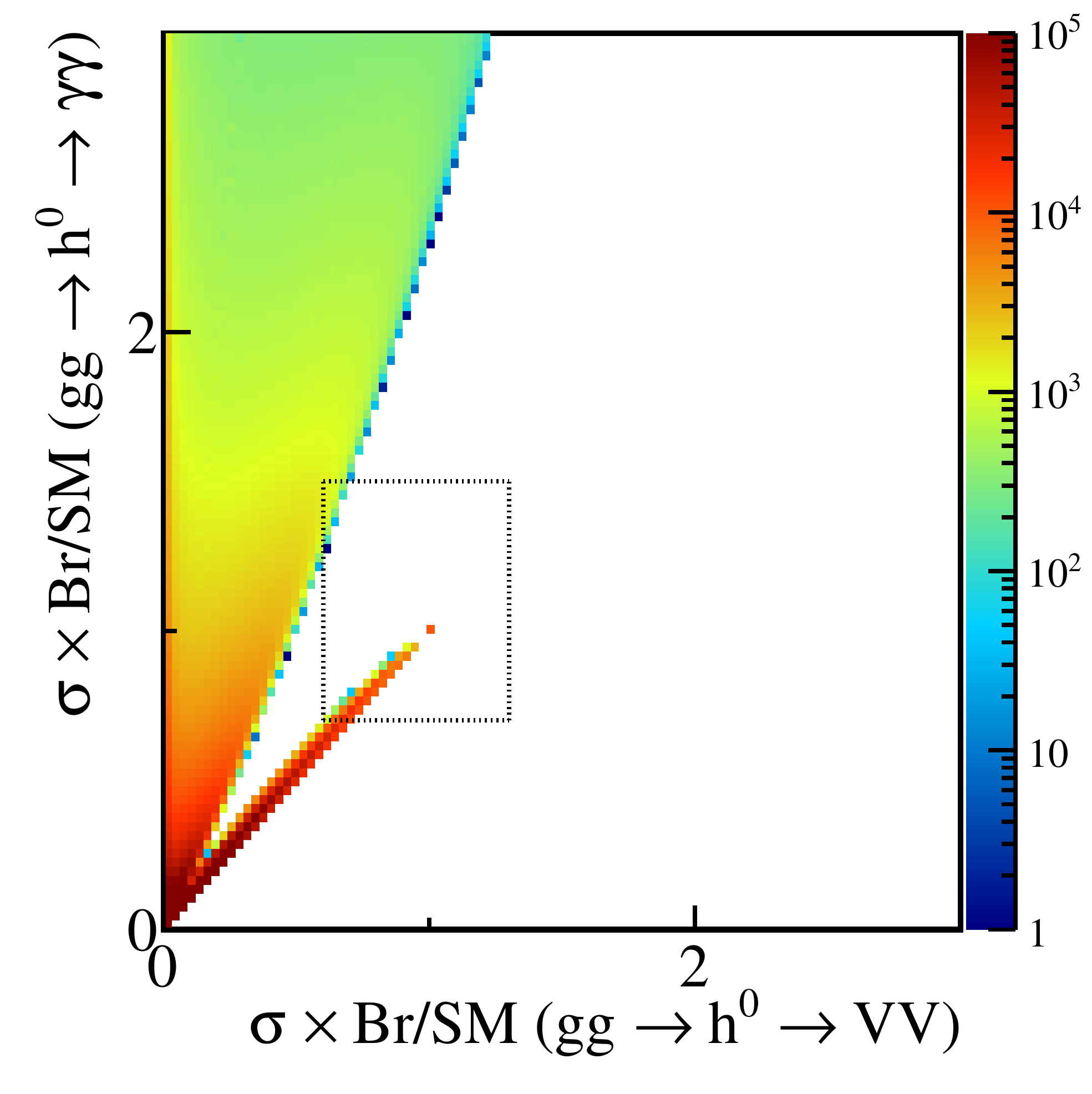}
	\includegraphics[scale=0.4]{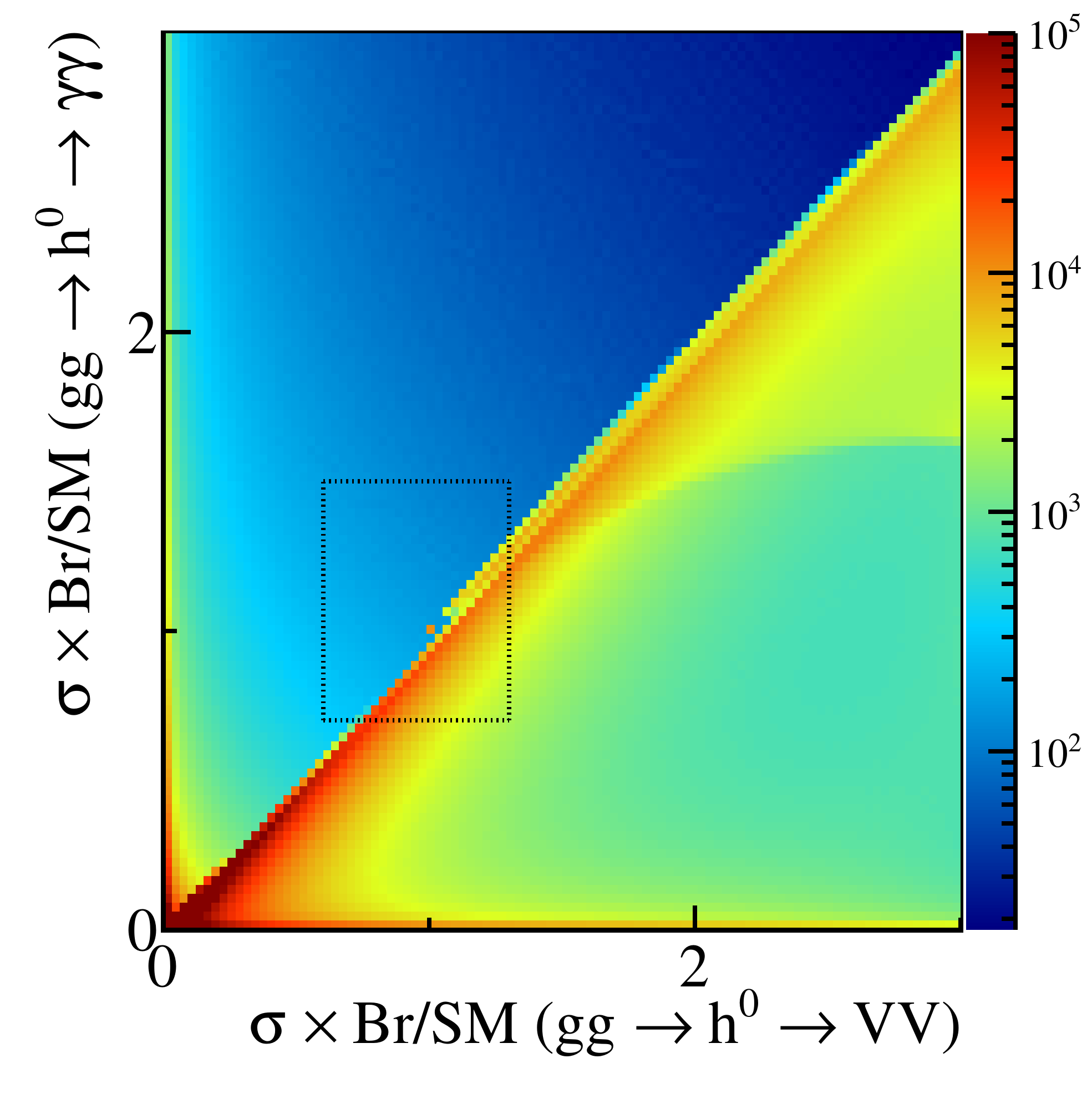}
\caption{$\sigma\times{\rm Br} /  {\rm SM}$ for  $gg\rightarrow \h \rightarrow \gamma \gamma$ versus $ gg\rightarrow \h \rightarrow VV$ for negative $\sin(\beta-\alpha)$ (left panel),  and  positive $\sin(\beta-\alpha)$ (right panel) in the $\h$-126 case.   Color map indicates the density of   points with red being the most dense region and blue being the least dense region.   Also indicated by the small rectangular box  is the normalized signal cross section range of $\gamma\gamma$ between 0.7 and 1.5, and $VV$ channels between 0.6 and 1.3 \cite{CMS-PAS-HIG-13-005,Aad:2013wqa}. } 
\label{fig:Scatter_Light1}
\end{figure}

In Fig.~\ref{fig:Scatter_Light1}, we show the correlations for $\sigma\times{\rm Br} /  {\rm SM}$ for the $\gamma\gamma$ channel against $VV$, for negative (positive) values of $\sin(\beta-\alpha)$ in the left (right) panel as a density plot.  Color coding is such that the points in red are the most dense (i.e., most likely) and points in blue are the least dense (i.e., less likely).    Also indicated by the small rectangular box  is the normalized signal cross section range  of $\gamma\gamma$ between 0.7 and 1.5, and $VV$ channels between 0.6 and 1.3, as given in Eq.~(\ref{eq:sigmabrrange}) \cite{CMS-PAS-HIG-13-005,Aad:2013wqa}.  Note that the corresponding signal windows in $\tan\beta$ versus $\sin(\beta-\alpha)$ plane are also sketched in Fig.~\ref{fig:Contour_Channel_Light} as the shaded gray regions.  
For negative $\sin(\beta-\alpha)$, there are two branches: the one along the diagonal line with $\gamma\gamma:VV\sim 1$ and $\sigma_{\gamma\gamma} \lesssim 1$, which can be mapped on to the  $\sin(\beta-\alpha)=-1$ branch  in Fig.~\ref{fig:Contour_Channel_Light}.  The other branch in the upper-half plane where 
$\gamma\gamma:VV  \gtrsim 2$ and $\sigma_{\gamma\gamma}$ extends to 2 or larger is strongly disfavored given the current observed Higgs signal region. 

For positive values of $\sin(\beta-\alpha)$, the diagonal region is the most probable, with $\gamma\gamma:VV \lesssim 1$ and $\sigma_{\gamma\gamma}$ possibly extending over a relatively large range around 1. Branches with $\sigma_{\gamma\gamma}$ or $\sigma_{VV} \sim 0$ along the axes are strongly disfavored given the current observation of the Higgs signal.

Thus we see that for 
all values of $\sin(\beta-\alpha)$, the $VV$ and $\gamma\gamma$ channels are  positively correlated\footnote{This 
agrees with the results of \cite{Drozd:2012vf}.}.   Most of the points falls into $\gamma\gamma:VV \sim 1$ with the cross section of both around the SM strength.    This means that an excess in the $\gamma\gamma$ channel should most likely be accompanied by an excess in the $ZZ$ and $WW$ channels, and this fact serves as an important piece of 
discrimination for this model as more data is accumulated.

The above analysis illustrates the cross section and decay branching fraction behavior of the  light CP-even Higgs  when it is interpreted as the observed 126 GeV SM-like Higgs, using the approximate formulae in Eqs.~(\ref{eq:sigma_ggh}) - (\ref{eq:BR}).    Note that we have only included the usual SM Higgs decay channels in $\Gamma_{total}$ in Eq.~(\ref{eq:BR}).  While it is a valid approximation in most regions of the parameter space,  it  might break down when  light states in the spectrum  open up new decay modes or introduce large loop contributions to either $gg\rightarrow h^0$ or $h^0 \rightarrow \gamma\gamma$.   In our full analysis presented below with scanning over the parameter spaces, we   used the program 2HDMC, which takes into account all the   decay channels of the Higgs, as well as other loop corrections to the gluon fusion production or Higgs decays to $\gamma\gamma$.  
 
%%%%%%%%%%%%%%%%%%
\subsection{Parameter spaces}
%%%%%%%%%%%%%%%%%%
\begin{figure}[htbp]
	\includegraphics[scale=0.4]{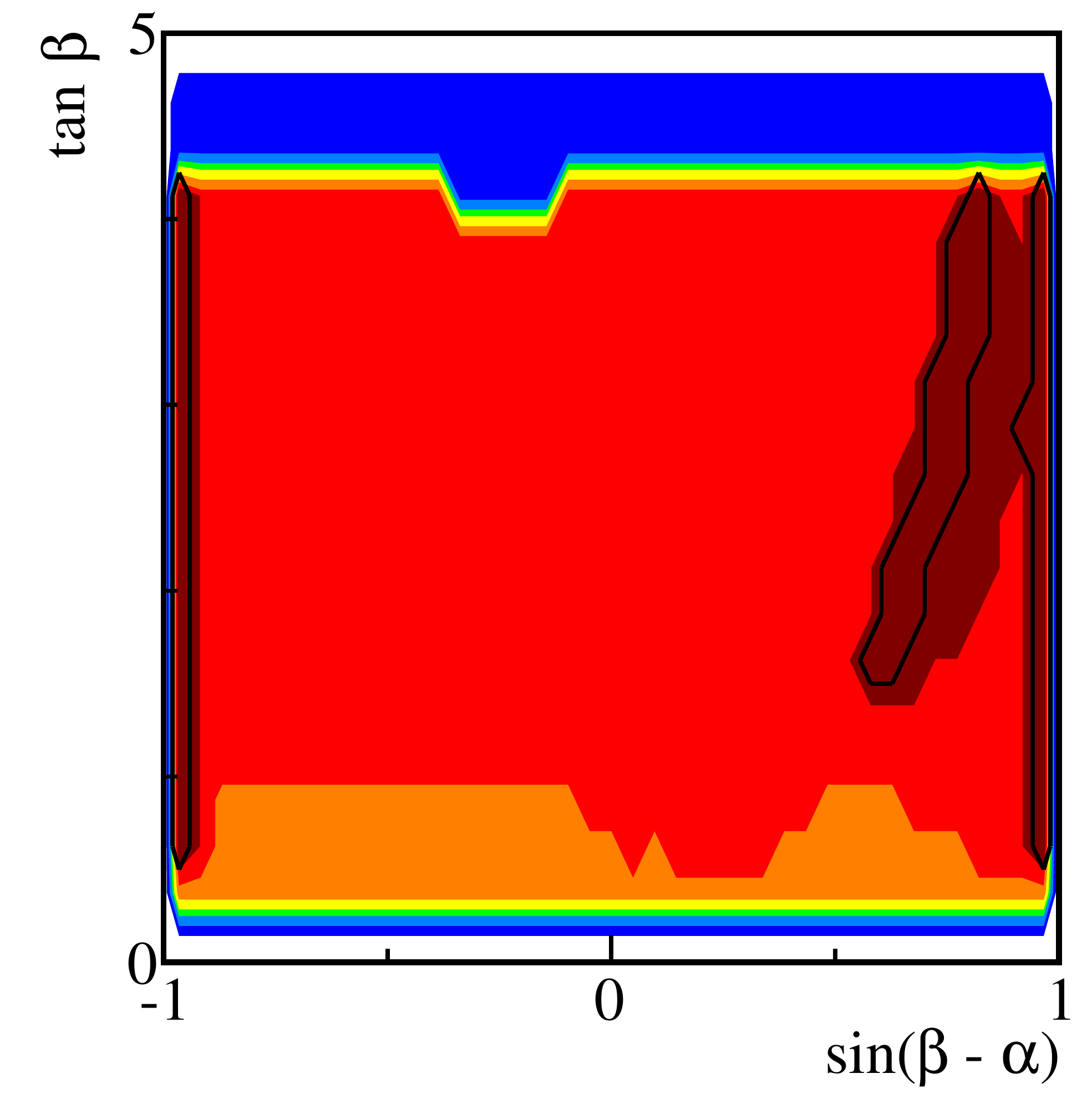}
	\includegraphics[scale=0.4]{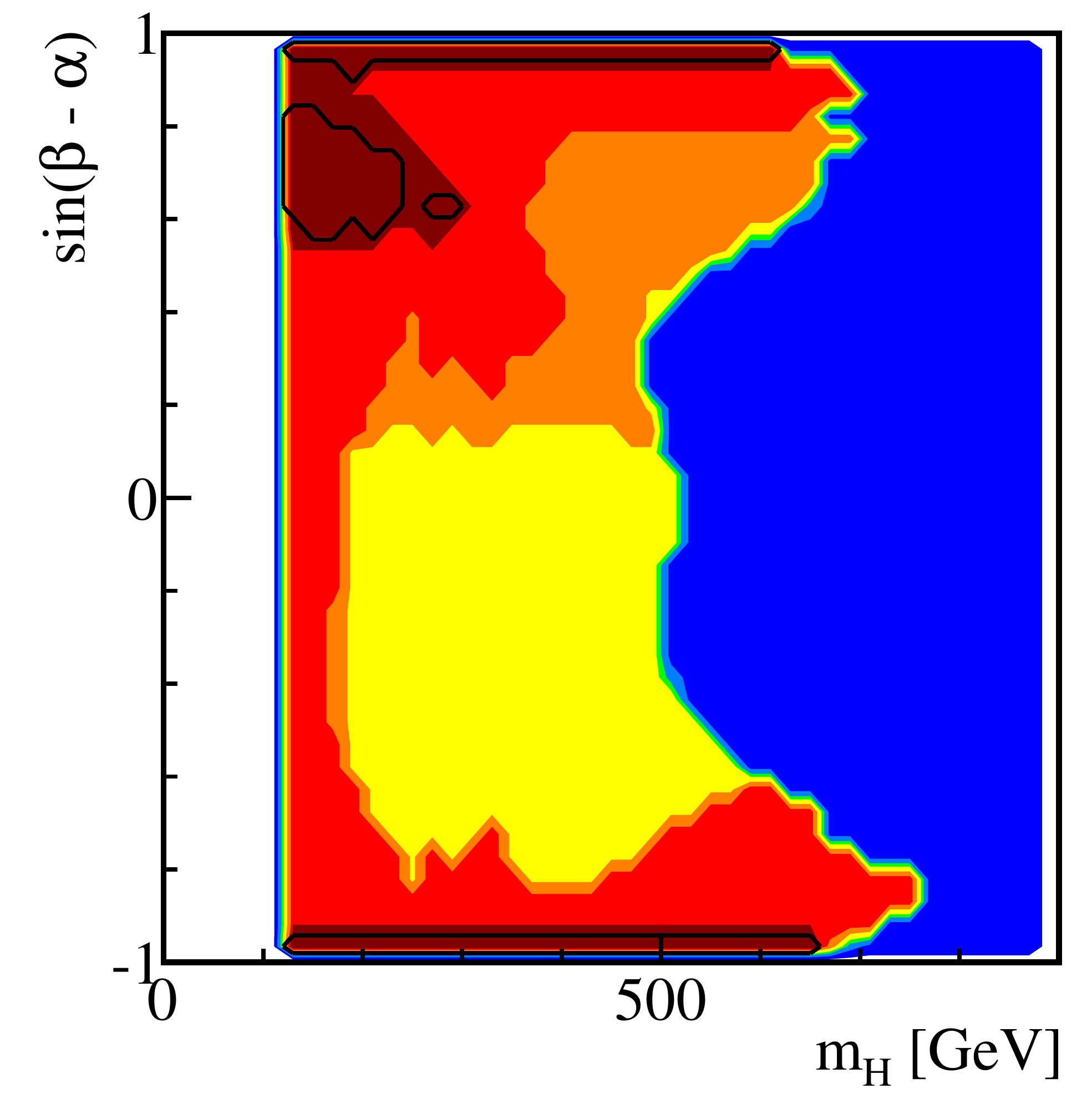}
 	\caption{Parameter regions in the $\h$-126 case   for $\tan\beta$ versus $\sin(\beta-\alpha)$ (left panel) and  $\sin(\beta-\alpha)$ versus $m_H$  (right panel).  We show regions excluded by stability, unitarity and perturbativity (dark blue),  $S$ and $T$ (light blue), LEP results (green), Tevatron and LHC results (yellow), and $R_b$ (orange).  Regions that survive all the theoretical and experimental constraints are shown in red.  Also shown in dark red are regions consistent with the light CP-even Higgs interpreted as the observed    126 GeV scalar resonance, satisfying the cross section requirement of Eq.~(\ref{eq:sigmabrrange}) for $gg \rightarrow \h \rightarrow \gamma \gamma , WW/ZZ$.  Regions enclosed by the black curves are the ones that survive the flavor constraints.  }  
	
	\label{fig:sbatb-light}
 \end{figure}

 Fixing $m_h=126$ GeV still leaves us with five parameters: three masses, $m_H, m_A, m_{H^\pm}$,  and two angles $\tan\beta$ and $\sin(\beta-\alpha)$.  Varying those parameters in the ranges given in Eqs.~(\ref{eq:range1})-(\ref{eq:range4}), we now study the remaining parameter regions satisfying all the theoretical and experimental constraints as well as regions that are consistent with the observed Higgs signal.     
  
The left panel of Fig.~\ref{fig:sbatb-light} shows the viable regions in  $\tan\beta$ versus  $\sin(\beta-\alpha)$ plane when  various theoretical constraints and  experimental bounds are imposed sequentially.   The red regions are those that satisfy all the  constraints.   Also shown in dark red are regions consistent with the light CP-even Higgs interpreted as the observed   126 GeV scalar particle, satisfying the cross section requirement of Eq.~(\ref{eq:sigmabrrange}) for $gg \rightarrow \h \rightarrow \gamma \gamma , WW/ZZ$.   
The signal regions (two narrow regions at $\sin(\beta-\alpha) = \pm1$, and one extended region with $0.55 < \sin(\beta-\alpha) < 0.9$) agree well with the shaded region in Fig.~\ref{fig:Contour_Channel_Light}.  The small region around $\sin(\beta-\alpha) \sim 0.3$, however, disappeared, due to the $R_b$ constraint \cite{Logan:1999if}. Regions with $\tan\beta \gtrsim$ 4 are excluded by perturbative bounds since one of $\lambda_{1,2}$ becomes non-perturbative for larger value of $\tan\beta$ ($\cos\beta \rightarrow 0$), as shown in Eq.~(\ref{eq:lambda12}).    Consequently, the bottom   loop contribution to the gluon fusion production cross section \cite{review} is not a major factor for the $\h$-126 case.  

To further explore the flavor constraints, we show in Fig.~\ref{fig:sbatb-light} the regions enclosed by the black curves 
being those that survive the flavor bounds.   As can clearly be seen, flavor bounds do not significantly impact the surviving signal regions.   

The right panel of Fig.~\ref{fig:sbatb-light} shows the allowed region in the $\sin(\beta-\alpha)-m_H$ plane.   Imposing all the theoretical constraints, in particular, the perturbativity requirement,   translates into an  upper bound on $m_H$ of around  750 GeV.   Higgs search bounds from the LHC removes a large region in negative $\sin(\beta-\alpha)$, mostly from the stringent bounds from $WW$ and $ZZ$ channels for the heavy Higgs.   The positive $\sin(\beta-\alpha)$ region is less constrained since $gg\rightarrow \H \rightarrow WW/ZZ$ are much more suppressed.    $R_b$, in addition, excludes part of the positive $\sin(\beta-\alpha)$ region with relatively large $m_H$.
 Requiring $\h$ to fit the observed Higgs signal further narrows down the favored regions, as shown in dark red.  
For $\sin(\beta-\alpha)=\pm 1$,   $m_H$ could be as large as 650 GeV.     For $0.55 \lesssim \sin(\beta-\alpha) \lesssim 0.9$, $m_H$ is constrained to be less than 300 GeV.    The correlation between $m_H$ and $\sin(\beta-\alpha)$ indicates that   if  a heavy CP-even Higgs is discovered to be  between 300 and 650 GeV, $\sin(\beta-\alpha)$ is constrained to be very close to $\pm1$, indicating  the light  Higgs has SM-like  couplings to the gauge sector. 
 
\begin{figure}[htbp]
	\includegraphics[scale=0.4]{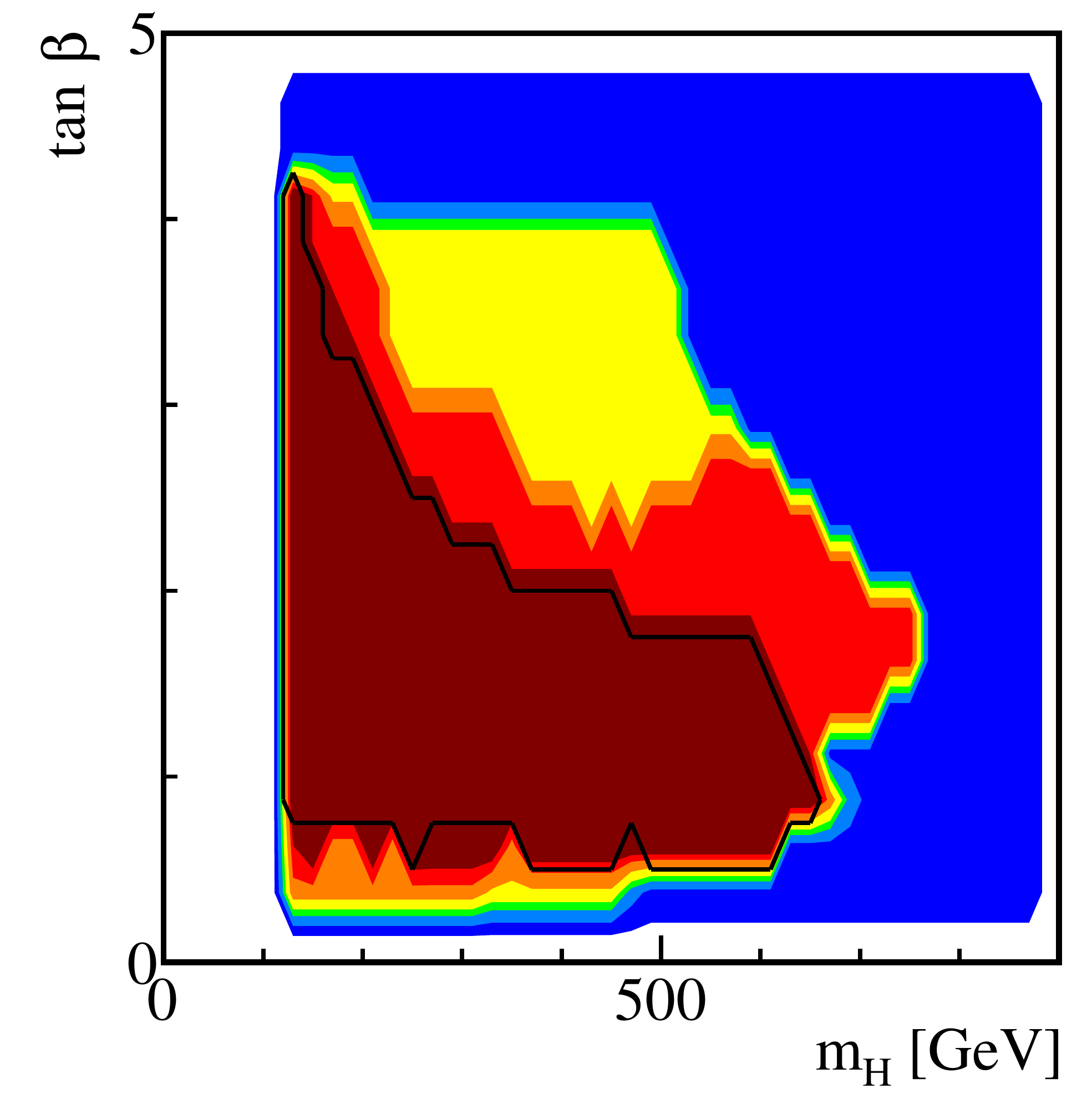}
	\includegraphics[scale=0.4]{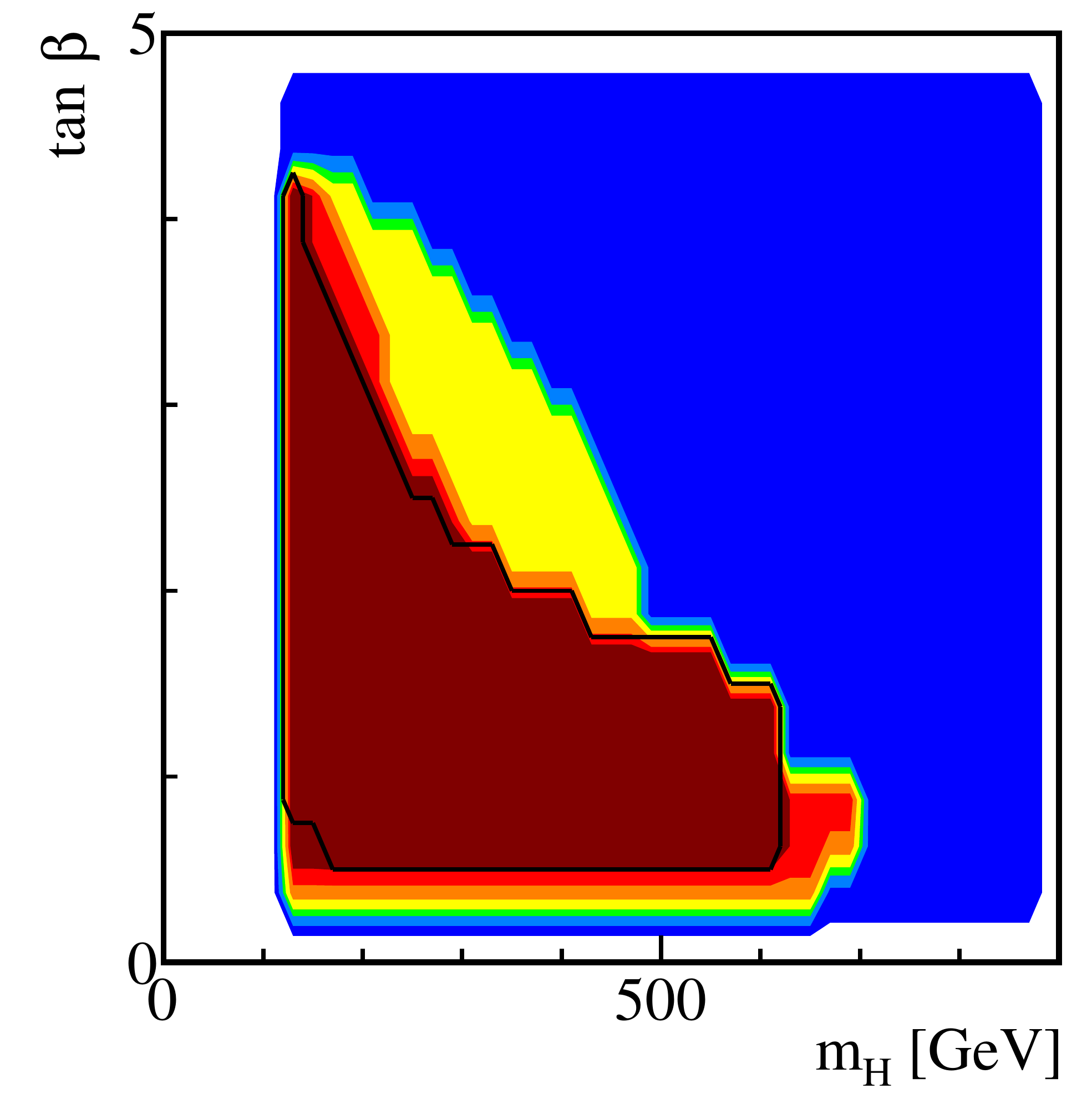}
	\caption{Parameter regions in the $\h$-126 case  for $\tan\beta$ versus $m_H$ with $\sin(\beta-\alpha)<0$ (left panel) and $\sin(\beta-\alpha)>0$ (right panel).  Color coding is the same as Fig.~\ref{fig:sbatb-light}. }
	\label{fig:lightmHtb}
\end{figure}

In  Fig.~\ref{fig:lightmHtb}, we present  the parameter regions   for $\tan\beta$ versus $m_H$ with $\sin(\beta-\alpha)<0$ (left panel) and $\sin(\beta-\alpha)>0$ (right panel).   Regions with large $m_H$ are typically realized for small $\tan\beta$   roughly between 1 and 2.    There are also noticeable difference for positive or negative $\sin(\beta-\alpha)$ for regions that survive all the experimental constraints (red regions).  Negative  $\sin(\beta-\alpha)$ allows larger values of $\tan\beta$ for a given mass of $m_H$.  
Small values of $\tan\beta$ is disfavored by the perturbativity of top Yukawa coupling \cite{Bijnens:2011gd}, $R_b$ \cite{Logan:1999if}, and the flavor constraints \cite{Mahmoudi:2009zx}.

\begin{figure}[htbp]
	\includegraphics[scale=0.4]{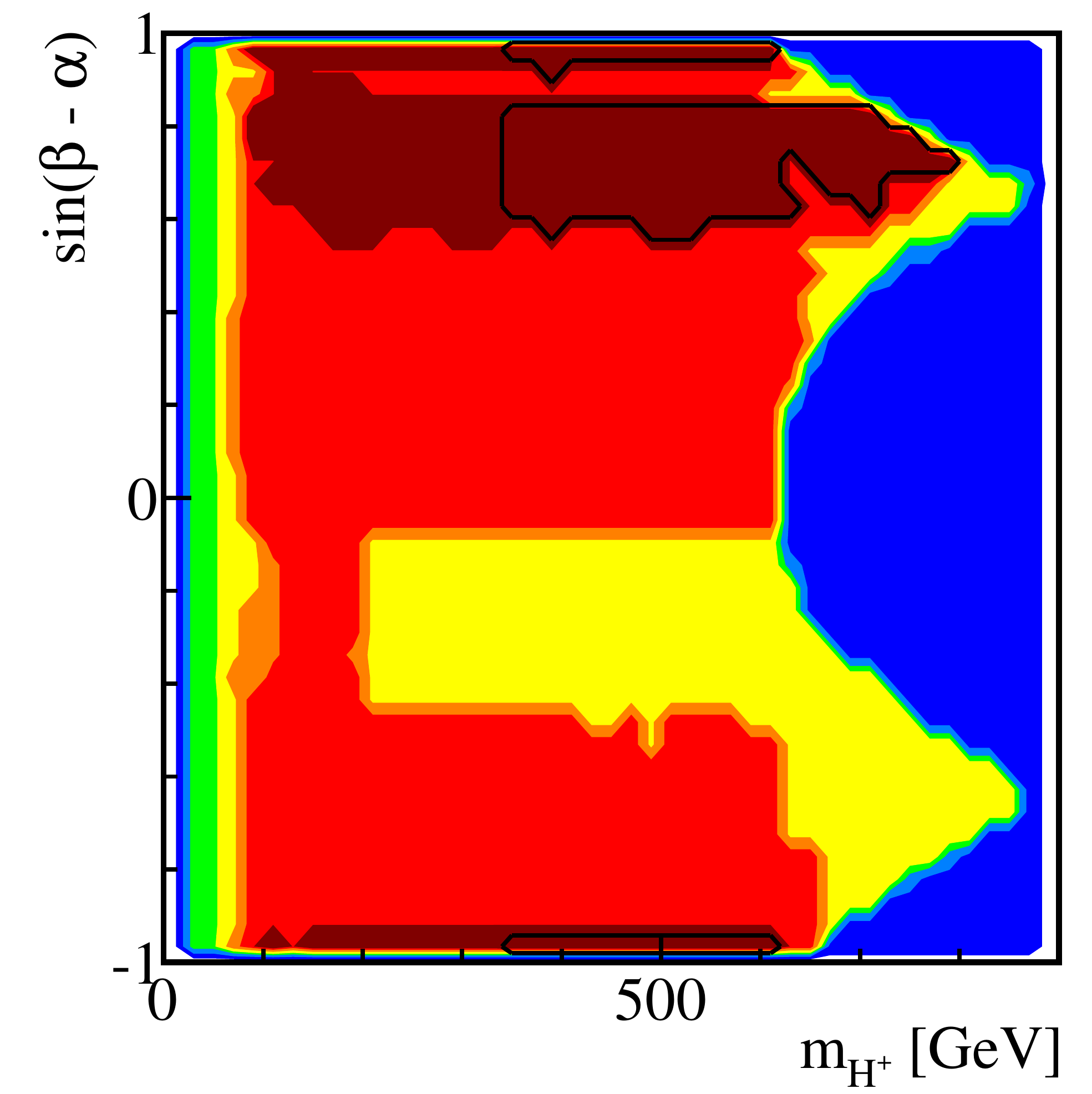}
	\includegraphics[scale=0.4]{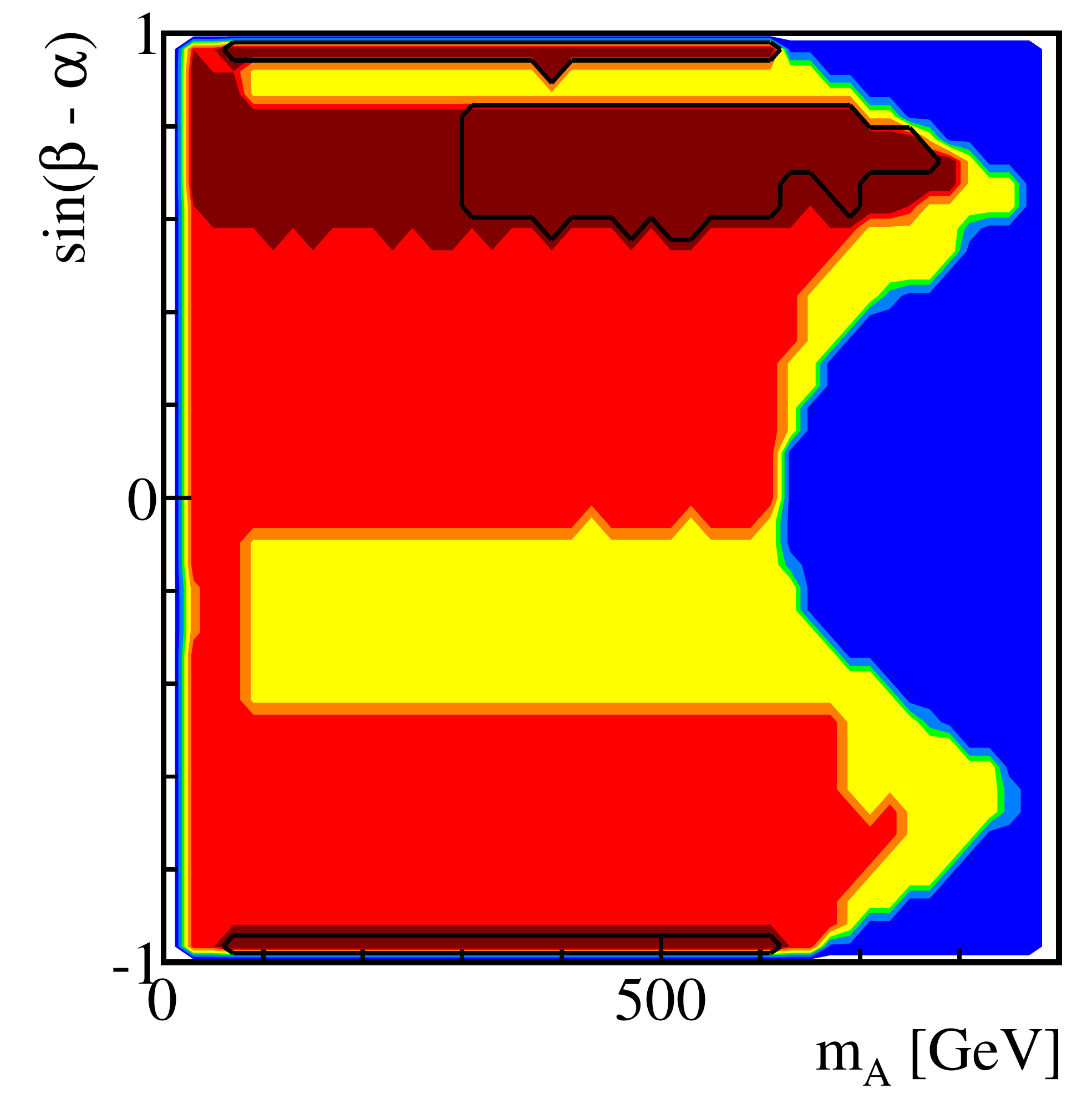}
	\caption{Parameter regions in the $\h$-126 case   for $\sin(\beta-\alpha)$  versus $m_{H^\pm}$ (left panel) and $m_{A}$ (right panel). Color coding is the same as Fig.~\ref{fig:sbatb-light}.	}	
	\label{fig:lightmCmAsinba}
\end{figure}

Fig.~\ref{fig:lightmCmAsinba} shows the parameter regions in $\sin(\beta-\alpha)$  versus $m_{H^\pm}$ (left panel) and $m_{A}$   (right panel).    For negative $\sin(\beta-\alpha)$ between  $-0.5$ to $-0.1$, only regions with  $ m_A<$ 60 GeV survive the LHC Higgs search bounds. This is because $\H\rightarrow \A\A$ opens
up in this region, which leads to the suppression of $\H \rightarrow WW/ZZ$ allowing it to escape the experimental constraints.  The corresponding surviving region in 120 GeV $<m_{H^\pm}<$ 200  GeV is   introduced by the correlation between $m_A$ and $m_{H^\pm}$ due to $\Delta\rho$ constraints.
Imposing the cross section requirement for $\h$ to satisfy the Higgs signal region results in three bands in both $m_A$ and $m_{H^\pm}$, with masses extending all the way to about 800 GeV.  Imposing the flavor constraints leaves regions with $m_{H^\pm}\gtrsim 300$ GeV viable for $\sin(\beta-\alpha)=\pm1$ or $\sin(\beta-\alpha)$ between 0.55 and  0.9, while even smaller values for $m_A$ remain viable at $\sin(\beta-\alpha)=\pm1$.

\begin{figure}[htbp]
	\includegraphics[scale=0.4]{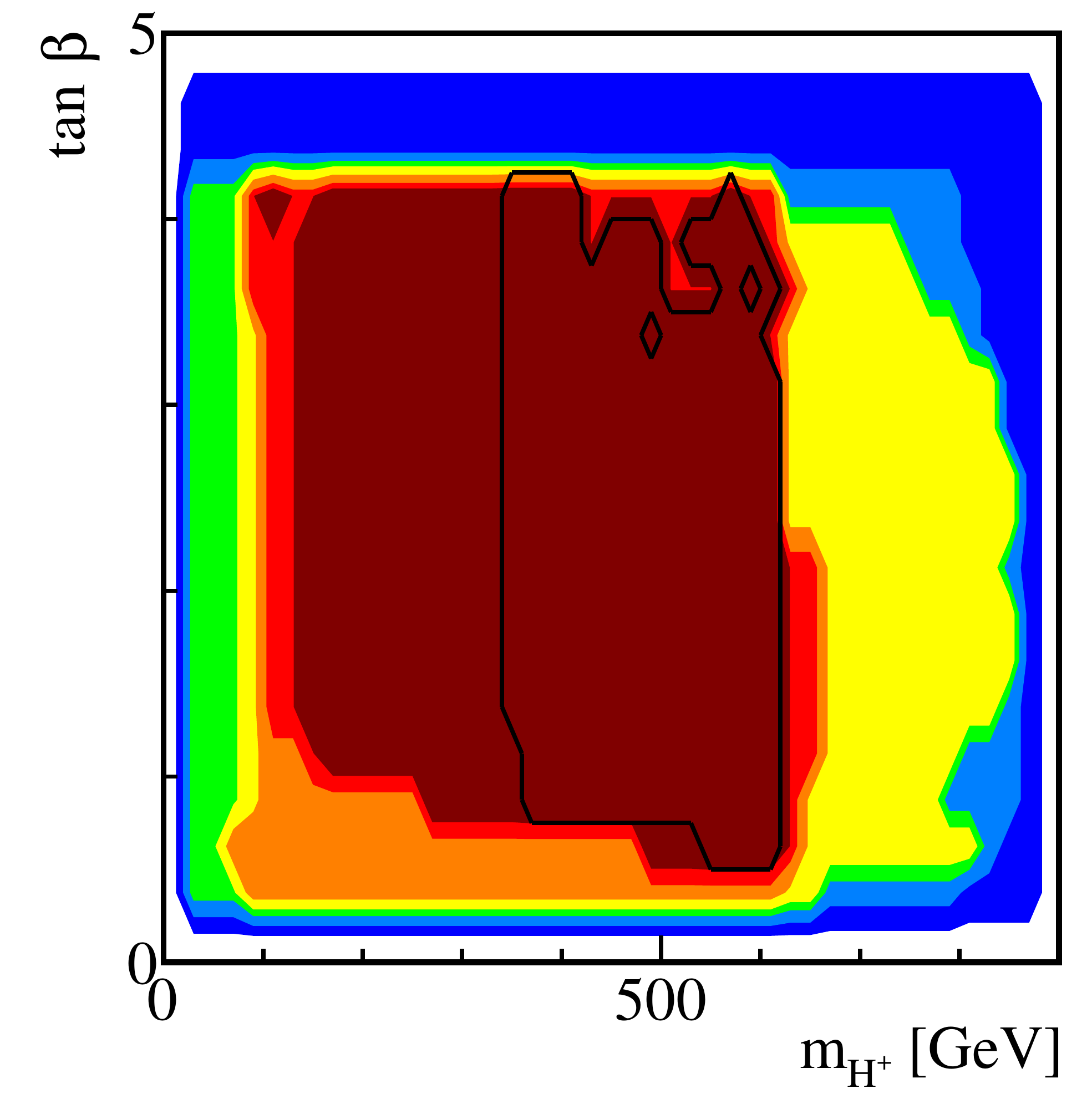}
	\includegraphics[scale=0.4]{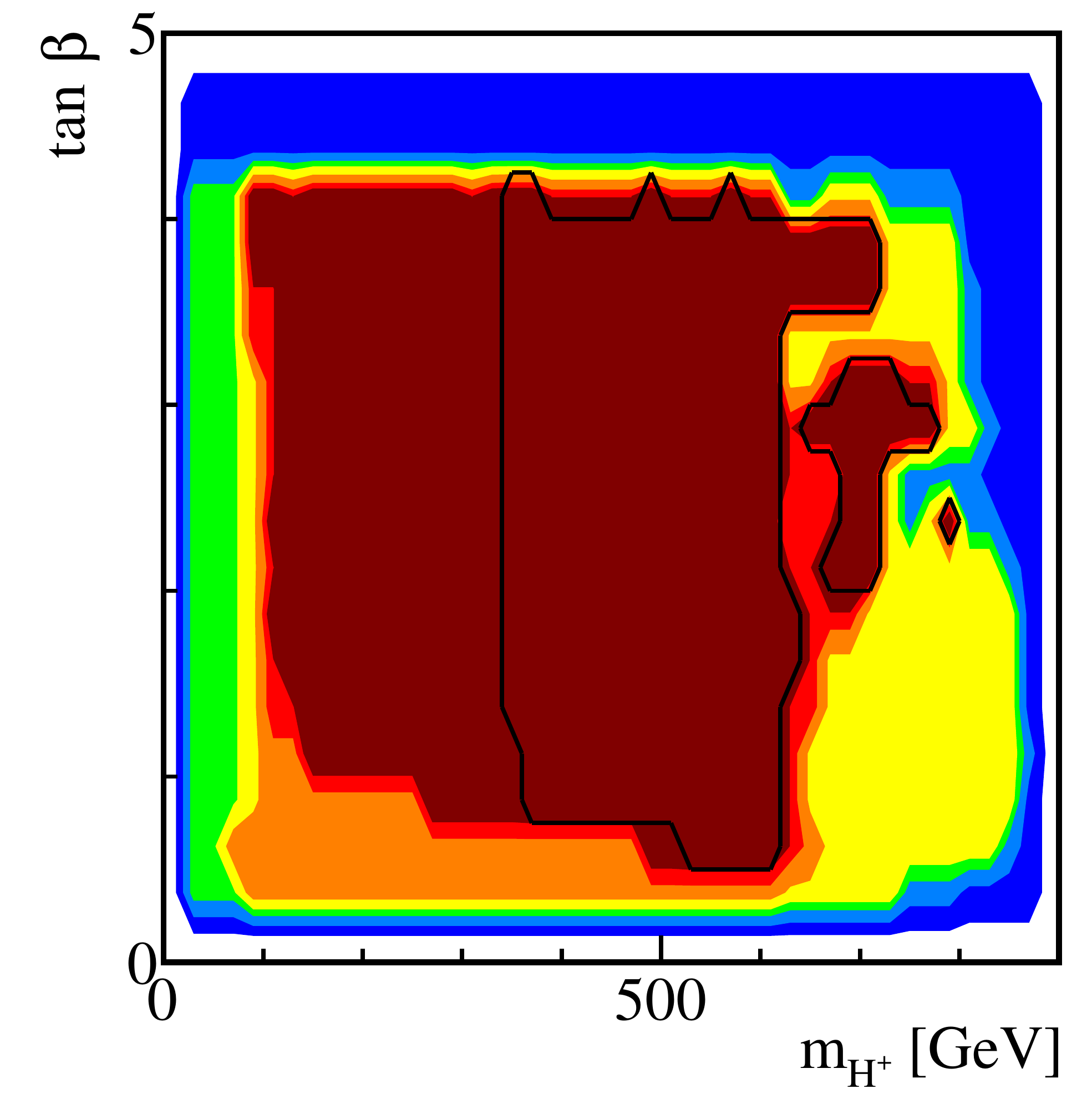}
	\includegraphics[scale=0.4]{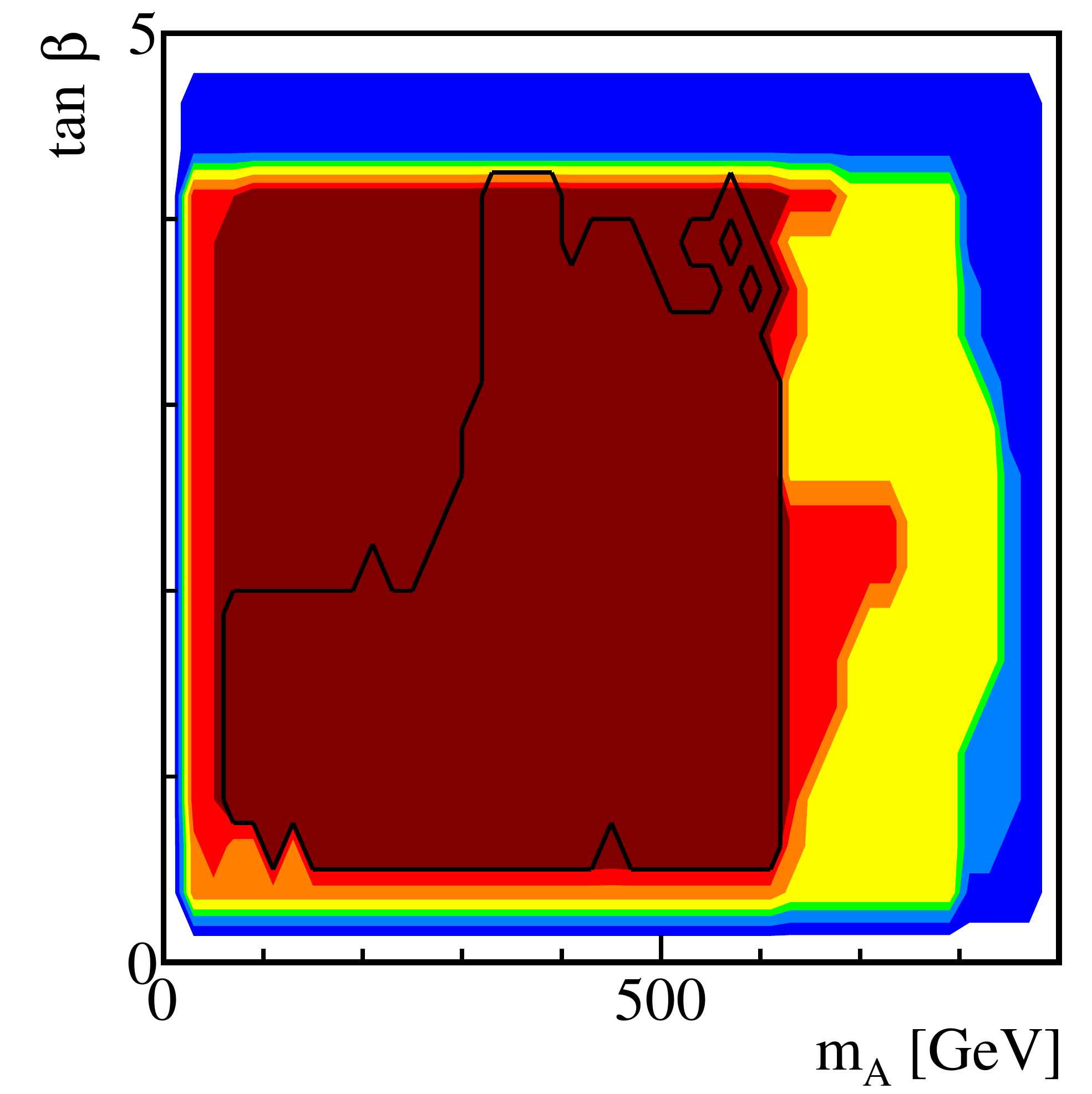}
	\includegraphics[scale=0.4]{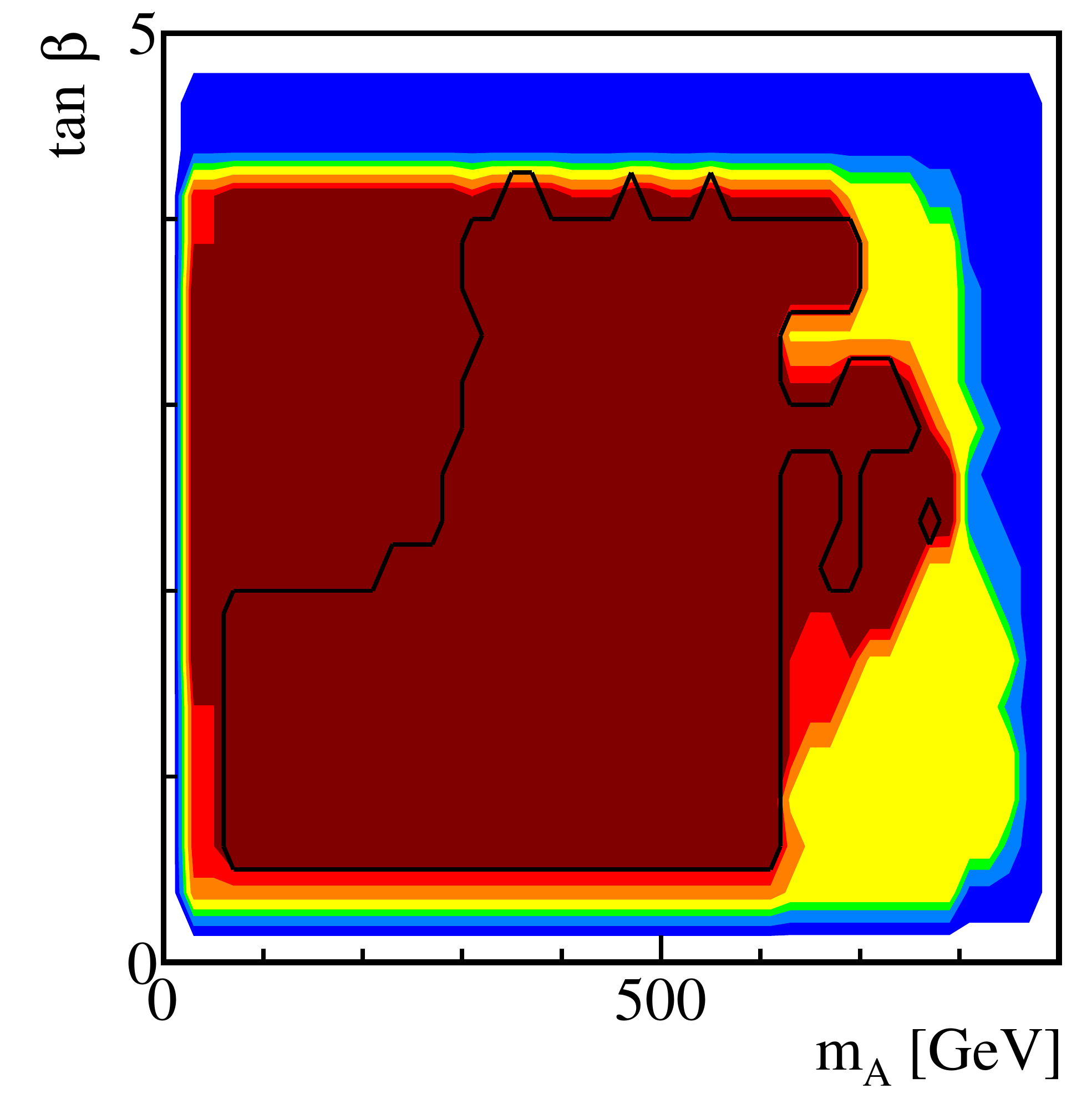}
	\caption{Parameter regions in the $\h$-126 case for    $\tan\beta$ versus $m_{H^\pm}$ (top panels) and $m_{A}$   (lower panels) with $\sin(\beta-\alpha)<0$ (left panels) and $\sin(\beta-\alpha)>0$ (right panels).  Color coding is the same as Fig.~\ref{fig:sbatb-light}.  
 }
	\label{fig:lightmCmAtb}
\end{figure}

The allowed regions in the $\tan\beta-m_{H^\pm}$ and $\tan\beta - m_A$ planes share similar features before flavor constraints are taken into account, which are shown  in Fig.~\ref{fig:lightmCmAtb}. The top two panels show the allowed regions in the $\tan\beta-m_{H^\pm}$ plane for negative and positive $\sin(\beta-\alpha)$, while the lower  two panels are for  $\tan\beta-m_A$. LEP places a lower bound on the charged Higgs mass around 80 GeV \cite{lepcharged0, lepcharged1}.  In the signal region for $\sin(\beta-\alpha)<0$, both $m_{H^\pm}$ and $m_A$ are less than about 600 GeV, while their masses could be extended to 800 GeV for $\sin(\beta-\alpha)>0$ and  $\tan\beta >2$.   The difference between the $m_A$ range for different signs of $\sin(\beta-\alpha)$ can be explained as follows: regions with $m_A>600$ GeV can only occur for $|\sin(\beta-\alpha)|$ between 0.4 and 0.8, as shown in  the right panel of Fig.~\ref{fig:lightmCmAsinba}.
The Higgs signal region of $\tan\beta$ versus $\sin(\beta-\alpha)$ (left panel of Fig.~\ref{fig:sbatb-light})  shows that 
to simultaneously satisfy both the $\tan\beta$ range and $\sin(\beta-\alpha)$ range, only positive $\sin(\beta-\alpha)$ case survives.

Flavor bounds, as expected, have a marked effect here ruling out any value of   $m_{H^\pm}\lesssim$ 300 GeV for all values of $\tan\beta$, mainly due to the $b \rightarrow s \gamma$ constraint.    For the CP-odd Higgs, only a corner of $\tan\beta > 2$ and $m_A < 300$ GeV is excluded, due to the combination of flavor and $\Delta\rho$ constraints.  As shown in Fig.~\ref{fig:lightmHtb}, only relatively light $m_H \lesssim 300$ GeV is allowed for $\tan\beta>2$.    The flavor constraints of  $m_{H^\pm}\gtrsim 300$ GeV is then translated to $m_A \gtrsim 300$ GeV since the difference between $m_A$ and $m_{H^\pm}$ is constrained by $\Delta \rho$ considerations when both $m_h$ and $m_H$ are relatively small.  For $\tan\beta<2$, $m_H$ could be relatively high, which cancels the large contribution  to $\Delta \rho$ from large $m_{H^\pm}$ while allowing $m_A$ to be light.
  
\begin{figure}[htbp]
	\includegraphics[scale=0.4]{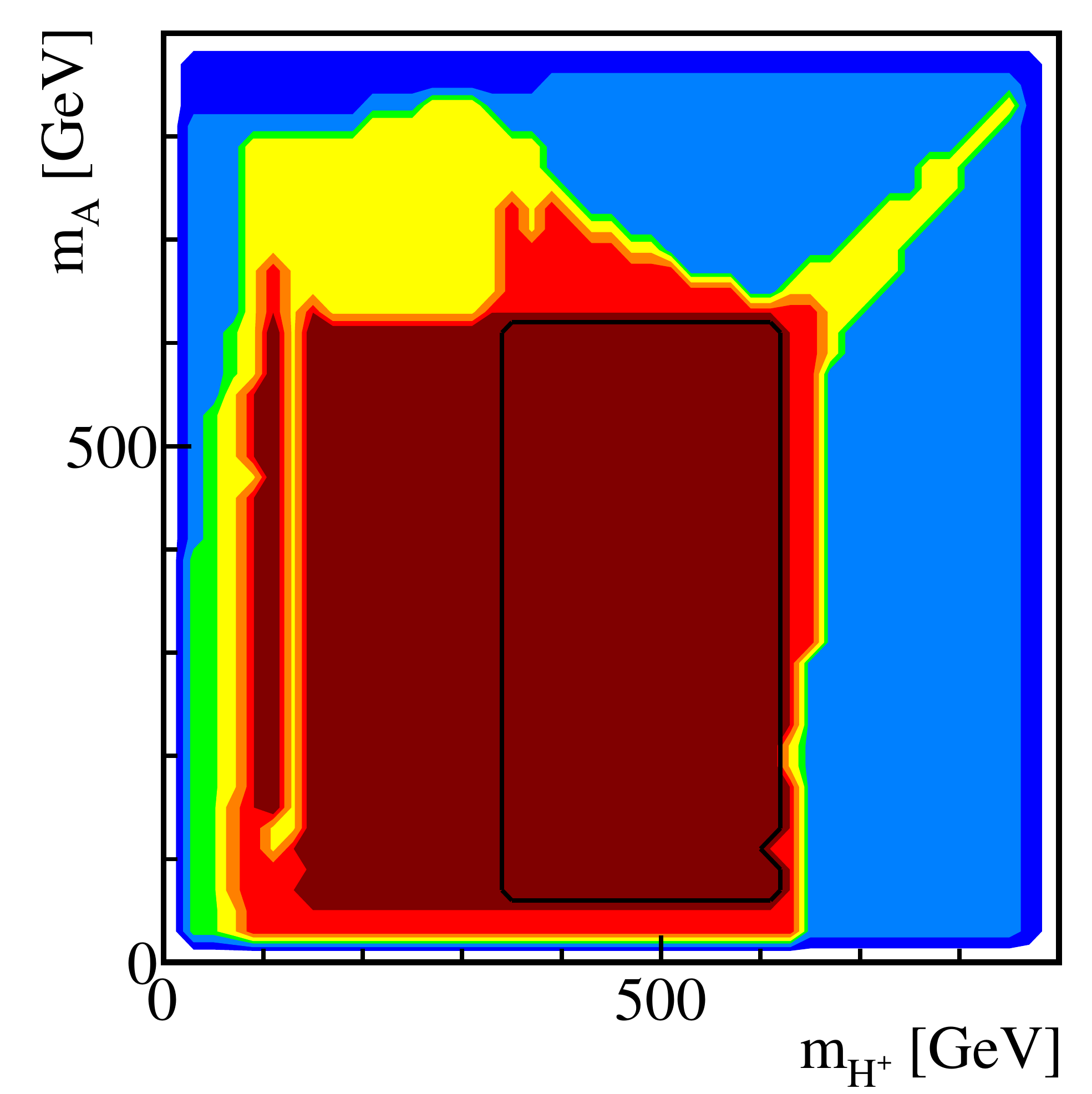}
	\includegraphics[scale=0.4]{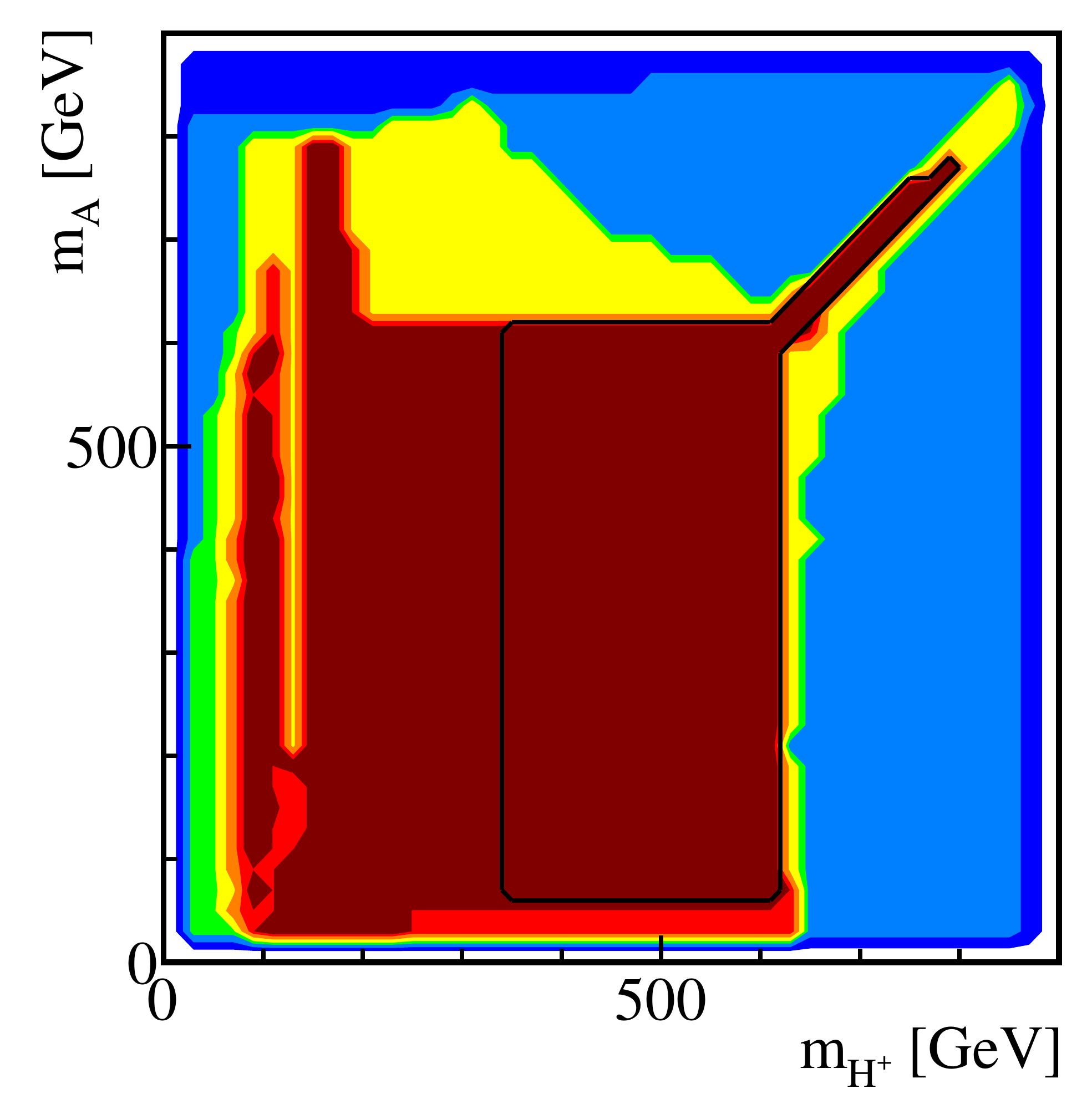}
	\caption{Parameter regions in the $\h$-126 case   for   $m_A$   versus $m_{H^\pm}$ with $\sin(\beta-\alpha)<0$ (left panel) and $\sin(\beta-\alpha)>0$ (right panel).   Color coding is the same as Fig.~\ref{fig:sbatb-light}.  }
	\label{fig:lightmAmC}
\end{figure}

In  Fig.~\ref{fig:lightmAmC}, we present the parameter regions in the $m_A-m_{H^\pm}$ plane  for negative and positive values of $\sin(\beta-\alpha)$. $m_A$ and $m_{H^\pm}$ are uncorrelated for most parts of the parameter space.  For $\sin(\beta-\alpha)>0$ when $m_{A, H^\pm}$ could reach values larger than 600 GeV, $\tan\beta$ is at least 2 or larger (see Fig.~\ref{fig:lightmCmAtb}).    $m_H$ is restricted to less than 300 GeV in this region, which results in  a strong correlation between $m_A$ and $m_{H^\pm}$  due to the $\Delta\rho$ constraints.

\begin{figure}[htbp]
	\includegraphics[scale=0.4]{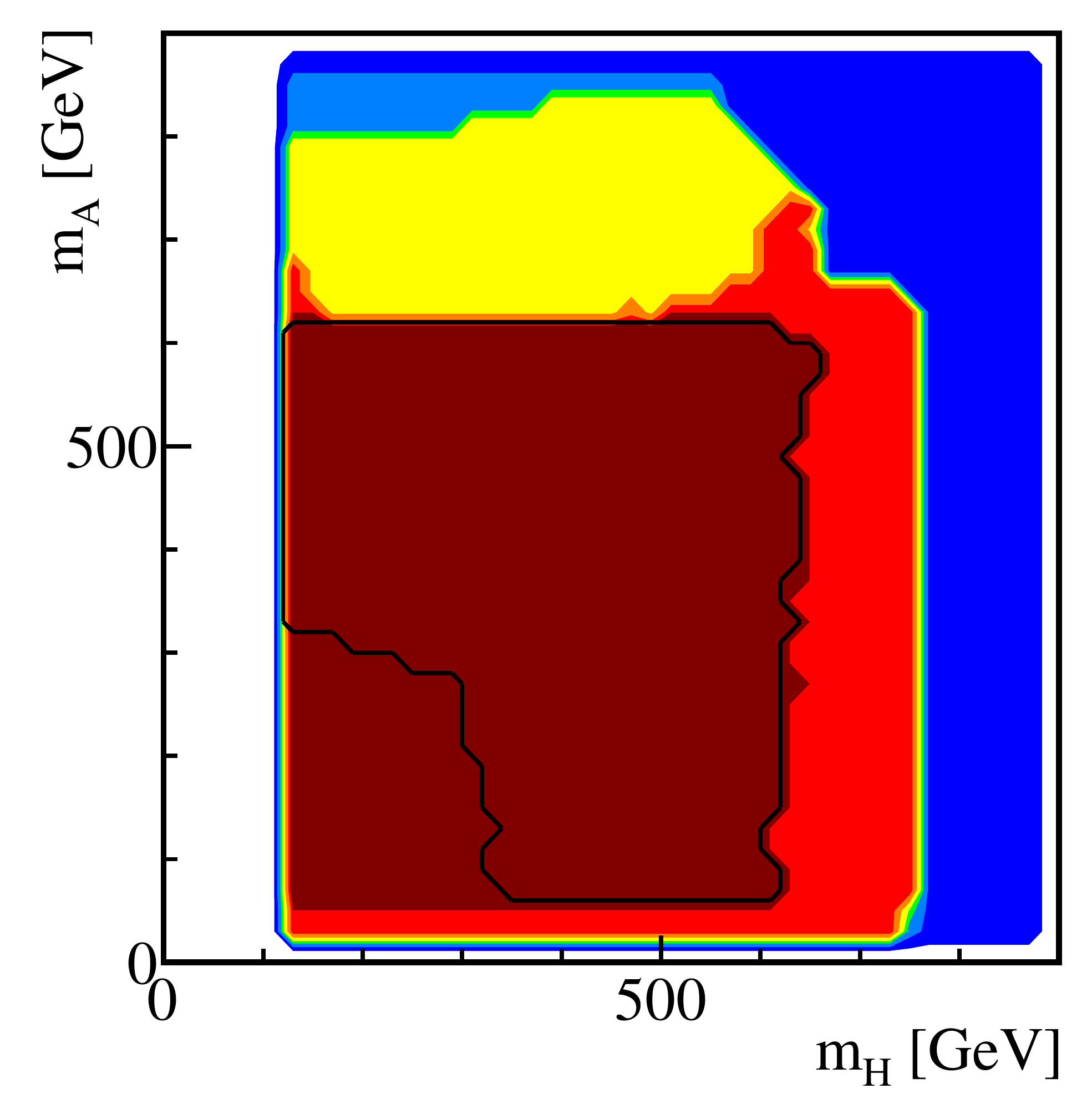}
	\includegraphics[scale=0.4]{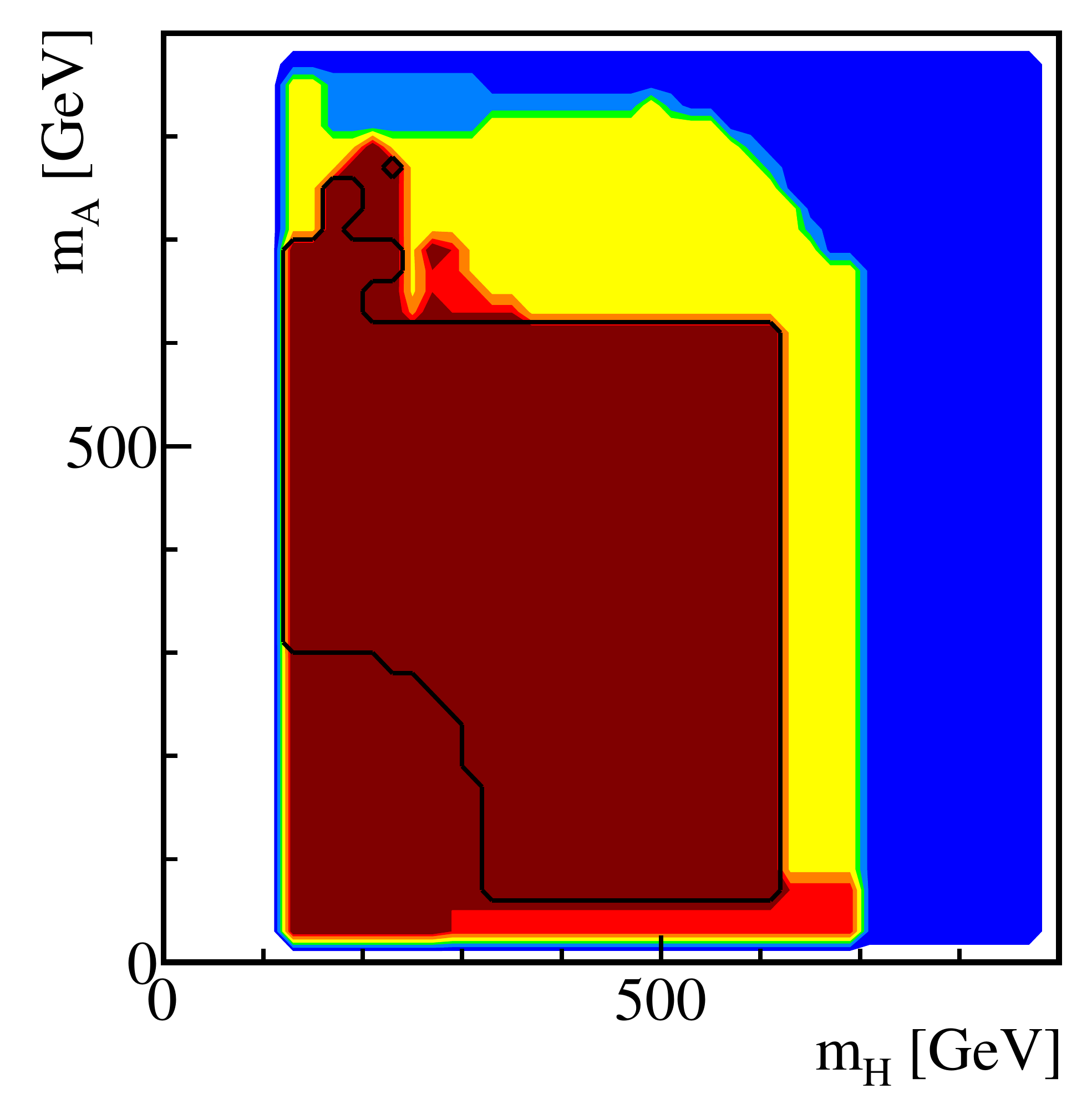}
	\caption{Parameter regions in the $\h$-126 case for  $m_{A}$ versus $m_H$   with $\sin(\beta-\alpha)<0$ (left panel) and $\sin(\beta-\alpha)>0$ (right panel).   Color coding is the same as Fig.~\ref{fig:sbatb-light}.	}
	\label{fig:lightmHmA}
\end{figure}

Fig.~\ref{fig:lightmHmA} shows the  parameter space in the $m_A - m_H$ plane for negative (left panel) and positive (right panel) $\sin(\beta-\alpha)$.    These two masses are largely uncorrelated  for either  sign of $\sin(\beta-\alpha)$.  Note that for  $\sin(\beta-\alpha)>0$, large $m_A$ between 600 $-$ 800 GeV is only possible for small values of  $m_H \lesssim$ 250 GeV.  This is because the corresponding $\tan\beta$ is larger than 2, which bounds $m_H$ from above.  The lower-left corners excluded by flavor constraints correspond to the upper-left corners in $m_A - \tan\beta$ plots in Fig.~\ref{fig:lightmCmAtb}, since at least  one of $m_A$ or $m_{H}$ would need to be relatively heavy to cancel the contribution to $\Delta\rho$ from $m_{H^\pm}>300$ GeV.
   
We conclude this section with the following comments:
\begin{itemize}
 \item{} If $\h$ is  the 126 GeV resonance, then the $\gamma\gamma$ channel is closely correlated with $WW/ZZ$.  Specifically, a moderate excess in $\gamma\gamma$ should be accompanied by a corresponding excess in $WW/ZZ$.
 
 \item{} The combination of all theoretical constraints requires $\tan\beta\lesssim 4$. Therefore,   the bottom-loop enhancement to the gluon fusion \cite{review} is never a major factor.    Regions of $\sin(\beta-\alpha)$ and $\tan\beta$ are highly restricted once we require the light CP-even Higgs to be the observed 126 GeV scalar particle: $\tan\beta$ between 0.5 to 4 for $\sin(\beta-\alpha) = \pm 1$, $\tan\beta$ between 1.5 to 4 for  $0.55<\sin(\beta-\alpha) < 0.9$. The masses of the other Higgses, $m_H$, $m_A$, and $m_{H^\pm}$, however, are largely unrestricted and uncorrelated, except for the region where $\sin(\beta-\alpha)>0$ and $m_{A, H^\pm}\gtrsim 600$ GeV, which exhibits a strong correlation between these two masses.   
 
 \item{}The discovery of any one of the extra scalars can largely narrow down the parameter space, in particular, if the masses of those particles are relatively high.  
 
 \item{}  Flavor bounds do not change  the allowed parameter space much except for the charged Higgs mass, which  is constrained to lie above 300  GeV.
\end{itemize} 
 
%%%%%%%%%%%%%%%%%%%%%%
\section{Heavy Higgs at 126 GeV}
\label{sec:H126}

\subsection{Cross sections and Correlations}
%%%%%%%%%%%%%%%%%%%%%%%%%
It is possible that the 126 GeV resonance discovered at the LHC corresponds to the heavier of the two CP-even Higgses, $\H$.   There are a few noticeable changes for  the heavy $\H$ being the SM-like Higgs boson.  First of all, since the coupling of the heavy Higgs to a gauge boson pair is scaled by a factor of $\cos(\beta-\alpha)$ as opposed to $\sin(\beta-\alpha)$, demanding SM-like cross sections for $\H$   forces us to consider $\sin(\beta-\alpha)\sim 0$, as opposed to $\sin(\beta-\alpha)\sim\pm1$ in the $\h$-126 case.   Secondly, as will be demonstrated below, the bottom contribution to the gluon fusion production could be significantly enhanced since  the range of $\tan\beta$ 
could be much larger compared to the $\h$-126 case.

\begin{figure}[htbp]
	\includegraphics[scale=0.4]{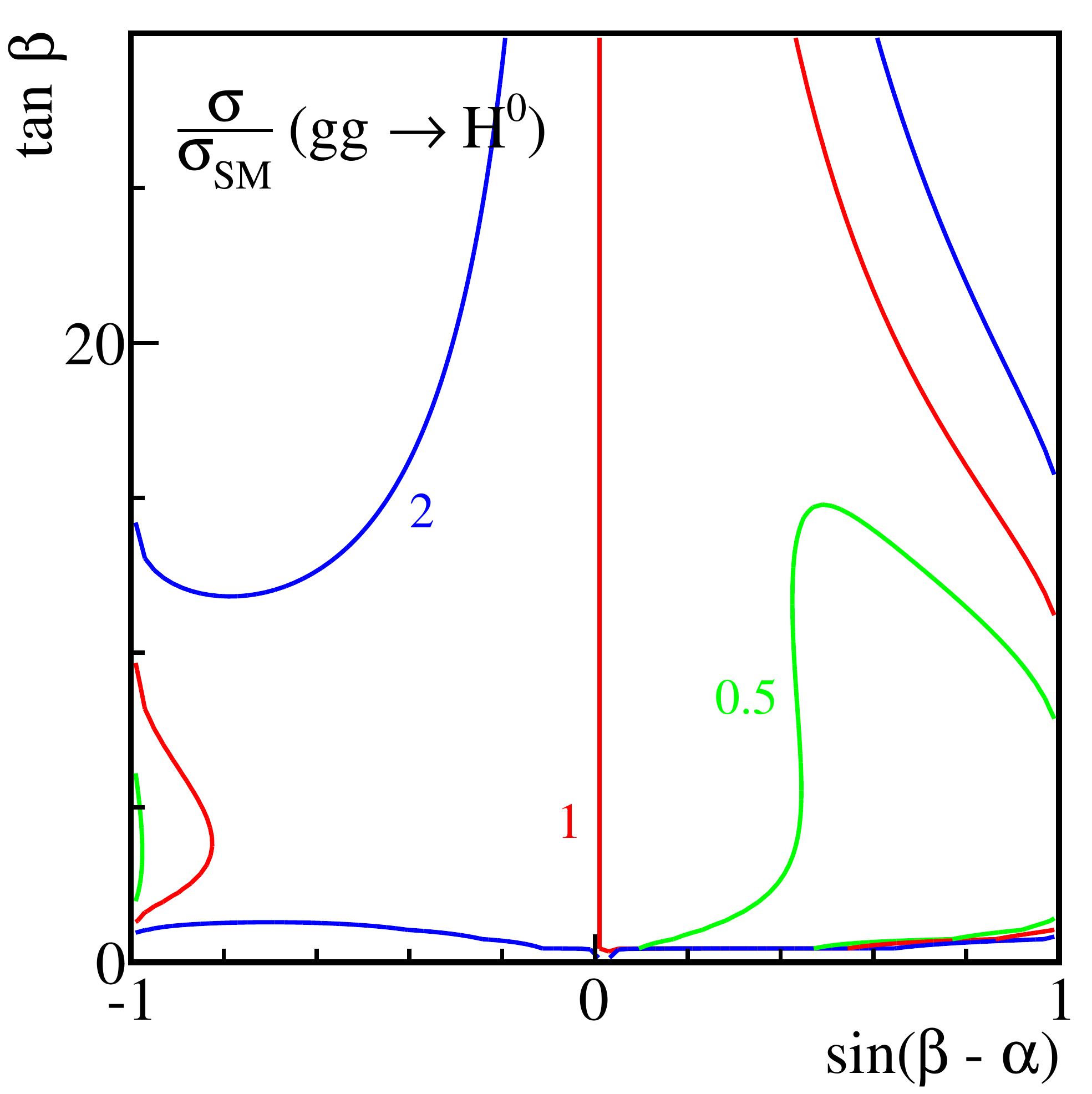}
	\includegraphics[scale=0.4]{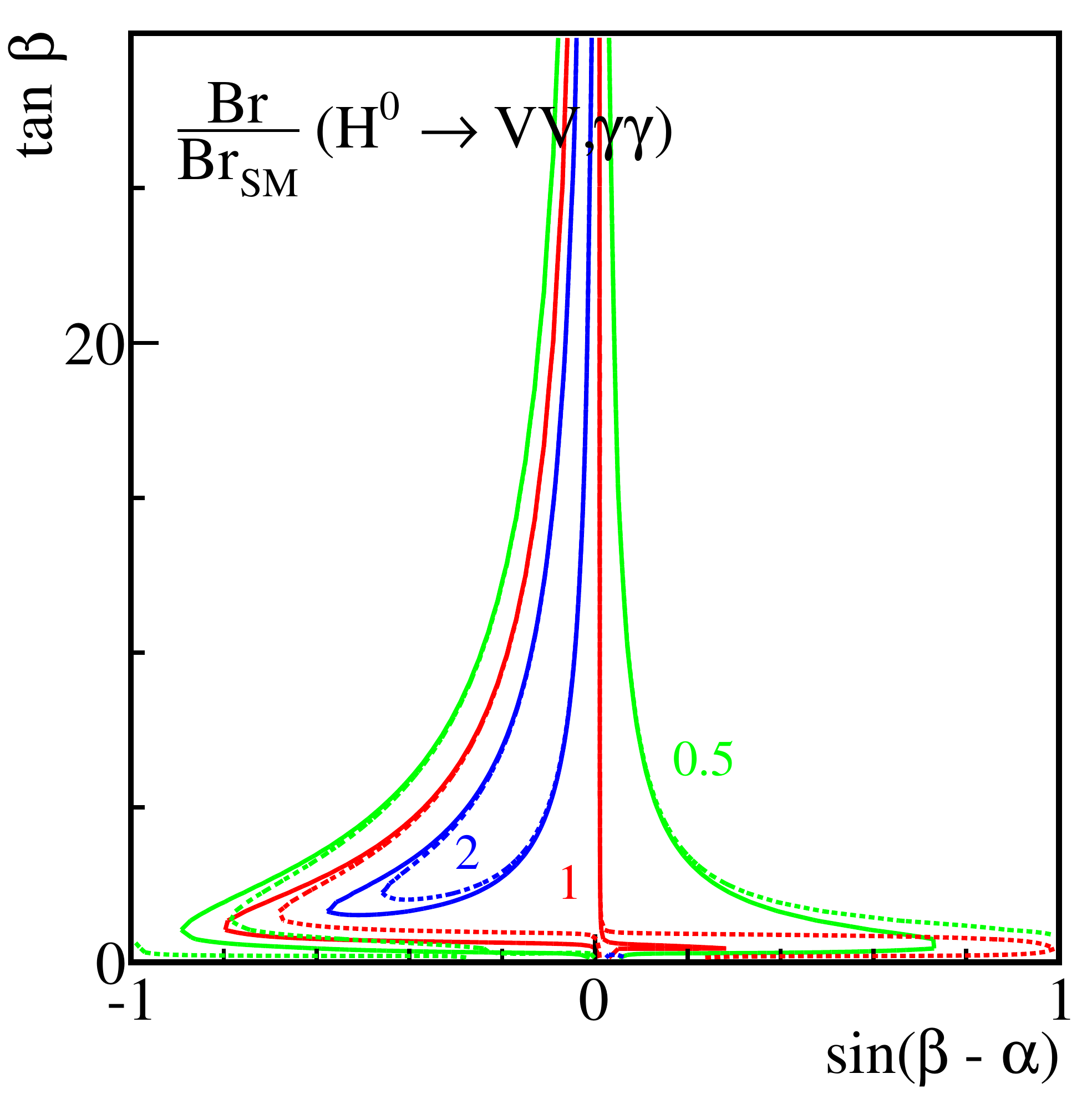}
 \caption{The normalized $gg\to \H$ production cross section contours (left panel) and $\H\to VV$ (solid lines of the right panel) and   $\H\to \gamma\gamma$  (dashed lines of the right panel) branching fractions in the $\H$-126 case.   The contour lines are $\sigma / \sigma_{\rm SM},\ {\rm Br}/{\rm Br}_{\rm SM}  = 0.5$ (green), 1 (red), and 2 (blue). }
\label{fig:sigma_Br_sep_Heavy}
\end{figure}

Similar to Eqs.~(\ref{eq:sigma_ggh}) and (\ref{eq:sigma_ggh_detail}) in Sec.~\ref{sec:h126}, the   ratios of the gluon fusion cross sections normalized to the SM can be written approximately as:  
\begin{eqnarray}
&&\frac{\sigma(gg\rightarrow \H)}{\sigma_{\rm SM}}=\frac{\sin^2\alpha}{\sin^2\beta}+\frac{\cos^2\alpha}{\cos^2\beta}\frac{|A_{1/2}(\tau_b)|^2}{|A_{1/2}(\tau_t)|^2}
\label{eq:sigma_ggh_Heavy}
\\
&&=\left[  \frac{\sin(\beta-\alpha)}{\tan\beta} -\cos(\beta-\alpha) \right]^2 + \left[  {\sin(\beta-\alpha)}{\tan\beta} +\cos(\beta-\alpha) \right]^2\frac{|A_{1/2}(\tau_b)|^2}{|A_{1/2}(\tau_t)|^2}.  
\label{eq:sigma_ggh_detail_Heavy}
\end{eqnarray}
Contours of $\sigma/\sigma_{\rm SM}(gg\rightarrow \H)=$ 0.5 (green), 1 (red), and 2 (blue) are shown in the left panel of Fig.~\ref{fig:sigma_Br_sep_Heavy}.  $\H$ couples exactly like the SM Higgs for $\sin(\beta-\alpha)=0$, while deviations from the SM values occur for  $\sin(\beta-\alpha)$ away from zero.  For $\sin(\beta-\alpha)<0$, $\sigma/\sigma_{\rm SM}(gg\rightarrow \H)$ is almost always larger than 1 (except for a small region around $\sin(\beta-\alpha)\sim -1$ and $\tan\beta \lesssim 10$)  while a suppression of the gluon fusion production is possible for positive values of $\sin(\beta-\alpha)$.  This is due to cancellations between the $\sin(\beta-\alpha)$ and  $\cos(\beta-\alpha)$ terms in the top Yukawa coupling, in particular, for low $\tan\beta$.   The bottom loop contributes significantly when $\tan\beta$ is large, which enhances the gluon fusion production cross section.

${\rm Br}(\H\rightarrow VV, \gamma\gamma)/{\rm Br}_{\rm SM}$ can also be expressed similar to Eq.~(\ref{eq:BR}): 
\begin{equation}
\frac{\textrm{BR}(\H\rightarrow XX)}{\textrm{BR}(h_{\rm SM}\rightarrow XX)}= \frac{\Gamma_{XX}}{\Gamma_{total}}\times \frac{\Gamma_{total}^{\rm SM}}{\Gamma_{XX}^{\rm SM}}
=\left\{
\begin{tabular}{c}
$\frac{\cos^2(\beta-\alpha)}
{\cos^2(\beta-\alpha){\rm Br}(h_{\rm SM}\rightarrow VV)
+\frac{\cos^2\alpha}{\cos^2\beta}{\rm Br}(h_{\rm SM}\rightarrow bb)+ \ldots}$\\
$\frac{\Gamma(\H\rightarrow \gamma\gamma)/\Gamma(h_{\rm SM}\rightarrow \gamma\gamma)}{\cos^2(\beta-\alpha){\rm Br}(h_{\rm SM}\rightarrow VV)+\frac{\cos^2\alpha}{\cos^2\beta}{\rm Br}(h_{\rm SM}\rightarrow bb) + \ldots} $
 \end{tabular}
 \right. , 
 \label{eq:BR_Heavy}
\end{equation}
with the contour lines given in the right panel of Fig.~\ref{fig:sigma_Br_sep_Heavy}.   A relative enhancement of the branching fractions over the SM values are   observed in extended region of negative   $\sin(\beta-\alpha)$, while it is mostly suppressed for positive  $\sin(\beta-\alpha)$.

\begin{figure}[htbp]
	\includegraphics[scale=0.4]{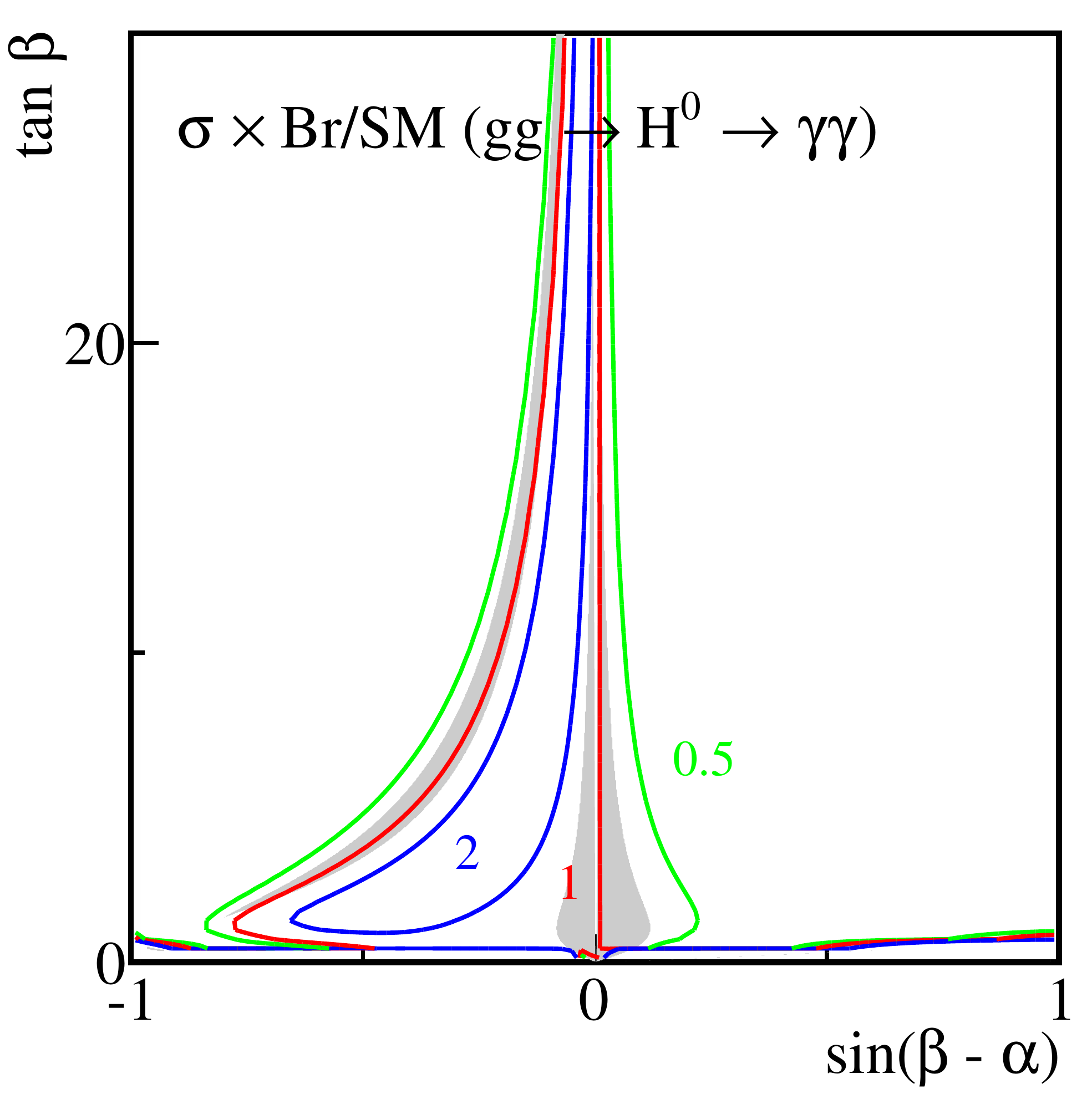}
	\includegraphics[scale=0.4]{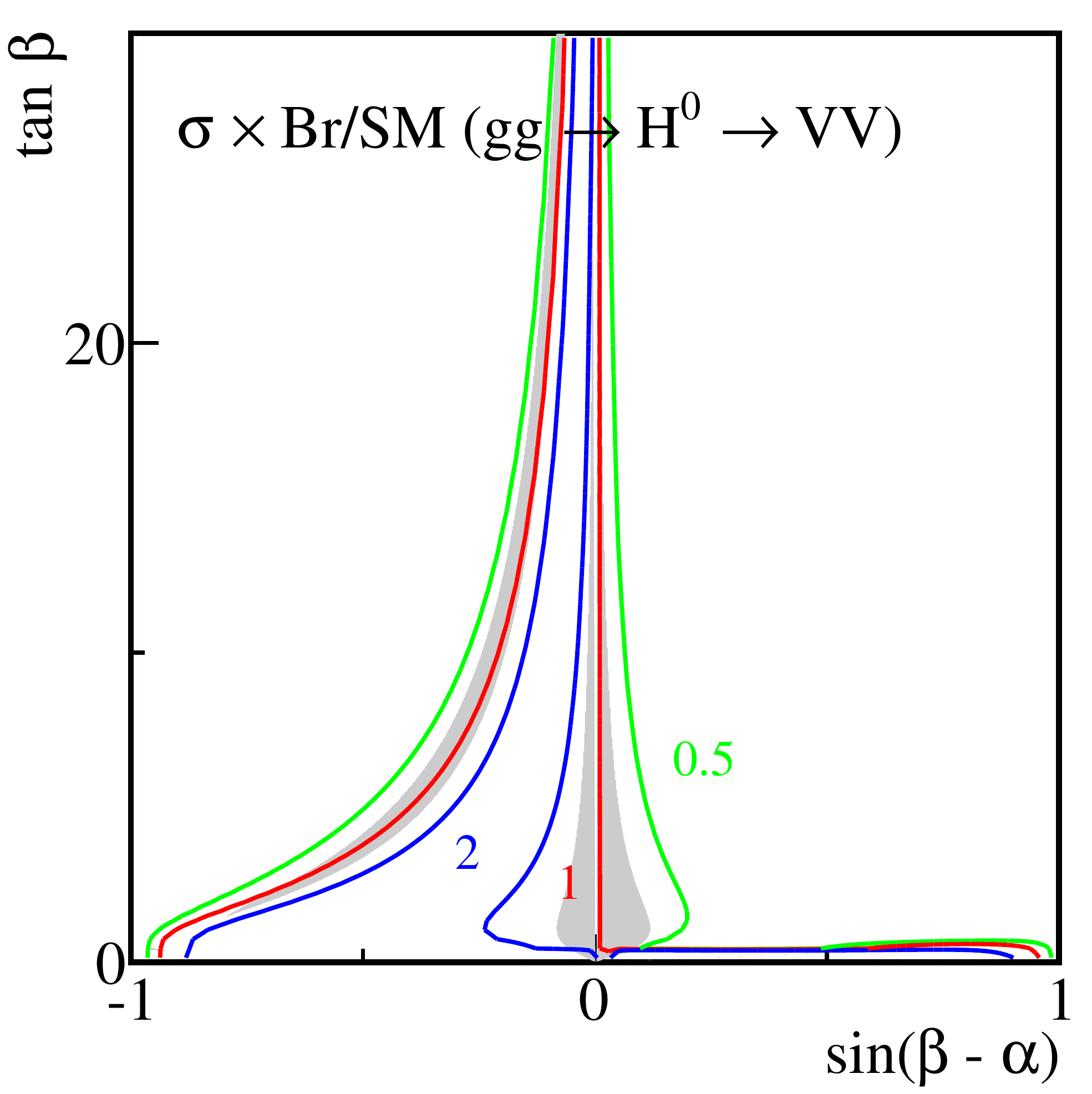}
 \caption{$\sigma \times{\rm Br} / {\rm SM}$ for the processes $gg \rightarrow \H \rightarrow \gamma \gamma$ (left), and $gg \rightarrow \H \rightarrow WW/ZZ $ (right) in the $\H$-126 case. The contour lines are $\sigma \times{\rm Br} / {\rm SM} = 0.5$ (green), 1 (red), and 2 (blue).  The regions where cross sections of $\gamma\gamma$ and $WW/ZZ$ channels satisfy Eq.~(\ref{eq:sigmabrrange}) are shaded gray.   }
\label{fig:Contour_Channel_Heavy}
\end{figure}

Combining   the production cross sections and the decay branching fractions,   contours   of $gg\rightarrow \H \rightarrow XX$ are given in Fig.~\ref{fig:Contour_Channel_Heavy} for $\gamma\gamma$ (left panel) and $WW/ZZ$ channels (right panel).  Requiring the cross section to be consistent with the observed Higgs signal:  0.7 $-$ 1.5 for the $\gamma\gamma$ channel and 0.6 $-$ 1.3 for the $WW/ZZ$ channel, results in two distinct regions: a   region close to $\sin(\beta-\alpha)\sim 0$, and an extended region of $-0.8 \lesssim  \sin(\beta-\alpha) \lesssim -0.05$.  

\begin{figure}[htbp]
	\includegraphics[scale=0.4]{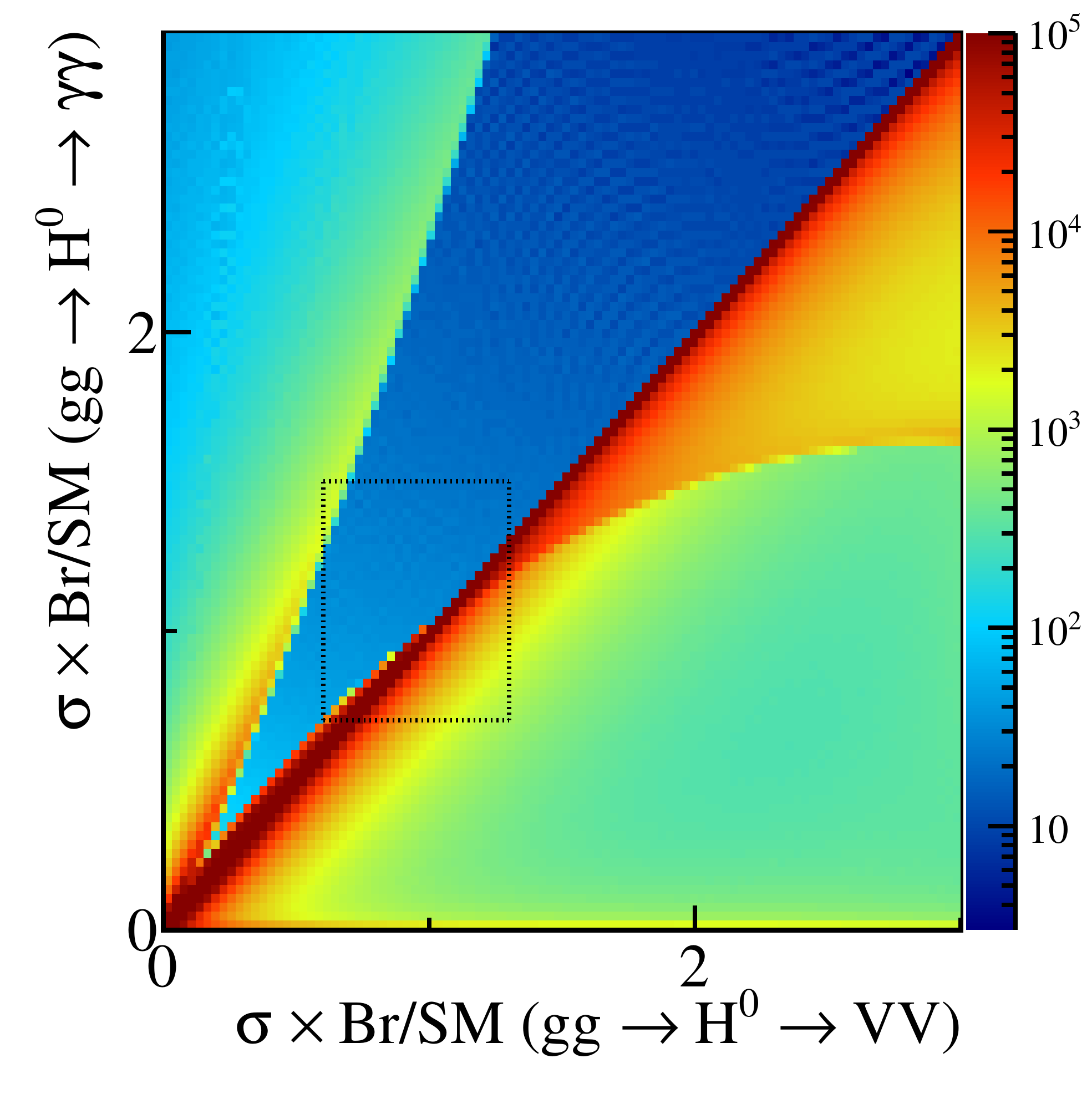}
 \caption{$\sigma\times{\rm Br} / {\rm SM}$ for $gg\rightarrow \H \rightarrow \gamma \gamma$ versus $ gg\rightarrow \H \rightarrow VV$ in the $\H$-126 case.  Color coding is the same as in Fig.~\ref{fig:Scatter_Light1}.  Also indicated by the small rectangular box  is the normalized signal cross section range of $\gamma\gamma$ between 0.7 and 1.5, and $VV$ channels between 0.6 and 1.3 \cite{CMS-PAS-HIG-13-005,Aad:2013wqa}. }
\label{fig:Scatter_Heavy}
\end{figure}

Fig.~\ref{fig:Scatter_Heavy} shows the correlation between the $\gamma\gamma$ and $VV$ channels.  Most of the points lie along the diagonal: $\gamma\gamma : VV \sim 1$.    A second branch of  $\gamma\gamma : WW \sim 2$ also appears, which corresponds to the very low $\tan\beta < 1$ region in Fig.~\ref{fig:Contour_Channel_Heavy}.  This region is strongly constrained by $R_b$ and  flavor bounds, and is therefore not considered further in our study.  

%%%%%%%%%%%%%%%
\subsection{Parameter Spaces}
%%%%%%%%%%%%%%%
\begin{figure}[htbp]
	\includegraphics[scale=0.4]{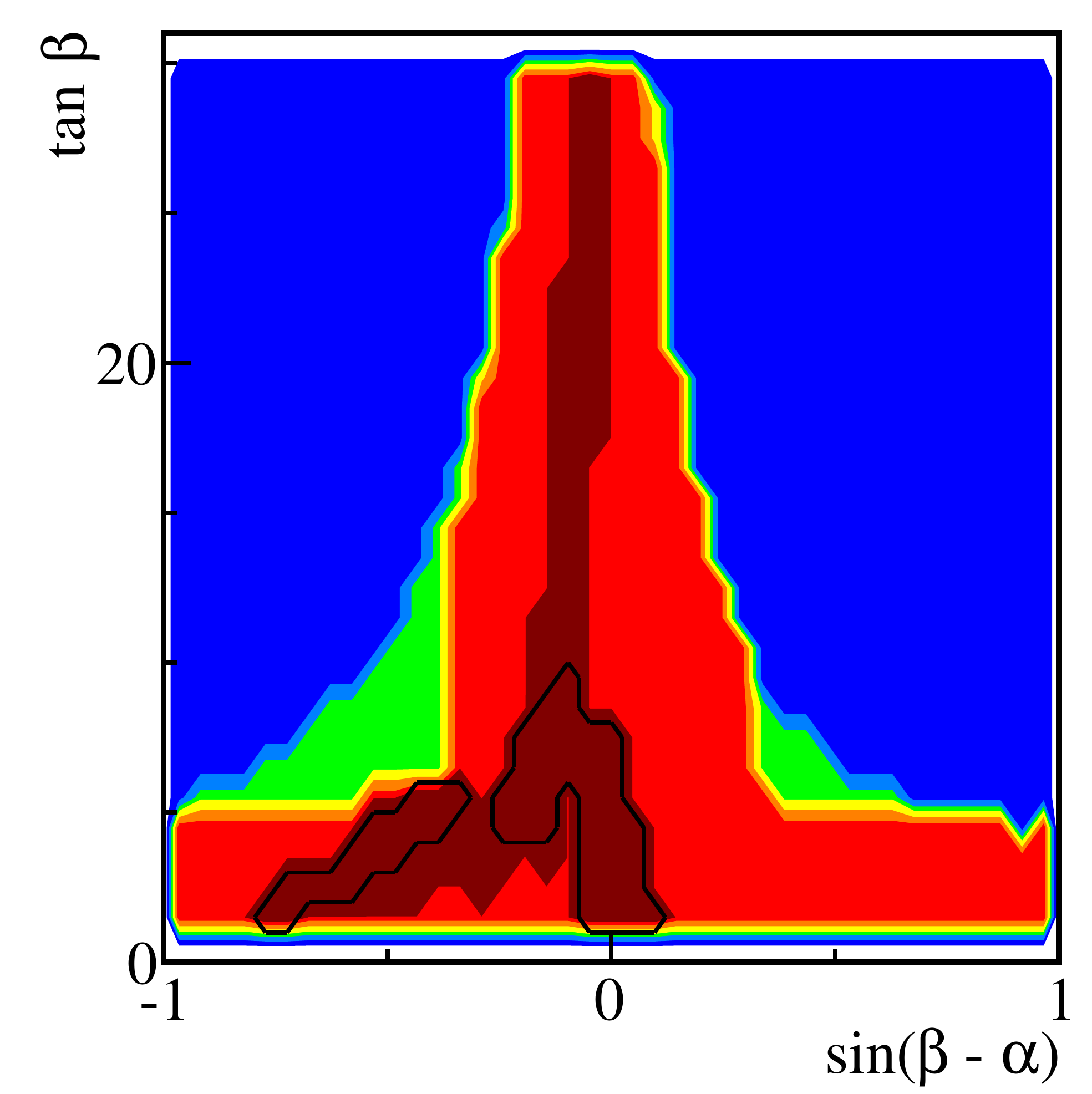}
	\caption{Parameter regions in the $\H$-126 case   for $\tan\beta$ versus $\sin(\beta-\alpha)$. Color coding is the same as Fig.~\ref{fig:sbatb-light} except that the dark red regions are the ones   consistent with the heavy CP-even Higgs interpreted as the observed Higgs signal.   }
	\label{fig:sbatb-heavy}
\end{figure}

We now present the results for $\H$-126 case with the full parameter scan, including all the theoretical and experimental constraints.  Fig.~\ref{fig:sbatb-heavy} presents the parameter regions in $\tan\beta$ versus $\sin(\beta-\alpha)$.  The color coding is the same as in Fig.~\ref{fig:sbatb-light}, except that the signal regions in dark red are those with the  heavy CP-even Higgs $\H$ interpreted as the observed 126 GeV scalar.    

Requiring the heavy CP-even Higgs to satisfy the cross section ranges of the observed Higgs signal results in two  signal regions: one region near $\sin(\beta-\alpha)\sim0$ and  an extended region of  $-0.8 \lesssim  \sin(\beta-\alpha) \lesssim -0.05$,  consistent with   Fig.~\ref{fig:Contour_Channel_Heavy}.   Note however that the   region around $\sin(\beta-\alpha) \sim 0$ is actually reduced to   $\tan\beta \lesssim 8$.  This is because larger values of $\tan\beta$ leads to smaller $m_h$ such that $m_h < m_H/2$ (see right panel of Fig.~\ref{fig:Heavy-mh} below).  The opening of $\H\rightarrow \h\h$ channel reduces the the branching fractions of $\H \rightarrow WW/ZZ, \gamma\gamma$ forcing it outside the signal cross section region. 
Regions surviving the flavor bounds are  the ones  enclosed by   black curves.  Larger values of $\tan\beta \gtrsim 10$ are disfavored.
 
\begin{figure}[htbp]
	\includegraphics[scale=0.4]{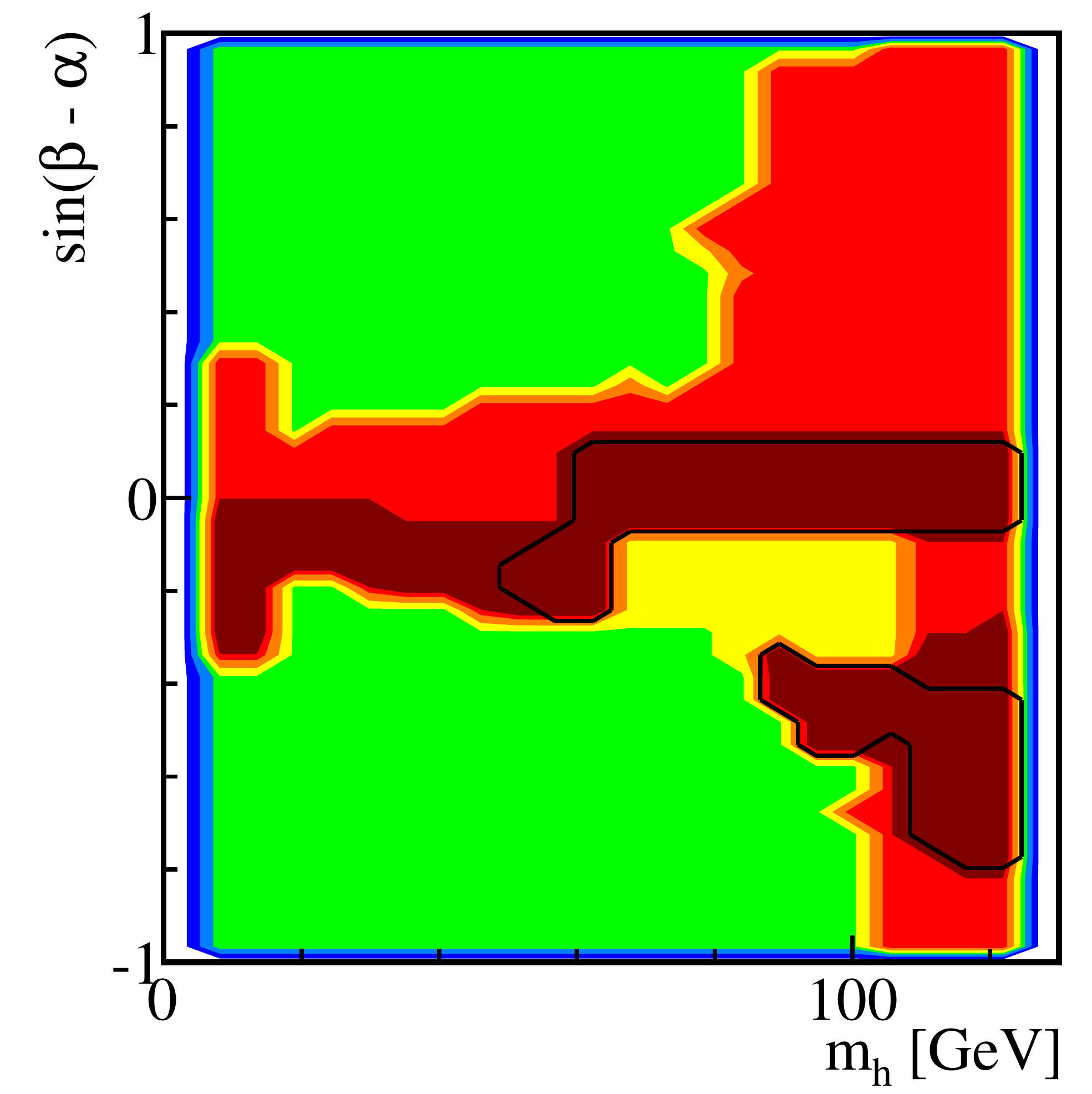}
	\includegraphics[scale=0.4]{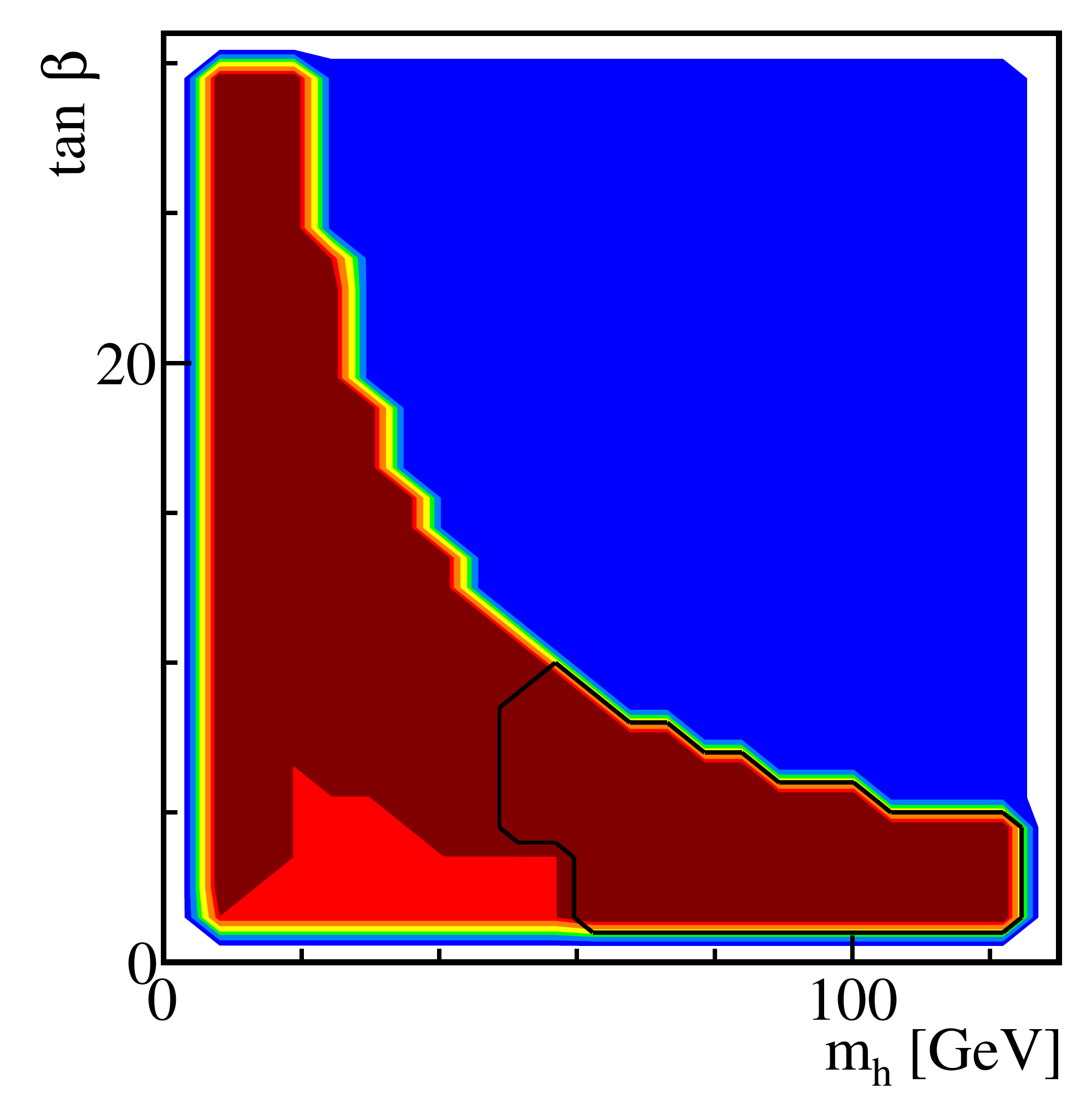}
\caption{Parameter regions in the $\H$-126 case  for  $\sin(\beta-\alpha)$ versus $m_h$ (left panel) and 
  $\tan\beta$ versus $m_h$ (right panel).  Color coding is the same as Fig.~\ref{fig:sbatb-heavy}.}
\label{fig:Heavy-mh}
\end{figure}

Fig.~\ref{fig:Heavy-mh} shows the parameter region in $\sin(\beta-\alpha)$ versus $m_h$ (left panel) and 
  $\tan\beta$ versus $m_h$ (right panel).   Within the narrow region around $\sin(\beta-\alpha)\sim 0$, $m_h$ can take all values up to 126 GeV.   For  $-0.8 \lesssim   \sin(\beta-\alpha) \lesssim -0.35$, when the $\H WW, \H ZZ$ couplings could   significantly deviate from the SM value while $\h WW$, $\h ZZ$ couplings are sizable, the light CP-even Higgs mass is constrained  to be larger than about 80 GeV from  LEP Higgs searches \cite{lep98, lep98b}.  This is  the interesting region where the two Higgses are close to being degenerate,  with both $\h$ and $\H$ showing significant  deviation of their couplings to gauge bosons from the SM value.

The right panel of Fig.~\ref{fig:Heavy-mh} shows the parameter region of  $\tan\beta$ versus $m_h$.   Larger values of $\tan\beta$ is only allowed for small values of $m_h$.   The red region where $m_h<60$ GeV and    $\tan\beta \lesssim 5$ can not satisfy the Higgs signal cross section requirement due to the opening of $\H\rightarrow \h\h$ mode, which corresponds to the $m_h<60$ GeV, $\sin(\beta-\alpha)\sim0$ red region in the    $\sin(\beta-\alpha)$ versus $m_h$ plot (left panel of  Fig.~\ref{fig:Heavy-mh}).  
 Imposing the  flavor bounds further rules out regions with light $m_h$ below about 50 GeV,  mainly due to the  process $B_s\to\mu^+\mu^-$, as shown in the right panel of Fig.~\ref{fig:Flav_new}.   Large values of  $\tan\beta \gtrsim 10$ are excluded  correspondingly.

\begin{figure}[htbp]
	\includegraphics[scale=0.4]{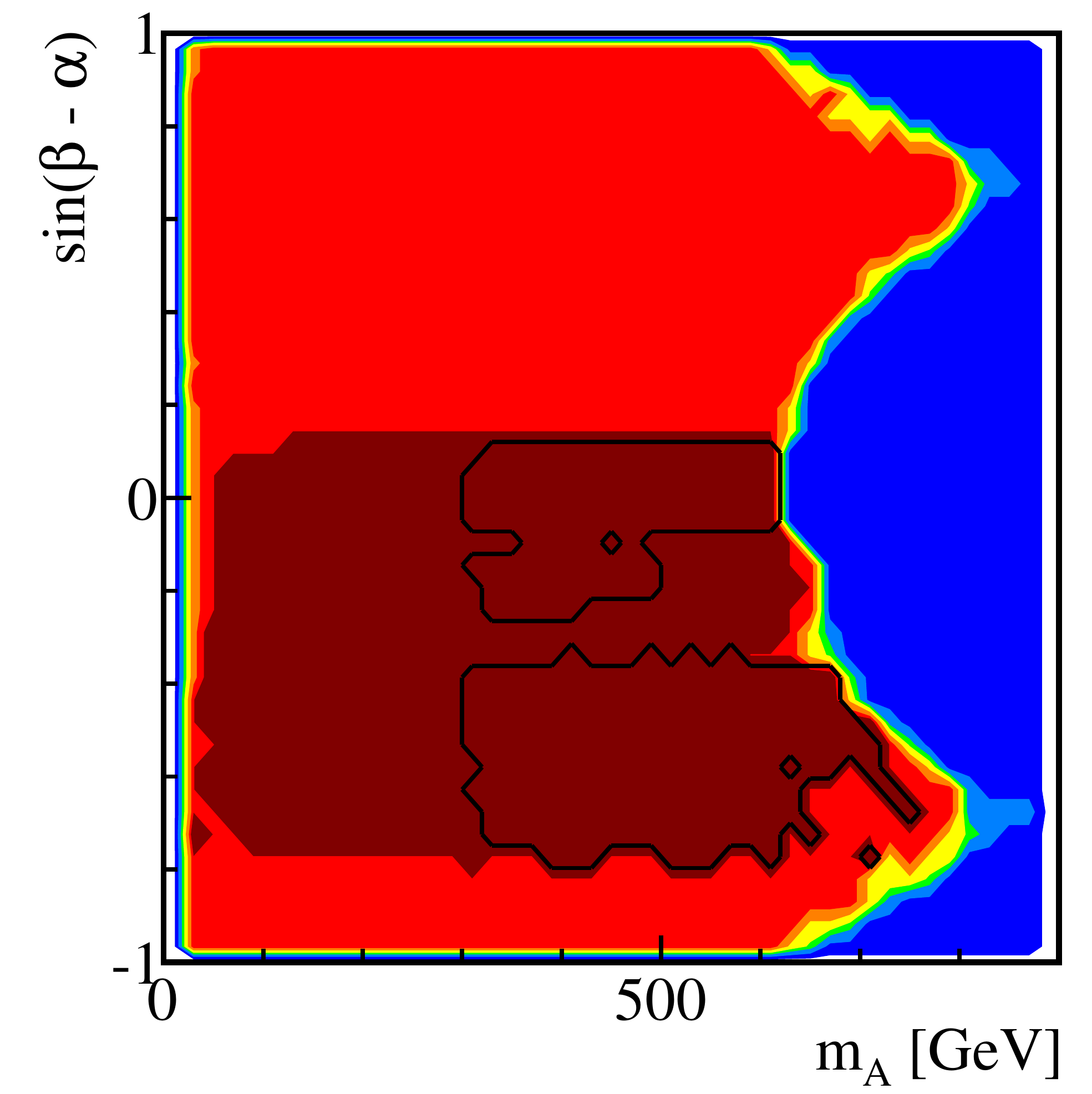}
	\includegraphics[scale=0.4]{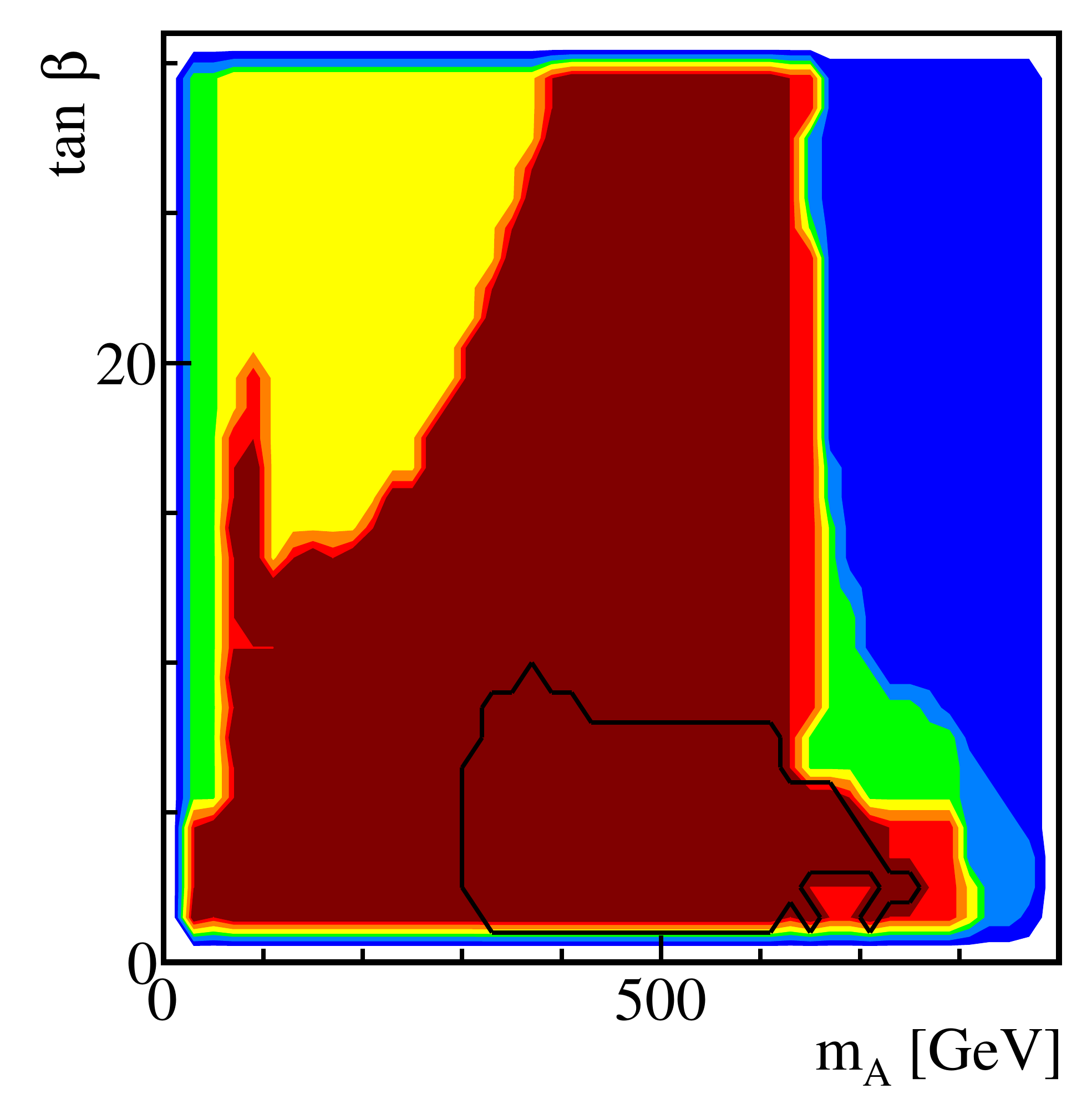}
	\includegraphics[scale=0.4]{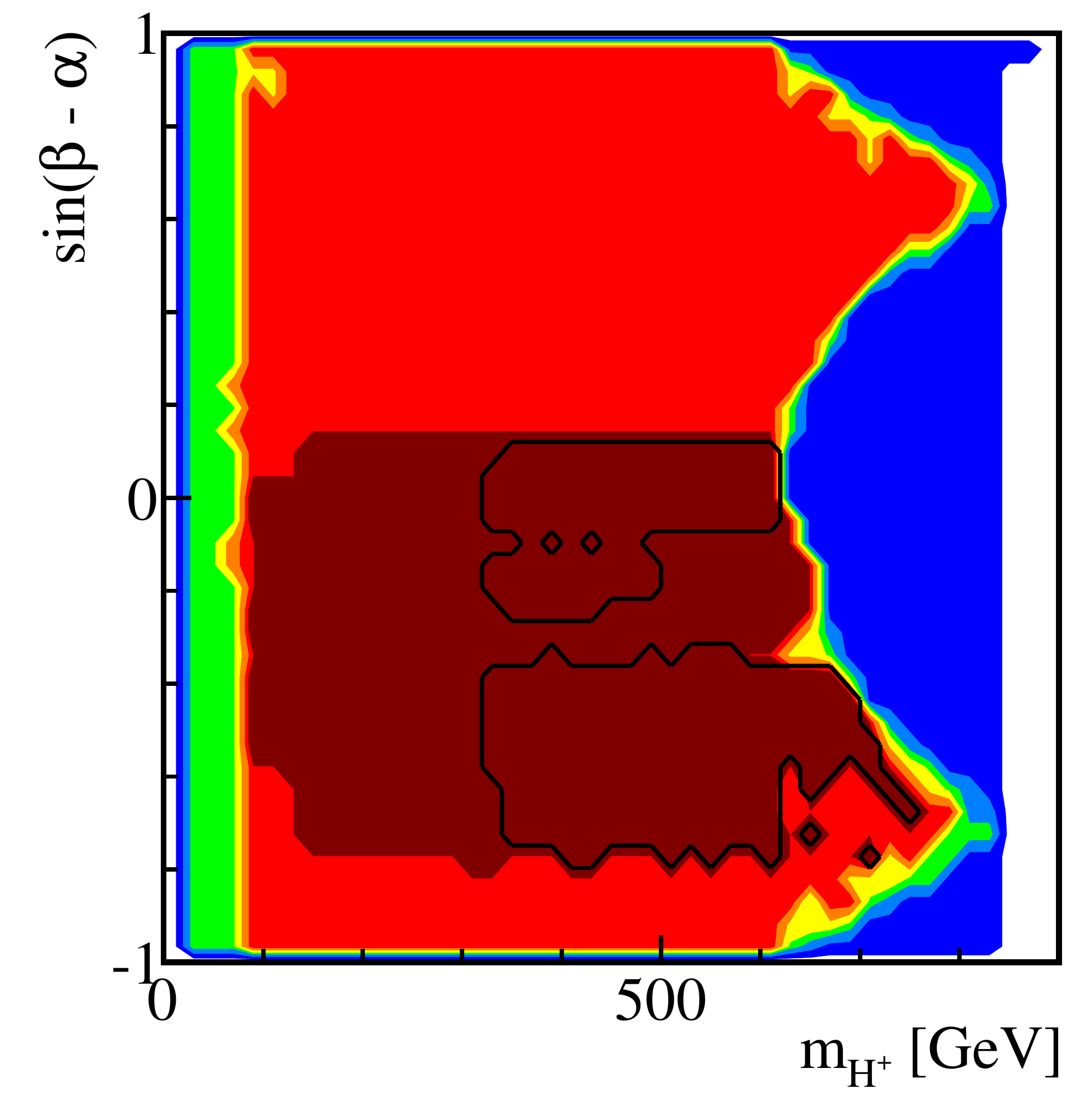}
	\includegraphics[scale=0.4]{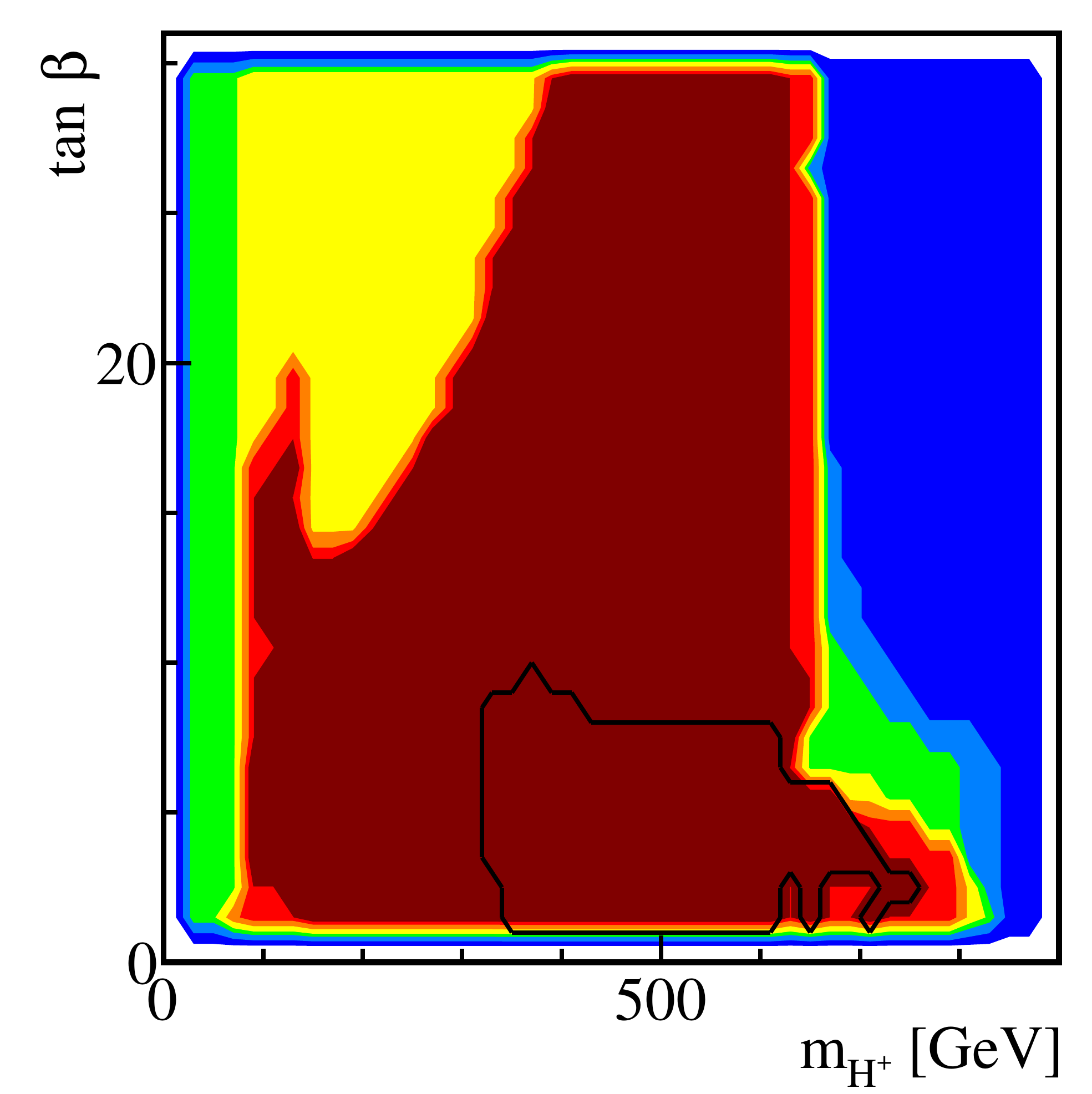}
\caption{Parameter regions in the $\H$-126 case  for    $\sin(\beta-\alpha)$ versus $m_A$ (upper left panel) and  $\tan\beta$ versus $m_A$  (upper  right panel), as well as similar plots for $m_H^\pm$ (lower panels). Color coding is the same as Fig.~\ref{fig:sbatb-heavy}.   }
\label{fig:Heavy-mCmA}
 \end{figure}
 
Fig.~\ref{fig:Heavy-mCmA} shows $\sin(\beta-\alpha)$  versus   $m_{A, H^\pm}$   (left panels) and  $\tan\beta$  versus   $m_{A, H^\pm}$   (right panels).  The plots for $m_A$ and $m_{H^\pm}$ are very similar, except for very low masses.    Very large values of $m_{A, H^\pm} \gtrsim 800$ GeV  are excluded by theoretical considerations, similar to   the $\h$-126 case.   $m_A \lesssim 60$ GeV and $\tan\beta \gtrsim 5$ are excluded by the LEP Higgs search \cite{lep98}, while the triangle region of  $130 \lesssim m_A \lesssim 400$ GeV and $\tan\beta \gtrsim 13$ is excluded by the LHC searches for the CP-odd Higgs in $\tau\tau$ mode \cite{ATLAS_MSSM, CMS_MSSM}.   For the charged Higgs,  small values of $m_{H^\pm} \lesssim 80$ GeV are  ruled out by LEP searches on charged Higgs \cite{lepcharged0, lepcharged1}.  Tevatron and the LHC charged Higgs searches  \cite{ATLAS_MSSM, CMS_MSSM}: $t \rightarrow H^\pm b \rightarrow \tau \nu_\tau b$ further rule out regions of $m_{H^\pm} \lesssim 150$ GeV and $\tan\beta \gtrsim 17$.  
The   triangle in  $m_{H^\pm}$ versus $\tan\beta$ plot for 150 GeV $\lesssim m_{H^\pm} \lesssim 400$ GeV and $\tan\beta \gtrsim 13$ is translated from the corresponding region in $\tan\beta$ versus $m_{A}$, due to the correlation between $m_A$ and $m_{H^\pm}$ introduced by $\Delta\rho$, as shown below in Fig.~\ref{fig:Heavy-mass}.  Imposing the flavor constraints further limits $m_A \gtrsim 300$ GeV, $m_{H^\pm} \gtrsim 300$ GeV and $\tan\beta \lesssim 10$.

\begin{figure}[htbp]
	\includegraphics[scale=0.4]{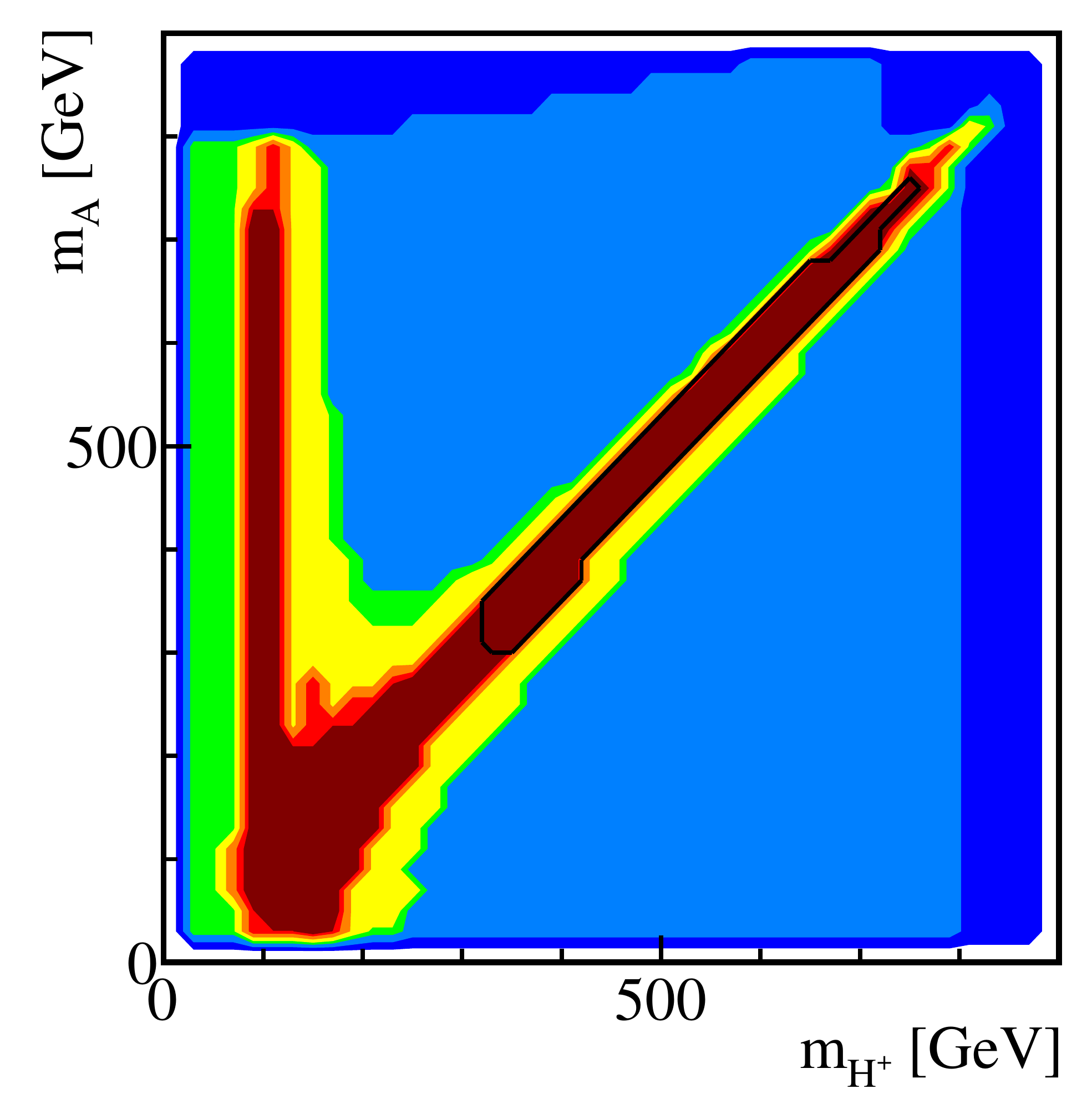}
	\includegraphics[scale=0.4]{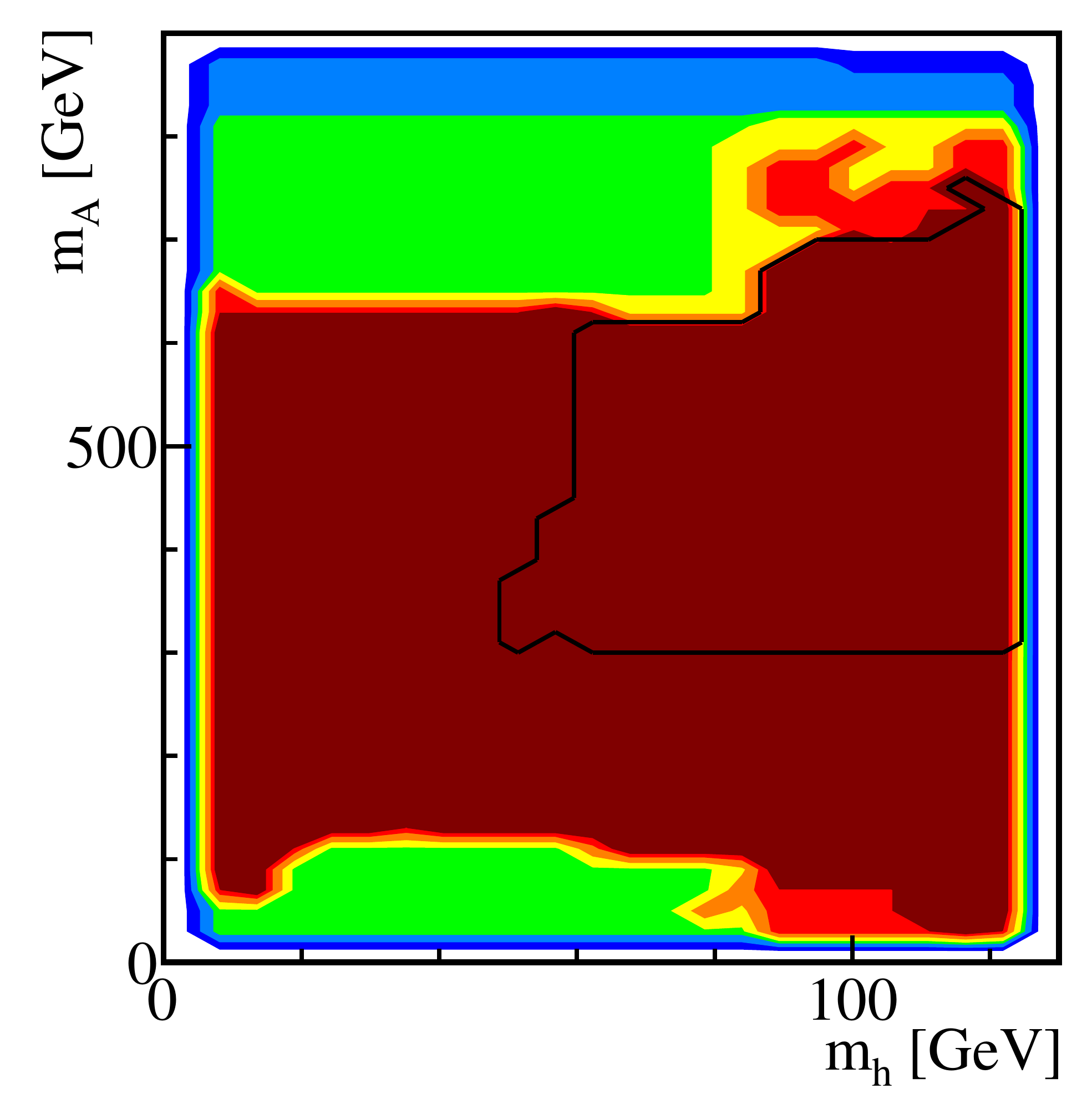}
\caption{Parameter regions in the $\H$-126 case  for    $m_A$ versus  $m_{H^\pm}$ (left panel) and  $m_h$ (right panel). Color coding is the same as Fig.~\ref{fig:sbatb-heavy}.  }
\label{fig:Heavy-mass}
\end{figure}

 $m_A$ and $m_{H^\pm}$ exhibit a much stronger correlation in the $\H$-126 case, mostly due to the the $\Delta\rho$ constraints, as shown in the left panel of Fig.~\ref{fig:Heavy-mass}.  Comparing with the $\h$-126 case, in which $m_H$ could be large with a relaxed constraints on $m_A$ and $m_{H^\pm}$ mass correlation, in the $\H$-126 case, both $m_h$ and $m_H$ are relatively small.   $m_A$ and $m_{H^\pm}$  should  therefore be  highly correlated in order to avoid large custodial symmetry breaking in the Higgs sector.     However, there is a small strip of allowed region at $m_{H^\pm}\sim100$ GeV with $m_A$   between 200 $-$ 700 GeV.  This region escapes the $\Delta\rho$ constraint since for $m_{H^\pm} \sim m_h \sim m_H$, the contribution to $\Delta\rho$ introduced by the large mass difference between $m_A$ and $m_{H^\pm}$ is cancelled by the $(\h,\ \A)$ loop and $(\H,\ \A)$ loop.   Imposing the flavor constraints again limits $m_{H^\pm}$ to be larger than 300 GeV.  $m_A$ is constrained to be more than 300 GeV as well due to the correlations.

The right panel of   Fig.~\ref{fig:Heavy-mass} shows the parameter region of $m_A$ versus $m_h$, which does not show much  correlation.   For $m_h \lesssim$ 90 GeV, low values of $m_A \lesssim 100$ GeV is excluded by LEP searches of $\h \A$ channel \cite{lep98}.  High values of $m_A \gtrsim$ 600 GeV are excluded for $m_h<90$ GeV. This is because such a large value of $m_A$ can only be realized for $|\sin(\beta-\alpha)|>0.3$ (see the upper-left panel of Fig.~\ref{fig:Heavy-mCmA}).  Such regions of $|\sin(\beta-\alpha)|>0.3$ and $m_h<90$ GeV are excluded by the LEP Higgs search of $\h Z$ channel \cite{lep98b}, as shown clearly in the $m_h$ versus $\sin(\beta-\alpha)$ plot (left panel of Fig.~\ref{fig:Heavy-mh}).  Such excluded regions for large $m_A$ (and large $m_{H^\pm}$ due to correlation) also appears in the $\tan\beta$ versus $m_A$ $(m_{H^\pm})$ plots  in Fig.~\ref{fig:Heavy-mCmA}.

We end the section with the following observations:
\begin{itemize}
 \item Contrary to the $\h$-126 case, fixing the heavy CP-even  Higgses to be the 126 GeV resonance forces us into a small narrow region of $\sin(\alpha-\beta)\sim 0$ with $\tan\beta\lesssim 8$  or  an extended region of $-0.8 \lesssim \sin(\alpha-\beta) \lesssim -0.05$ with   less restrictions  on  $\tan\beta$.  
 \item The light CP-even Higgs  can have mass of any value up to 126 GeV, with smaller $m_h$ only allowed for $\sin(\beta-\alpha)\sim 0$.
 Note that the case of nearly degenerate $\h$ and $\H$ is allowed, as studied in detail  in Ref.~\cite{Ferreira:2012nv}.   \item $m_A$ and $m_{H^\pm}$ exhibit a strong correlation: $m_A \simeq m_{H^\pm}$,   due to $\Delta \rho$ constraints.
 \item Flavor bounds impose the strong constraints:  $\tan\beta \lesssim 10$, $m_h > 50$ GeV, and $m_{H^\pm}>$ 300 GeV.  $m_A$ is also constrained to be more than 300 GeV due to the correlation between $m_A$ and $m_{H^\pm}$.
 \end{itemize}

%%%%%%%%%%%%%%%%
\section{Other Higgs Channels}
\label{sec:future}
%%%%%%%%%%%%%%%%
Thus far, we have concentrated on the gluon fusion production mechanism and the dominant $\gamma\gamma$, $ZZ$ and $WW$ decay channels for the Higgs.    The vector boson fusion    channel is another important production channel for the CP-even Higgses.   For certain Higgs decay channels, for example, $\tau\tau$ mode, VBF production is the one  that provides the dominant sensitivity due to the excellent discrimination of the backgrounds using the two forward tagging jets and the central jet-veto \cite{Rainwater:1998kj}.  
Other production channels, $VH$ and $ttH$ associated production, can also be of interest for Higgs decay to $bb$. 
   In this section, we discuss the cross sections  in other search channels for both $\h$ and $\H$ when they are interpreted as the observed 126 GeV scalar.
   
\begin{figure}[htbp]
	\includegraphics[scale=0.4]{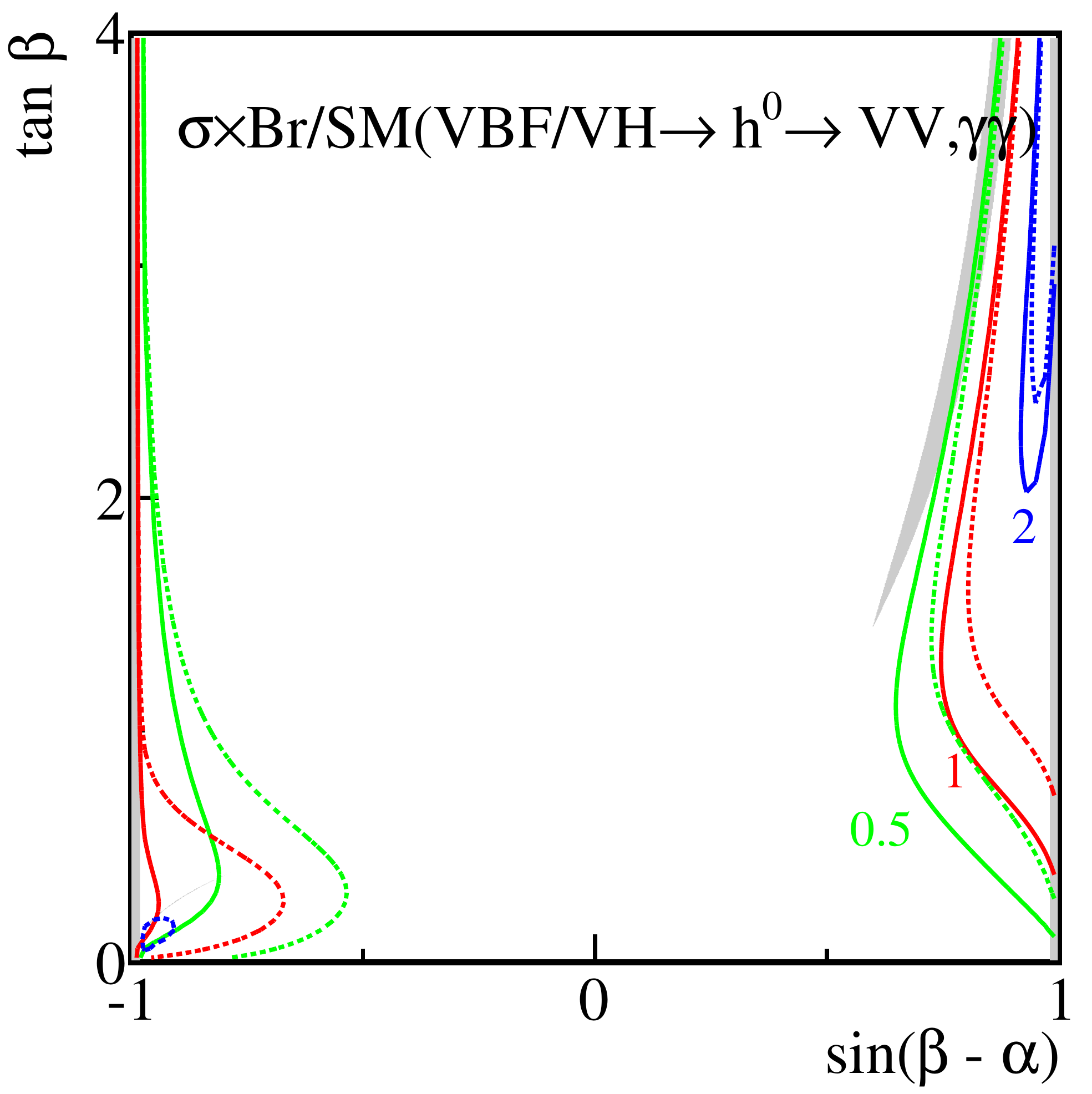}
	\includegraphics[scale=0.4]{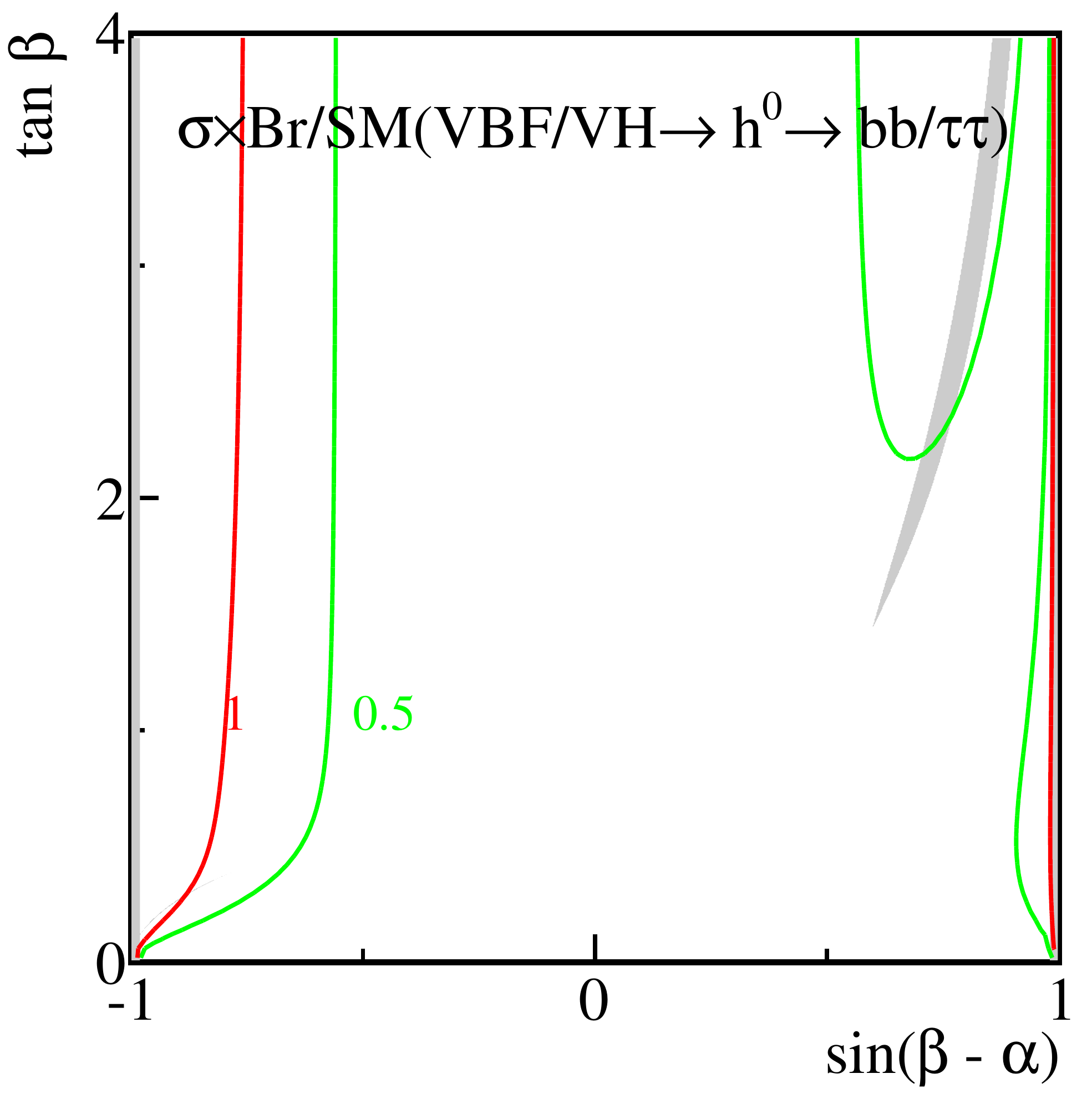}
\caption{$\sigma\times{\rm Br} / {\rm SM}$ for   $VBF/VH \rightarrow \h \rightarrow WW/ZZ$ (solid curves in left panel),  $\gamma\gamma $ (dashed curves in left panel)  and $VBF/VH \rightarrow \h \rightarrow bb/\tau\tau$ (right panel) for the $\h$-126 case. The contour lines show $\sigma\times{\rm Br} / {\rm SM}= 0.5$(green), 1 (red) and 2 (blue).  The shaded gray regions correspond to the signal regions where cross sections of $\gamma\gamma$ and $WW/ZZ$ channels satisfy Eq.~(\ref{eq:sigmabrrange}) as well as $R_b$. }      
\label{fig:Contour_VVChannel_Light}
\end{figure}

In Fig.~\ref{fig:Contour_VVChannel_Light}, we show the  normalized cross sections    for the $WW/ZZ$, $\gamma\gamma$ (left panel) and $bb/\tau\tau$ (right panel)    final states via VBF or $VH$ associated production (both production cross sections are controlled by $\h VV$ coupling) in the $\tan\beta$ versus $\sin(\beta-\alpha)$ plane   for the $\h$-126 case. 
For $VBF/VH \rightarrow \h \rightarrow WW/ZZ$, both the production and decay are proportional to $\sin(\beta-\alpha)$, resulting in regions highly centered around $\sin(\beta-\alpha) \sim \pm 1$ for any enhancement above the SM value.
For the currently preferred gray Higgs signal   regions, $VBF/VH \rightarrow \h\rightarrow WW/ZZ$ is typically in the range of 0.5 $-$ 1 of the SM value.
  
The current observation of the Higgs signal has been fitted into the signal strength in both the gluon fusion channel and VBF channel for $\gamma\gamma$, $WW$ and $ZZ$ final states \cite{ATLAS_spin_coupling,Aad:2013wqa,CMS-PAS-HIG-13-005}.  Imposing the 95\% C.L. contours of the  $\mu_{ggF+ttH}\times {\rm B}/{\rm B_{SM}}$ versus   $\mu_{VBF+VH}\times {\rm B}/{\rm B_{SM}}$ on top of the one-dimensional gluon fusion signal regions as given in Eq.~(\ref{eq:sigmabrrange})  does not lead to additional reduction of the signal parameter space, given the VBF channel is relatively loosely constrained.

For $VBF/VH\rightarrow \h \rightarrow bb/\tau\tau$, the cross section is suppressed for most of the regions, except in the neighborhood of $\sin(\beta-\alpha)=\pm 1$ where SM rates can   be achieved.  The current preferred signal regions typically have a suppression of 0.5 or stronger for this $bb/\tau\tau$ channel.
There is also a strong inverse correlation between the $WW/ZZ$ and $bb/\tau\tau$ channels, since an increase in  $bb$ decay branching fraction can only occur at the expense of $WW$.    Given the relatively loose bounds on the signal strength in the $bb$ and $\tau\tau$ channels from   the LHC and the Tevatron experiments \cite{CMS-PAS-HIG-13-005,CMS-PAS-HIG-13-019, ATLAS-CONF-2013-079,ATLAS-CONF-2012-135,Aaltonen:2013kxa},  imposing the current search results for $bb$ and $\tau\tau$ channels does not lead to further reduction of the signal parameter space.

\begin{figure}[htbp]
	\includegraphics[scale=0.4]{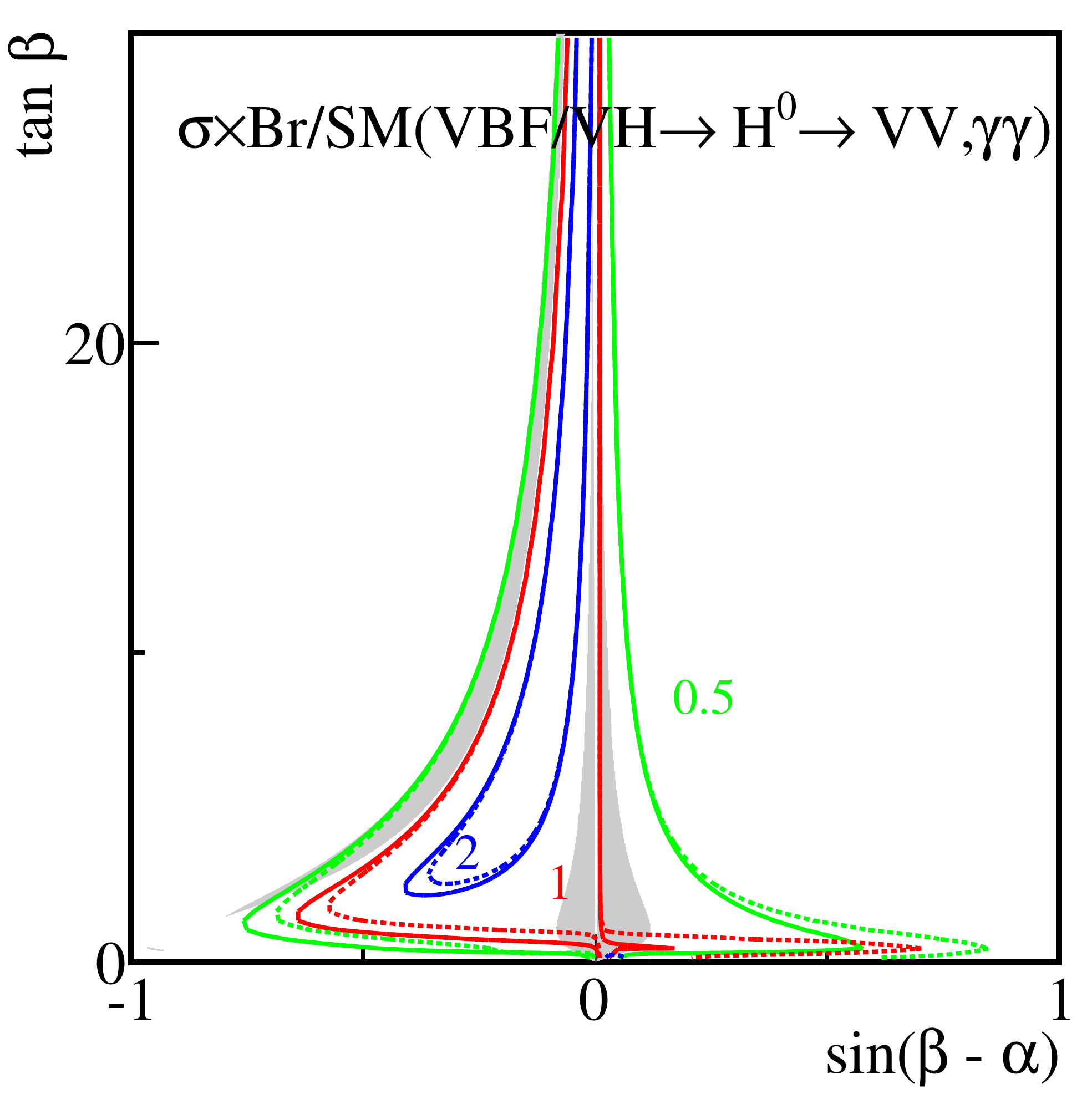}
	\includegraphics[scale=0.4]{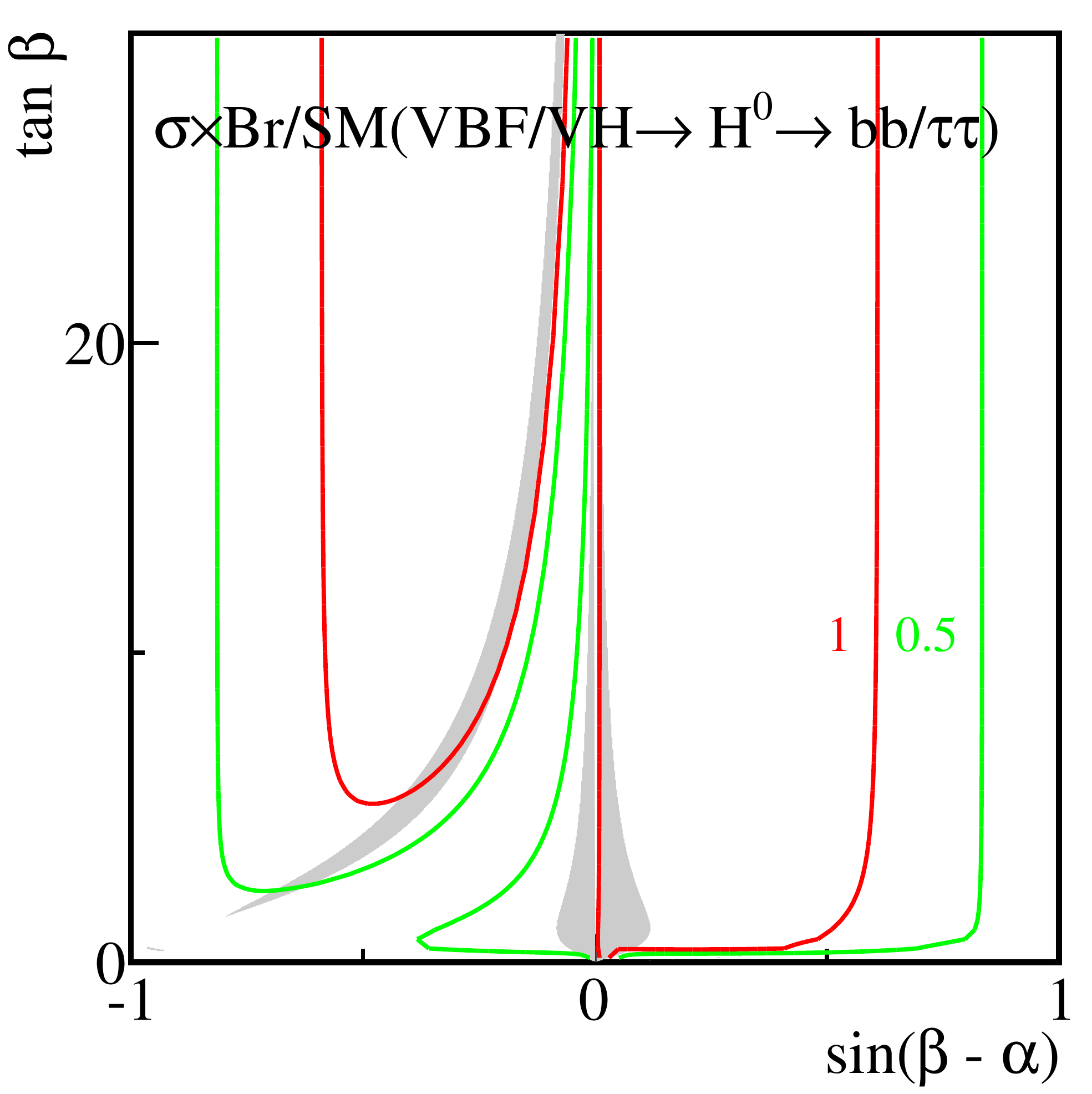}
\caption{ $\sigma\times{\rm Br}/{\rm SM}$ for   $VBF/VH \rightarrow \H \rightarrow WW/ZZ, \gamma\gamma $ (left) and $VBF/VH \rightarrow \H \rightarrow bb/\tau\tau$ (right) for the $\H$-126 case.   Color coding is the same as in Fig.~\ref{fig:Contour_VVChannel_Light}.   }
\label{fig:Contour_VVChannel_Heavy}
\end{figure}

Fig.~\ref{fig:Contour_VVChannel_Heavy} show the   $\sigma\times{\rm Br}/{\rm SM}$ plots for  $VV$, 
  $\gamma\gamma$,  and $bb/\tau\tau$ channel via VBF/$VH$ production for the $\H$-126 case. The qualitative features of the $VV$, $\gamma\gamma$ plot is the same as that of Fig.~\ref{fig:Contour_Channel_Heavy}.  The currently favored gray  signal   regions typically correspond to  a normalized  cross section of $VBF/VH \rightarrow \H \rightarrow WW/ZZ$ around 1 as well.  

The $bb/\tau\tau$ channel, however, exhibits  a very different  behavior.  For two regions of  $-0.6\leq\sin(\beta-\alpha)\leq -0.1 $ and $0 \leq\sin(\beta-\alpha)\leq 0.6 $ (regions enclosed by   the red curves in the right panel of Fig.~\ref{fig:Contour_VVChannel_Heavy}), a normalized cross section of at least the SM signal strength can be achieved.   A strong suppression, sometime as small as 0.1,  can be obtained in the other regions.   The currently favored gray signal region near $\sin(\beta-\alpha)\sim 0$
corresponds to $\sigma/\sigma_{\rm SM}$ of order 1 for $VBF/VH \rightarrow \H \rightarrow bb/\tau\tau$ channel, while a suppression as large as 0.5 is possible for  the extended regions in negative $\sin(\beta-\alpha)$.  The inverse correlation between $bb/\tau\tau$ and $WW$ channels also appears in the $\H$-126 case.   
Similar to the $\h$-126 case, imposing the 95\% C.L. range for the VBF process for $\gamma\gamma$ and $WW/ZZ$ channel, as well as the signal strength obtained from the $bb$ and $\tau\tau$ modes does not lead to further reduction of the signal region.

We also studied $gg\rightarrow \h, \H \rightarrow bb/\tau\tau$ channel for both the $\h$-126 and $\H$-126 cases,  and noticed that for the currently favored Higgs signal regions, a factor of 2 enhancement could be realized. 
%%%%%%%%%%%%
\section{Conclusions}
\label{sec:conclusions}
%%%%%%%%%%%%%%
In this paper, we presented a detailed analysis of  the Type II 2HDM (with an imposed $Z_2$ symmetry)  parameter space,  identifying either the light or the heavy CP-even Higgs as the recently discovered resonance at   126 GeV.   
We scanned the remaining five parameters $\sin(\beta-\alpha)$, $\tan\beta$, $m_A$, $m_{H^\pm}$,  and $m_H$ or $m_h$ while fixing either $m_h $ or $m_H$ to be 126 GeV.    
We took into  account all the theoretical constraints, precision measurements, as well as current experimental search limits on the Higgses.  We further studied the implications on the parameter space once flavor constraints are imposed.  We found unique features in each of these two cases. 

 In the $\h$-126 case, we are forced into   regions of parameter space where $\sin(\beta-\alpha) = \pm 1$ with $\tan\beta$ between 0.5 to 4, or an extended region of $0.55 < \sin(\beta-\alpha) < 0.9$, with $\tan\beta$  constrained to be in the range of 1.5 to 4.
There is, however, a wide range of   values  that are still allowed for the masses of the heavy CP-even, pseudo scalar and charged Higgses.   The Higgs masses are typically not correlated,   except when $m_{A, H^\pm}\gtrsim 600$ GeV and $\sin(\beta-\alpha)>0$ where there is a strong correlation between $m_A$ and $m_{H^\pm}$ because of  the $\Delta \rho$ constraint. 
Imposing flavor constraints further restricts   $m_{H^\pm} > 300$ GeV.

In the $\H$-126 case, we are forced into an orthogonal region of parameter space where $\sin(\beta-\alpha) \sim 0$, $\tan\beta\lesssim 8$  or  an extended region of  $-0.8 \lesssim \sin(\alpha-\beta) \lesssim -0.05$ with less restricted  $\tan\beta$.   $m_A$ and $m_{H^\pm}$ exhibit strong correlations: $m_A \simeq m_{H^\pm}$,   due to the $\Delta \rho$ constraint.  The interesting scenario of the light CP-even Higgs being close to 126 GeV still survives.  Imposing  flavor bounds further shrinks the  parameter space   considerably: $\tan\beta \lesssim 10$, $m_h > 50$ GeV,   $m_{H^\pm}>300$ GeV, and $m_A > 300$ GeV. 

 Note that in both cases, the extended region in $ \sin(\beta-\alpha)$ is of particular interest, since a deviation of the Higgs coupling to $WW$ and $ZZ$ can be accommodated   for the observed Higgs signal at 126 GeV.

We find that in either of these scenarios, one can identify regions of parameter space that pass all theoretical and experimental bounds and still allow a slightly higher than SM rate to diphotons.   $\gamma\gamma$ and $WW/ZZ$ rates are most likely strongly correlated: $\gamma\gamma : VV  \sim 1$ for the normalized cross sections.     
 
We further studied the implication for the Higgs production via VBF or $VH$ process, and decays  to $bb$, $\tau\tau$ channels.   We found that in the $\h$-126 case, both $VBF/VH \rightarrow \h \rightarrow bb/\tau\tau, WW/ZZ$ could be significantly suppressed in the Higgs signal region. For the  $\H$-126 case, $VBF/VH \rightarrow \H \rightarrow WW/ZZ$ channel is almost the SM strength.  Possible suppression of $bb/\tau\tau$ channel up to 0.5 is possible for the extended signal regions in negative $\sin(\beta-\alpha)$.  Future observation of the $bb$ and $\tau\tau$ modes can provide valuable information for the parameter regions of  the type II 2HDM.
 
Comparing to the MSSM, with its Higgs sector being a restricted type II 2HDM and the tree level Higgs spectrum completely determined by $m_A$ and $\tan\beta$, the parameter regions of the general Type II 2HDM is much more relaxed.  Unlike the MSSM in which the  $\h$-126 case corresponds to the decoupling region where $m_A \gtrsim 300$ GeV, and the $\H$-126 GeV case corresponds to the non-decoupling region where $m_A \sim 100 - 130$ GeV \cite{Christensen:2012ei}, the value of $m_A$ in the general Type II 2HDM could vary over the entire viable region up to about 800 GeV.   The MSSM relation of $m_A \sim m_{H^\pm} \sim m_H$ in the decoupling region is also much more relaxed in the Type II 2HDM.   No obvious correlation is observed between $m_A$,  $m_{H^\pm}$, and  $m_H$ for the $\h$-126 case, except for  the region with large $m_{A, H^\pm} \gtrsim 600$ GeV.  Note also that in the Type II 2HDM with $Z_2$ symmetry (such that $m_{12}=0$) that we are considering, with the additional perturbativity and unitarity constraints imposed, there is an upper limit of about 800 GeV for the mass of $H^0$, $A^0$ and $H^\pm$.  The presence of an upper bound on the heavy Higgs masses reiterates our point that unlike the MSSM, there is no sensible decoupling limit in this case where  only one light SM-like Higgs appears in the low energy spectrum with   other Higgses heavy and decouple. 

Observations of extra Higgses in the future would further pin down the Higgs sector beyond the SM.  While the conventional decay channels of Higgses to SM particles continue to be important channels to search for extra Higgses, novel decay channels of a heavy Higgs into light Higgses or light Higgs plus gauge boson could also appear.  Future work along the lines of collider phenomenology of multiple Higgs scenarios is definitely warranted.  

\begin{acknowledgements}
We thank L. Carpenter for her participation at the beginning of this project.  We would also like to thank David Lopez-Val   for useful discussions   and Oscar St\aa l for sharing the 2HDMC package.  This work was supported by  the Department of Energy under  Grant~DE-FG02-04ER-41298.
\end{acknowledgements}

\end{document}